\documentclass[12pt,letterpaper,a4paper]{article}
\usepackage[includeheadfoot,
marginratio={1:1,2:3},
width=525pt,
height=750pt,]{geometry}

\usepackage{amsmath}
\usepackage{amsfonts}
\usepackage{amssymb}
\usepackage{graphicx}
\usepackage{cancel}
\usepackage{empheq}
\usepackage{color}
\usepackage{hyperref}

\numberwithin{equation}{section}
\usepackage{float}
\restylefloat{table}
\restylefloat{figure}
\usepackage[utf8]{inputenc}
\usepackage{amsmath, amsthm, amssymb, amsfonts}
\usepackage[font=small, labelfont=bf]{caption}
\usepackage{amsmath, amsthm, amssymb, amsfonts}
\usepackage{multirow}
\usepackage{graphicx}
\usepackage{float}
\usepackage[numbers,sort&compress]{natbib}
\usepackage{enumerate}
\usepackage{amsmath}
\usepackage{amsfonts}
\usepackage{amssymb}
\usepackage{graphicx}
\usepackage{cancel}
\usepackage{empheq}
\usepackage{color}
\usepackage{colortbl}
\definecolor{green2}{cmyk}{0, 1, 0.5, 0.3}
\definecolor{green3}{cmyk}{1, 0.75, 1.0, 0.0}

\definecolor{lightgreen}{cmyk}{0.2, 0, 0.2, 0.2}
\definecolor{lightgray}{cmyk}{0.1,0.2,0,0.1}
\definecolor{lightgray2}{cmyk}{0.4,0.4,0,0.8}
\definecolor{black}{cmyk}{1.0,1.0,1.0,1.0}

\usepackage{hyperref}
\usepackage{float}
\restylefloat{table}
\restylefloat{figure}
\usepackage{longtable}
\usepackage{lipsum} 

\usepackage{amsfonts}
\usepackage{amssymb}
\usepackage{comment}
\usepackage{graphicx}
\usepackage{amsmath}
\usepackage{tabu}
\usepackage{dsfont}
\usepackage{float}
\usepackage{amsthm}
\usepackage{slashed}
\usepackage{setspace}
\usepackage[labelformat=simple]{subcaption}

\usepackage{pgfplots}
\usepackage{tikz}
\usepackage{tikz-cd}
\usetikzlibrary{shapes.misc,shadows,fit,decorations.pathmorphing,positioning,trees,decorations.markings,decorations.pathreplacing,calc,shapes,patterns,arrows,positioning,arrows.meta}
\usepackage{xcolor}
\usepackage{pgfplots}
\usepackage{pgfplotstable}
\usepackage{xcolor}
\usepackage{makecell}
\usepackage{diagbox}
\usepackage{environ}
\usepackage[utf8]{inputenc}
\usepackage{amsmath, amsthm, amssymb, amsfonts}
\usepackage{amsmath, amsthm, amssymb, amsfonts}
\usepackage{multirow}
\usepackage{graphicx}
\usepackage{float}
\usepackage[numbers,sort&compress]{natbib}
\usepackage{enumerate}
\usepackage{enumitem}
\usepackage[capitalize,sort]{cleveref}
\usepackage{hhline}
\usepackage{mathtools}
\usepackage{longtable}
\usepackage{makecell}
\usepackage{tabularx}
\usepackage{cellspace}
\usepackage{alphalph}
\usepackage{lscape}
\usepackage{slashed}
\usepackage{setspace}
\usepackage{pifont}

\crefname{figure}{Figure}{Figures}
\crefname{table}{Table}{Tables}

\def\be{\begin{equation}}
\def\ee{\end{equation}}
\def\bea{\begin{eqnarray}}
\def\eea{\end{eqnarray}}
\def\bes{\begin{subequations}}
	\def\ees{\end{subequations}}

\newcommand{\bmat}{\left(\begin{array}}
	\newcommand{\emat}{\end{array}\right)}

\def\ov{\overline}

\def\ov{\overline}
\def\1{{\bf 1}}
\def\2{{\bf 2}}
\def\3{{\bf 3}}
\def\4{{\bf 4}}
\def\6{{\bf 6}}

\newcommand{\nn}{\nonumber}

\newcommand{\beq}{\begin{equation}}
\newcommand{\eeq}{\end{equation}}

\def\ov{\overline}

\allowdisplaybreaks[2]
\numberwithin{equation}{section}

\def\be{\begin{equation}}
\def\ee{\end{equation}}
\def\bea{\begin{eqnarray}}
\def\eea{\end{eqnarray}}
\def\bes{\begin{subequations}}
	\def\ees{\end{subequations}}

\usepackage{multicol}
\usepackage{float}
\usepackage{caption}
\allowdisplaybreaks[2]
\numberwithin{equation}{section}

\begin{document}
	{\hfill
		%
		\hfill
		arXiv:2405.06738}

	\vspace{1.0cm}
	\begin{center}
		{\Large
			Inflating in perturbative LVS: Global Embedding and Robustness}
		\vspace{0.4cm}
	\end{center}

	\vspace{0.35cm}
	\begin{center}
		Swagata Bera$^\dagger{^\ast}$, Dibya Chakraborty$^\ddagger$, George K. Leontaris$^\Diamond$, Pramod Shukla$^\dagger$ \footnote{Email: iswagata78@gmail.com, dibyac@physics.iitm.ac.in, leonta@uoi.gr, pshukla@jcbose.ac.in}
	\end{center}

\vspace{0.1cm}
\begin{center}
{$^\dagger$ Department of Physical Sciences, Bose Institute,\\
Unified Academic Campus, EN 80, Sector V, Bidhannagar, Kolkata 700091, India}\\
\vspace{0.3cm}
{$^\ast$ Department of Integrated Science Education and Research Center,\\
Visva Bharati University, Santiniketan 731235, India}\\
\vspace{0.3cm}
{$^\ddagger$ Centre for Strings, Gravitation and Cosmology, Department of Physics, Indian Institute of Technology Madras,
Chennai 600036, India}\\
\vspace{0.3cm}
{$^\Diamond$ Physics Department, University of Ioannina, University Campus, \\
Ioannina 45110, Greece}
\end{center}
\vspace{1cm}

\abstract{
The perturbative LARGE volume scenario (LVS) is a promising moduli stabilisation scheme in which the overall volume modulus of the compactifying Calabi-Yau (CY) threefold is dynamically stabilised to exponentially large values via using only perturbative corrections. In this article, using an orientifold of a K3-fibred CY threefold, we present the global embedding of an inflationary model proposed in the framework of perturbative LVS, in which the overall volume modulus acts as the inflaton field rolling on a nearly flat potential induced by a combination of the ${\alpha^\prime}^3$-corrections and the so-called {\it log-loop} effects. Given that having a concrete global construction facilitates explicit expressions for a set of sub-leading corrections, as a next step, we present a detailed analysis investigating the robustness of the single-field inflationary model against such corrections, in particular those arising from the winding-type string loop corrections and the higher derivative F$^4$-corrections. 
}

\clearpage

\tableofcontents


\section{Introduction}
\label{sec_intro}

Integrating the scenario of cosmological inflation into effective field theories derived from superstrings is one of the biggest challenges of string model building today. The basic mechanism of inflation is based on the existence of a scalar field -dubbed the inflaton field- rolling towards the minimum of its potential while causing an exponential spatial  expansion of our Universe which follows  just after the Big Bang.  Remarkably,  four-dimensional superstring constructions emerging after compactifying the extra dimensions, predict the existence of (scalar) moduli fields, where eventually some of those could act as natural candidates for the inflaton field.

Such a top-down derived field theory model, however, must be compatible with a series of other observable phenomena. First of all, since we know that the present state of the Universe undergoes an accelerated  expansion, a de-Sitter (dS) vacuum  must be ensured compatible with the present value of the cosmological constant. Furthermore, all moduli fields predicted in a certain (chosen) compactification procedure, should be stabilized at this positive vacuum. During the last two decades, several scenarios have been proposed to implement the cosmic inflation while at the same time remedying  those issues, e.g. see \cite{Cicoli:2023opf,McAllister:2023vgy} and references therein. It is noteworthy that string theory provides sufficient tools (such as  Dp-branes pierced with magnetic fluxes) that can be useful in achieving a physical and natural solution to the above problems.

The first step towards string model building starts with the need to stabilize the moduli. Note  that, while the  IIB perturbative superpotential depends on the axio-dilaton and the complex structure moduli, however, it is independent of the K\"ahler moduli $T_\alpha$ leading to the so-called `no-scale structure', and thus $T_\alpha$ remain undetermined at this stage. Nevertheless, by including non-perturbative K\"ahler moduli dependent contributions to the superpotential \cite{Witten:1996bn,Green:1997di,Blumenhagen:2009qh} and ${\alpha'}$ corrections \cite{Becker:2002nn} to the K\"ahler potential, it was shown that the K\"ahler  moduli can be stabilized, and the robust $N=1$ supersymmetric effective theories with positive vacuum can be engineered which can possibly accommodate all observable data associated with slow roll inflation. In the context of moduli stabilization, the two main and the most successful schemes proposed within the framework of type IIB superstring compactifications, are the so-called KKLT scheme \cite{Kachru:2002sk} and the LARGE Volume Scenario (LVS) \cite{Balasubramanian:2005zx,Conlon:2005ki}. One of the most attractive features in the LVS scheme of moduli stabilization is the fact that it can dynamically stabilise the overall volume ${\cal V}$ of the compactifying CY threefold to exponentially large values via using a combination of the perturbative ${\alpha^\prime}^3$ (BBHL) corrections to the K\"ahler potential ($K$) \cite{Becker:2002nn}, and the non-perturbative corrections to the superpotential ($W$) \cite{Witten:1996bn,Green:1997di,Blumenhagen:2009qh}. Moreover, the underlying CY threefold should  possess a rigid diagonal del-Pezzo divisor to facilitate the so-called `Swiss-cheese' structure in the volume-form, which means that the triple intersection numbers ($k_{\alpha\beta\gamma}$) of the CY needs to be such that ${\cal V} \equiv \frac{1}{3!}k_{\alpha\beta\gamma}t^\alpha t^\beta t^\gamma =\gamma_b \tau_b^{3/2} -\gamma_k \tau_k^{3/2}$, where the four-cycle volume moduli $\tau_\alpha$ and the two-cycle volume moduli $t^\alpha$ are related as $\tau_\alpha=\partial_{t^\alpha}{\cal V}$. We also note that the minimal LVS model is a two-field model realized with a CY threefold having $h^{1,1}_+({\rm CY}) = 2$ which fixes the overall volume ${\cal V}$ and the volume of the rigid four-cycle at leading order volume dependent pieces, and therefore models with $h^{1,1}_+ \geq 3$ have been used to drive inflation in LVS where the third modulus which remain flat in the prescription of the minimal LVS moduli stabilisation can serve as inflaton field rolling down a nearly flat potential induced by sub-leading corrections. This has lead to three classes of inflationary models in LVS, namely Blow-up inflation \cite{Conlon:2005ki,Blanco-Pillado:2009dmu,Cicoli:2017shd}, Fibre inflation \cite{Cicoli:2008gp,Cicoli:2016chb,Cicoli:2016xae,Cicoli:2017axo} and poly-instanton inflation \cite{Cicoli:2011ct,Blumenhagen:2012ue,Gao:2013hn,Gao:2014fva}. In addition, a new class of models, namely the loop-blowup inflation, has been recently proposed in \cite{Bansal:2024uzr} where a blow-up modulus serves as the inflaton field rolling down a potential induced by the string-loop effects..

The main underlying idea behind KKLT and standard LVS is the fact that the K\"ahler moduli, which remain flat due to the so-called `no-scale structure' even after turning-on the background three-form ($F_3/H_3$) fluxes, are stabilised by including non-perturbative superpotential contributions \cite{Witten:1996bn,Green:1997di}. However such non-perturbative corrections may not be generically available in a given concrete construction, as these are very specific to the underlying CY geometry and one needs to ensure several constraints, e.g. Witten's unit arithmetic genus condition \cite{Witten:1996bn} showing that a rigid divisor can be relevant in this regard. In addition, sometimes the inclusion of certain fluxes in a very specifically engineered manner can `rigidify' a non-rigid divisor such that $E3$-instanton or gaugino condensation effects may arise via wrapping with such rigidified divisors \cite{Bianchi:2011qh,Bianchi:2012pn,Louis:2012nb}. Moreover, while attempting to realize chiral visible sector, some charged zero modes can also forbid such non-perturbative corrections to contribute to the holomorphic superpotential \cite{Blumenhagen:2007sm,Blumenhagen:2008zz,Blumenhagen:2009qh,Cvetic:2012ts, Blumenhagen:2012kz}. Therefore, it would be advantageous to have alternative moduli stabilisation schemes, in particular those which could fix all the K\"ahler moduli by perturbative effects. In this regard, perturbative string-loop corrections \cite{Berg:2004ek,vonGersdorff:2005bf, Berg:2005ja, Berg:2005yu, Cicoli:2007xp, Gao:2022uop} or higher derivative F$^4$-corrections \cite{Ciupke:2015msa} can be used to induce volume moduli dependent scalar potential pieces and facilitate moduli stabilisation. In addition to these schemes, K\"ahler moduli can also be stabilised by tree level effects via including non-geometric fluxes, e.g. see \cite{Aldazabal:2006up,deCarlos:2009fq,deCarlos:2009qm, Blumenhagen:2015kja,Shukla:2016xdy,Plauschinn:2020ram,Damian:2023ote}.

However, the aforementioned alternative attempts of K\"ahler moduli stabilization without non-perturbative superpotential contributions do not lead to large volume minimisation in the sense of dynamically realising an exponential large VEV of the overall volume modulus -- the way it has been beautifully incorporated in the standard LVS. In this regard an interesting alternative proposal has been made in \cite{Antoniadis:2018hqy,Antoniadis:2019rkh,Antoniadis:2020ryh} where it was shown that for a geometric configuration of three  sets of mutually intersecting space-filling D7 brane stacks and three K\"ahler moduli, stabilisation with large $\langle {\cal V} \rangle$ can be attainable even in the absence of non-perturbative corrections  to the superpotential, provided  that a recently proposed novel source of perturbative one-loop corrections to the K\"ahler potential  is taken into account. More specifically, the origin of these perturbative corrections are due to a ten-dimensional string action  augmented by an $R^4$ term, the latter being  the result of higher derivative couplings generated  by multi-graviton scattering associated with higher derivative terms in string theory. Subsequent dimensional reduction of the string action to four dimensions gives rise to a novel localised Einstein-Hilbert (EH)  term ${\cal R}_{\mathfrak{b}}$ in the bulk with a multiplicative coefficient proportional to the Euler characteristic of the compactification manifold. Then, in the presence of three  D7-brane stacks KK-graviton modes propagating in the 2-dimensional space transverse to D7-stacks  and towards the ${\cal R}_{\mathfrak{b}}$-vertex  give rise to K\"ahler moduli-dependent  logarithmic corrections incorporated by an appropriate  shift of the internal volume.  This setup  proves to be sufficient to stabilize the three K\"ahler moduli. In addition, the dS vacuum is obtained by virtue  of   positive D-term contributions,  associated with the universal $U(1)$ factors of the intersecting D7-brane stacks. Moreover, the possibility of implementing a successful cosmological inflationary scenario in the above framework of perturbative LVS has been initiated in a series of papers \cite{Antoniadis:2018ngr,Antoniadis:2019doc,Antoniadis:2020stf,Antoniadis:2021lhi}.

This toroidal based proposal has been further implemented in a concrete type IIB model using a K3-fibred CY orientifold in \cite{Leontaris:2022rzj} as a global embedding of the perturbative LVS. It has been shown that one can have $\langle {\cal V} \rangle \propto e^{c_1/g_s^2}$ where $g_s$ is the string coupling and $c_1$ is some positive order one constant. This exponential behavior is similar to the standard LVS, and the two schemes are mainly distinguished by the fact that standard LVS utilises a combination of BBHL corrections to the K\"ahler potential and the non-perturbative corrections to the superpotential while the perturbative LVS uses perturbative string-loop effects of {\it log-loop} type along with the BBHL-correction. Extending the global embedding program initiated in \cite{Leontaris:2022rzj}, we plan to study the inflationary aspects in the current work. For that purpose, we aim to investigate along the following three points:

\begin{itemize}
\item
First we revisit the inflationary proposal of \cite{Antoniadis:2019doc,Antoniadis:2020stf} to detail the various insights of the single-field inflation. After invoking a shift in the canonical modulus corresponding to the overall volume ${\cal V}$ which serves as inflaton, we find that it acquires an effective scalar potential whose shape is controlled by a single parameter $x$, whilst there is an overall factor which depends on the string coupling $g_s$ and the superpotential $|\langle W_0 \rangle|$. A numerical investigation follows and a benchmark model is presented with a sufficiently flat plateau to realize the inflationary scenario with adequate number of e-foldings and predict successfully all other cosmological observables.

One of the main goals of this revisit was to set the constraints under which the single-field approximation remains valid, in the sense that there is a mass-hierarchy between the inflaton modulus and the remaining two moduli which are stabilized by leading order D-term effects. Focusing on the region of the parameter space where the internal volume ${\cal V}$ is the lightest modulus, it is shown that a  single field inflation with ${\cal V}$ being the inflaton field is an admissible and successful  scenario.

\item
Secondly, we embed the inflationary scenario \cite{Antoniadis:2020stf} in an explicit K3-fibred CY orientifold with properties close to those of the toroidal case used to stabilise the K\"ahler moduli and build the inflationary potential. To this end, scanning the Kreuzer-Skarke database of reflexive polytopes a CY threefold possessing three K\"ahler moduli $T_\alpha$ and a toroidal-like volume of the form ${\cal V}\propto \sqrt{\tau_1\tau_2\tau_3}$ has been identified \cite{Leontaris:2022rzj}. Using this CY threefold, we extend our previous investigations,  and explore the possibility of realising a similar inflationary potential as proposed in \cite{Antoniadis:2019doc,Antoniadis:2020stf}.

\item
Finally, after having the global model with concrete CY example we examine the robustness of the stabilisation procedure and the cosmic inflationary scenario against various possible quantum corrections. As we will demonstrate, the intersections of D7-brane stacks taking place on non-shrinkable two-tori, result to string-loop effects of the winding-type \cite{Berg:2004ek,vonGersdorff:2005bf, Berg:2005ja, Berg:2005yu, Cicoli:2007xp} while the implications of corrections stemming from higher derivative F$^4$ terms \cite{Ciupke:2015msa} are also investigated.  Both are relatively suppressed to the previously included ones and their role is examined in detail for moduli stabilisation as well as for the inflationary scenario.

\end{itemize}
So the simple plan is the following: first we analyse the minimal formulation of the inflationary potential $V_{\rm inf}$ in a global CY orientifold model to revisit the three-field moduli stabilisation and ${\cal V}$ driven inflation of \cite{Antoniadis:2019doc,Antoniadis:2020stf}. Subsequently we study the inflationary dynamics through a detailed analysis using the full potential $V = V_{\rm inf} + V_{\rm corr}$ with the inclusion of a set of corrections encoded in $V_{\rm corr}$ which are explicitly known in our global construction. We also quantify the range of parameters controlling these sub-leading corrections under which the inflationary model remains robust and viable.

The paper is organised as follows: In section \ref{sec_preliminaries}, we review the relevant preliminaries about moduli stabilisation in the framework of the LARGE Volume Scenario (LVS), including both the schemes, namely the {\it standard LVS} and the {\it perturbative LVS}.  
In section \ref{sec_revisit} we investigate the inflationary dynamics via computing the single field potential from our generic master formula, giving its global embedding with details on orientifold and brane-setting, and subsequently collecting all the (known) sub-leading corrections which can possibly affect the inflationary dynamics. In section \ref{sec_robustness} we examine the robustness of the inflationary dynamics against the various types of corrections and provide several benchmark models. Finally, in section \ref{sec_conclusions} we present our conclusions and discuss the future directions.


\section{LARGE volume scenarios (LVS): a brief review}
\label{sec_preliminaries}


The low energy dynamics of the four-dimensional effective supergravity theory arising from the type IIB superstring compactifications on CY orientifolds can be captured by a holomorphic superpotential ($W$) and a real K\"ahler potential ($K$) and the gauge kinetic function $(g)$. These quantities depend on the various chiral coordinates obtained by complexifying  various moduli with a set of RR axions. Let us start by fixing the conventions. We will be using the following definitions of such chiral variables:
\bea
\label{eq:chiral-variables}
& & U^i = v^i - i\, u^i, \qquad S = c_0 + i\, s, \qquad T_\alpha = c_\alpha - i\, \tau_\alpha~,
\eea
where $s$ is the dilaton dependent modulus, $u^i$'s are the complex structure saxions, and $\tau_\alpha$'s are the Einstein frame four-cycle volume moduli defined as $\tau_\alpha = \partial_{t^\alpha} {\cal V} = \frac12 k_{\alpha\beta\gamma} t^\beta t^\gamma$. In addition, the $c_0$ and $c_\alpha$'s are universal RR axion and RR four-form axions, respectively, while the complex structure axions are denoted by $v^i$. Here the indices $\{i, \alpha\}$ are such that $i \in h^{2,1}_-({\rm CY}/{\cal O})$ while $\alpha \in h^{1,1}_+({\rm CY}/{\cal O})$. Moreover, we assume that $h^{1,1} = h^{1,1}_+$ for simplicity, and hence, the so-called odd-moduli $G^a$  are not present in our analysis; we refer interested readers to \cite{Cicoli:2021tzt}.

The F-term contributions to the scalar potential are computed using the following well known formula,
\be
\label{eq:V_gen}
e^{- {K}} \, V = {K}^{{\cal A} \ov {\cal B}} \, (D_{\cal A} W) \, (D_{\ov {\cal B}} \ov{W}) -3 |W|^2 \equiv V_{\rm cs} + V_{\rm k}\,,
\ee
where:
\be
\label{eq:VcsVk}
V_{\rm cs} =  K_{\rm cs}^{i \ov {j}} \, (D_i W) \, (D_{\ov {j}} \ov{W}) \qquad \text{and}\qquad V_{\rm k} =  K^{{A} \ov {B}} \, (D_{A} W) \, (D_{\ov {B}} \ov{W}) -3 |W|^2\,.
\ee
Moduli stabilisation in 4D type IIB effective supergravity models follows a two-step strategy. First, one fixes the complex structure moduli $U^i$ and the axio-dilaton $S$ by the leading order flux superpotential $W_{\rm flux}$ induced by usual S-dual pair of the 3-form fluxes $(F_3, H_3)$ \cite{Gukov:1999ya}. This demands solving the following supersymmetric flatness conditions:
\bea
&& D_i W_{\rm flux} = 0 = D_{\ov {i}} \ov{W}_{\rm flux}, \qquad D_{S} W_{\rm flux} = 0 = D_{\ov {S}} \ov{W}_{\rm flux}.
\label{UStab}
\eea
After supersymmetric stabilization of axio-dilaton and the complex structure moduli, one has $\langle W_{\rm flux} \rangle = W_0$. At this leading order no-scale structure protects the K\"ahler moduli $T_\alpha$ which subsequently remain flat, and as a second step, they can be stabilized via including other sub-leading contributions to the scalar potential, e.g. those induced via the non-perturbative corrections in the holomorphic superpotential $W$ or the other (non-)perturbative corrections arising from the whole series of $\alpha^\prime$ and string-loop ($g_s$) corrections. 


\subsection{Standard LVS}
The LVS scheme of moduli stabilization considers a combination of perturbative $(\alpha^\prime)^3$ corrections to the K\"ahler potential ($K$) and a non-perturbative contribution to the superpotential $W$ which can be generated by using rigid divisors, such as shrinkable del-Pezzo 4-cycles, or by rigidifying non-rigid divisors using magnetic fluxes \cite{Bianchi:2011qh, Bianchi:2012pn, Louis:2012nb}. The minimal LVS construction includes two K\"ahler moduli corresponding to a so-called Swiss-cheese like volume form of the CY threefold given as\footnote{Ref. \cite{AbdusSalam:2020ywo} has shown that LVS moduli fixing can be realized also for generic cases where the CY threefold does not have a Swiss-cheese structure.}:
\be
{\cal V} = \frac{k_{bbb}}{6} \, (t^b)^3 + \frac{k_{sss}}{6} \, (t^s)^3 \, ,
\ee
where $k_{\alpha\beta\gamma}$ denotes the triple intersection number on the CY threefold, and the 2-cycle volume moduli $t^{\alpha}$ are related to the 4-cycle volume moduli $\tau_\alpha$ via $\tau_\alpha = \partial_{t^\alpha} {\cal V}$. Subsequently one has the following Swiss-cheese like volume form\footnote{Given that the CY threefold has a Swiss-cheese form, one can always find a basis of divisors such that the only non-vanishing intersection numbers are $k_{bbb}$ and $k_{sss}$, which leads to the relation $t^s = - \sqrt{2\tau_s/k_{sss}}$. Here, the minus sign is dictated from the K\"ahler-cone conditions because the so-called `small' divisor $D_s$ in this Swiss-cheese CY is an exceptional 4-cycle.}:
\be
{\cal V} =  \gamma_b \, \, \tau_{b}^{3/2} - \gamma_s\, \, \tau_{s}^{3/2} \, ,
\ee
where $\gamma_b$ and $\gamma_s$ are determined through the triple intersection numbers $k_{bbb}$ and $k_{sss}$. The K\"ahler potential including $\alpha'^3$ corrections takes the form \cite{Becker:2002nn}:
\be
\label{eq:K}
K = -\ln\left[-i\int \Omega\wedge\bar{\Omega}\right]-\ln\left[-\,i\,(S-\bar{S})\right]-2\ln\left[{\cal V}\,+\frac{\xi}{2}\left(\frac{S-\bar{S}}{2i}\right)^{3/2}\right], \nonumber
\ee
where $\Omega$ denotes the nowhere vanishing holomorphic 3-form which depends on the complex-structure moduli, while the CY volume ${\cal V}$ receives a shift through the $\alpha'^3$ corrections encoded in the parameter $\xi=-\frac{\chi(X)\,\zeta(3)}{2\,(2\pi)^3}$, where $\chi(X)$ is the CY Euler characteristic and $\zeta(3)\simeq 1.202$.

Furthermore, the presence of a `diagonal' del-Pezzo divisor corresponding to the so-called `small' $4$-cycle of the CY threefold induces the superpotential with a non-perturbative effect of the following form:
\be
W= W_0 + A_s\, e^{- i\, a_s\, T_s}\,,
\label{eq:Wnp-n}
\ee
where after fixing $S$ and the $U$-moduli, the flux superpotential can effectively be considered as constant: $W_0=\langle W_{\rm flux}\rangle$. In addition, the pre-factor $A_s$ can generically depend on the complex-structure moduli which after the first-step of the supersymmetric moduli stabilisation can be considered as a parameter. Moreover, without any loss of generality, we consider $W_0$ and $A_s$ to be a real quantity. Subsequently the leading order pieces in the large volume expansion are collected in three types of terms \cite{Balasubramanian:2005zx}:
\be
V \simeq \frac{\beta_{\alpha'}}{{\cal V}^3} + \beta_{\rm np1}\,\frac{\tau_s}{{\cal V}^2}\, e^{- a_s \tau_s} \cos\left(a_s \,c_s\right) \\
+ \beta_{\rm np2}\,\frac{ \sqrt{\tau_s}}{{\cal V}}\, e^{-2 a_s \tau_s},
\label{VlvsSimpl}
\ee
with:
\be
\beta_{\alpha'} = \frac{3 \,\kappa\,\hat\xi |W_0|^2}{4}\,, \quad \beta_{\rm np1} = 4 \, \kappa\, a_s |W_0| |A_s|\,, \quad \beta_{\rm np2} = 4 \, \kappa\, a_s^2 |A_s|^2 \sqrt{2 k_{sss}}, \quad \kappa = \frac{g_s}{8\pi}\,.
\ee
The minimal LVS scheme of moduli stabilisation fixes the CY volume ${\cal V}$ along with a small modulus $\tau_s$ controlling the volume of an exceptional del Pezzo divisor. Therefore any LVS models with 3 or more K\"ahler moduli, $h^{1,1}\geq 3$, can generically have flat directions at leading order. These flat directions are promising inflaton candidates with a potential generated at sub-leading order. Based on the geometric nature of the inflaton field and the source of inflaton potential, there are three popular inflationary models based on LVS mechanism to fix the overall volume of the internal CY threefold.


\subsection{Perturbative LVS}

With the inclusion of 1-loop effects-- also known as the log-loop corrections -- to the K\"ahler potential on top of the BBHL corrections used in the standard LVS, one arrives at an effectively modified overall volume ${\cal V}$ which we denote as ${\cal Y}$. It takes the following explicit form,
\bea
& & {\cal Y} = {\cal Y}_0 + {\cal Y} _1, 
\eea
where ${\cal Y}_0$ denotes the overall volume modified by $\alpha^\prime$ corrections appearing at string tree-level while ${\cal Y}_1$ is induced at string 1-loop level as given  below  \cite{Antoniadis:2018hqy,Antoniadis:2018ngr,Antoniadis:2019doc,Antoniadis:2019rkh,Antoniadis:2020ryh,Antoniadis:2020stf},
\bea
& & {\cal Y}_0 = {\cal V} +  \frac{\xi}{2} \, e^{-\frac{3}{2} \phi} = {\cal V} + \frac{\xi}{2}\, \left(\frac{S-\ov{S}}{2\,{\rm i}}\right)^{3/2} \,, \\
& & {\cal Y}_1 = e^{\frac{1}{2} \phi}\, f({\cal V}) = \left(\frac{S-\ov{S}}{2\,{\rm i}}\right)^{-1/2} \left(\sigma + \eta \, \ln{\cal V}\right)\,.\nonumber
\label{eq:defY}
\eea
Here one has the following correlations among the various coefficients, $\xi$, $\sigma$ and $\eta$,
\bea
\label{eq:def-xi-eta}
& & \hskip-1cm  \xi = - \frac{\chi({\rm CY})\, \zeta[3]}{2(2\pi)^3}~, \qquad \sigma  = - \frac{\chi({\rm CY})\, \zeta[2]}{2(2\pi)^3} = - \, \eta, \qquad \frac{\xi}{\eta} = -\frac{\zeta[3]}{\zeta[2]} \\
& & \hskip-1cm \hat\xi = \frac{\xi}{g_s^{3/2}}~, \qquad \hat\eta = g_s^{1/2}\, \eta~, \qquad \qquad \frac{\hat\xi}{\hat\eta} = -\frac{\zeta[3]}{\zeta[2]\,g_s^2}~. \nonumber
\eea
The K\"ahler derivatives can be subsequently found to take the following form,
\bea
\label{eq:derK}
& & K_S = \frac{{\rm i}}{2\, s\, {\cal Y}} \left({\cal V} + 2 \,\hat\xi \right) = - K_{\ov S}, \quad K_{T_\alpha} = - \frac{{\rm i} \, t^\alpha}{2 \, {\cal Y}} \left(1+ \frac{\partial {\cal Y}_1}{\partial {\cal V}} \right) = - K_{\ov{T}_\alpha}
\eea
Further, it turns out that the K\"ahler metric and its inverse generically admit the following explicit forms,
\bea
\label{eq:simpinvKij-1}
& & \hskip-1cm K_{S \ov{S}} = {\cal P}_1, \qquad K_{T_\alpha\, \ov{S}} = t^\alpha\, {\cal P}_2 = K_{S\,\ov{T}_\alpha}, \qquad  K_{T_\alpha \, \ov{T}_\beta} = (t^\alpha \, t^\beta)\, {\cal P}_3 \, - k^{\alpha\beta}\, {\cal P}_4~, \\
& & \hskip-1cm  K^{S \ov{S}} = \tilde{\cal P}_1, \qquad K^{T_\alpha\, \ov{S}} = k_\alpha\, \tilde{\cal P}_2 = K^{S\,\ov{T}_\alpha}, \qquad  K^{T_\alpha \, \ov{T}_\beta} = (k_\alpha \, k_\beta)\, \tilde{\cal P}_3 \, - k_{\alpha\beta}\, \tilde{\cal P}_4~, \nonumber
\eea
where the two sets of functions ${\cal P}_i$ and $\tilde{\cal P}_i$'s are
\bea
\label{eq:Pis}
& & {\cal P}_1 = \frac{1}{8\,s^2\, {\cal Y}^2}\, \left({\cal V} ({\cal Y} + {\cal V}) - 4 \hat\xi ({\cal Y} - {\cal V}) + 4 \hat\xi^2\right), \\
& & {\cal P}_2 = -\frac{1}{8\, s\, {\cal Y}^2} \left(\frac{3}{2}\, \hat\xi - s^{-\frac{1}{2}}\, (\sigma+\eta \ln {\cal V}) + \, s^{-\frac{1}{2}} \, ({\cal V} + 2\hat\xi) \frac{\eta}{{\cal V}} \right), \nonumber\\
& & {\cal P}_3 = \frac{1}{8\, {\cal Y}^2} \left(1 + s^{-\frac{1}{2}} \, \frac{\eta}{{\cal V}} +  {\cal Y} \,  s^{-\frac{1}{2}} \, \frac{\eta}{{\cal V}^2}\right), \nonumber\\
& & {\cal P}_4 = \frac{1}{4\, {\cal Y}} \left(1 + s^{-\frac{1}{2}} \, \frac{\eta}{{\cal V}} \right)~, \nonumber
\eea
and
\bea
\label{eq:tildePis}
& & \tilde{\cal P}_1 = \frac{{\cal P}_4-6 {\cal P}_3 {\cal V}}{{\cal P}_1 {\cal P}_4 + 6 {\cal P}_2^2 {\cal V}-6 {\cal P}_1 {\cal P}_3 {\cal V}}\,, \quad \tilde{\cal P}_2 = \frac{{\cal P}_2}{{\cal P}_1 {\cal P}_4 + 6 {\cal P}_2^2 {\cal V}-6 {\cal P}_1 {\cal P}_3 {\cal V}}\,, \\
& & \tilde{\cal P}_3 = \frac{{\cal P}_2^2-{\cal P}_1 {\cal P}_3}{{\cal P}_4\left({\cal P}_1 {\cal P}_4 + 6 {\cal P}_2^2 {\cal V}-6 {\cal P}_1 {\cal P}_3 {\cal V}\right)}\,, \quad \tilde{\cal P}_4 = ({\cal P}_4)^{-1}. \nonumber
\eea
This subsequently leads to the following form of the scalar potential,
\bea
\label{eq:masterV}
&&\hskip-1cm  V_{\alpha^\prime + {\rm log} \, g_s} = \frac{\kappa}{{\cal Y}^2} \biggl[\frac{3\, {\cal V} }{2\, {\cal Y}^2} \left(1 + \frac{\partial {\cal Y}_1}{\partial{\cal V}}\right)^2 \left(6 {\cal V} \tilde{\cal P}_3 - \tilde{\cal P}_4 \right) - 3 \biggr]\, |W_0|^2 = V_{\alpha^\prime +{\rm log} \, g_s}^{(1)} + \cdots~,
\eea
where we have set $e^{K_{cs}} =1$ and $\kappa = \left(\frac{g_s}{8 \pi}\right)$. Whilst by considering $\sigma = - \eta$ or equivalently $\hat\sigma = - \hat\eta$ as mentioned in Eq.~(\ref{eq:def-xi-eta}) we have the following pieces at leading order,
\bea
\label{eq:pheno-potV2}
& & V_{\alpha^\prime +{\rm log} \, g_s}^{(1)} = \frac{3\, \kappa\, \hat\xi}{4\, {\cal V}^3}\, |W_0|^2 + \frac{3 \, \kappa\, \hat\eta\, (\ln{\cal V} - 2)}{2{\cal V}^3}\,|W_0|^2.
\eea
This scalar potential results in an exponentially large VEV for the overall volume determined by the following approximate relation:
\bea
\label{eq:pert-LVS}
& & \langle {\cal V} \rangle \simeq e^{-\frac{\hat\xi}{2\, \hat\eta} + \frac{7}{3}} = e^{a/g_s^2 + b}, \qquad a = \frac{\zeta[3]}{2 \zeta[2]} \simeq 0.365381, \quad b = \frac73~\cdot
\eea
For a numerical estimate we note that using $g_s = 0.2$ in Eq.~(\ref{eq:pert-LVS}) results in $\langle {\cal V} \rangle = 95593.3$ while $g_s = 0.1$ leads to $\langle {\cal V} \rangle = 7.61463 \cdot 10^{16}$. Given that an exponentially large VEV of the overall volume $\cal V$ is obtained by using only the perturbative effects, this scheme is refereed as ``perturbative LVS". Moreover, similar to the standard LVS case, this solution also corresponds to an AdS minimum.


\subsection{On dS vacua in Perturbative LVS}

In addition to the anti-D3 uplifting \cite{Kachru:2002sk,Crino:2020qwk, Bento:2021nbb}, there can be various other ways of inducing an uplifting term which can result in de-Sitter solution. In this regard we consider the $D$-term potential associated with the anomalous $U(1)$'s living on the stack of $D7$-branes wrapping the $O7$-planes (say corresponding to divisor class $D_h$), which can be expressed as below \cite{Burgess:2003ic},
\bea
\label{eq:}
& & V_{D} = \frac{1}{2\, Re(f_{D7})} \left(\sum_i q_{\varphi_i} \frac{|\varphi_i|^2}{Re(S)} - \xi_{h}\right)^2.
\eea
Here $f_{D7} = T_h/(2\pi)$ denotes the holomorphic gauge kinetic function expressed in terms of complexified four-cycle volume of the divisor $D_h$, and $q_{\varphi_i}$ denotes the $U(1)$ charge corresponding to the matter field $\varphi_i$. These may correspond to, for example, the deformation of divisors wrapped by the respective $D7$-brane stacks and hence can be counted by $h^{2,0}(D)$ of a suitable divisor of the CY threefold. The FI-parameters $\xi_{h}$ are defined as,
\bea
\label{eq:FI-parameter}
& & \xi_{h} = \frac{1}{4\pi\,{\cal V}} \int_{D_h} {\cal F} \wedge J = \frac{1}{2\pi} \sum_\alpha \frac{q_{h\alpha}}{2} \frac{t^\alpha}{{\cal V}} = - \frac{i}{2\pi} \sum_\alpha q_{h\alpha} \frac{\partial K}{\partial T_\alpha},
\eea
where in the last equality the K\"ahler derivatives have been introduced using Eq.~(\ref{eq:derK}). Moreover, ${\cal F}$ denotes the gauge flux that is turned on the Divisor class $D_h$, and $J$ is the K\"ahler form expressed as $J = t^1 D_1 + t^2 D_2 + t^3 D_3$. The $U(1)$ charge corresponding to the closed string modulus $T_\alpha$ is denoted as $q_{h\alpha}$ and can be given as the following,
\bea
& & q_{h\alpha} = \frac{1}{l_s^4}\int_{D_h} \hat{D}_\alpha \wedge {\cal F},
\eea
where $l_s$ denoted the string length, parametrised as: $l_s = 2\pi \sqrt{\alpha^\prime}$.

To implement the de-Sitter solution we present two scenarios; one in which we introduce $D$-term uplifting via Fayet-Iliopoulos (FI) term assuming that matter fields receive vanishing VEVs and the second one being a scenario of the so-called $T$-brane uplifting  in which matter field have non-zero VEVs \cite{Cicoli:2015ylx}. Both of these scenarios present a different volume scaling in the scalar potential term inducing an uplift of the AdS, as discussed above.

\subsubsection*{Scenario 1: $D$-term uplifting via matter fields of vanishing VEVs}
Assuming that the matter fields receive vanishing VEVs along with each one of the $D7$-brane stack being appropriately magnetized by suitable gauge fluxes so as to  generate a moduli-dependent FI term, one can generically have the following $D$-term contributions to the scalar potential,
\bea
\label{eq:VDup0}
& & V_{\rm D} 
= \sum_{\alpha =1}^3 \frac{{d}_\alpha}{\tau_\alpha^{3}}~,
\eea
where ${d}_\alpha$'s are some uplifting parameters. This piece can facilitate an uplifting of the AdS vacua realized in the perturbative LVS. For this case, below we present a set of numerical parameters and the respective moduli VEVs corresponding to a nearly Minkowskian minimum \cite{Leontaris:2022rzj}:
\bea
& & g_s = 0.2, \quad \hat\xi = 6.06818, \quad \hat\eta = -0.332156, \quad d = 1.24711\cdot 10^{-8}, \, \\
& & \langle t^\alpha \rangle = 48.0191 \, \quad \forall \alpha, \quad \langle {\cal V} \rangle = 221447.96, \quad \langle V \rangle = 1.54074\cdot10^{-31}, \nonumber\\
& & {\rm Eigen}(V_{ij}) = \{5.04286\cdot 10^{-22},\, 5.04286\cdot 10^{-22}, \, 5.04286\cdot 10^{-22}\}\,. \nonumber
\eea

\subsubsection*{Scenario 2: $T$-brane uplifting via matter fields of non-vanishing VEVs}
In the presence of non-zero gauge flux on the hidden sector D7-branes, a non-vanishing FI term can be induced leading to the so-called $T$-brane configuration. It has been shown in \cite{Cicoli:2015ylx} that after expanding the D7-brane action around such $T$-brane background with three-form supersymmetry breaking fluxes, one can get a positive definite uplifting piece to the scalar potential.
Considering such an $T$-brane uplifting case, the matter fields $\varphi$ receive VEVs of the following kind \cite{Cicoli:2015ylx,Cicoli:2017shd,Cicoli:2021dhg},
\bea
& & |\varphi|^2 \simeq \frac{c_\varphi}{{\cal V}^{2/3}},
\eea
where $c_\varphi$ is a model dependent  quantity which involves $U(1)$ charges corresponding to the matter fields. This subsequently leads to an uplifting term to the scalar potential induced as a hidden sector supersymmetry breaking $F$-term contribution, achieved through the $D$-term stabilisation  of the matter fields. Subsequently  the soft-term arising as an $F$-term effect can be given as \cite{Cicoli:2015ylx,Cicoli:2017shd,Cicoli:2021dhg},
\bea
\label{eq:T-brane-pot}
& & \hskip-1cm V_{\rm T} = m_{3/2}^2 |\varphi|^2 = \frac{\kappa\, {\cal C}_{up}\, |W_0|^2}{{\cal V}^{8/3}} \geq 0,
\eea
where $m_{3/2}$ denotes the gravitino mass, and ${\cal C}_{up}$ denotes a model dependent coefficient which also involves the $U(1)$ charges corresponding to the matter fields. One can subsequently use this positive semidefinite piece to uplift the AdS solution of the perturbative LVS to a de-Sitter minimum.

For the case of $T$-brane uplifting piece $V_{\rm up} = V_{\rm T}$ as given in Eq.~(\ref{eq:T-brane-pot}), one obtains the following dS minimum with isotropic VEVs for all the three K\"ahler moduli \cite{Leontaris:2022rzj}:
\bea
& & g_s = 0.3, \quad \hat\xi = 3.3031, \quad \hat\eta = -0.406806, \quad {\cal C}_{up} = 0.0814039, \, \\
& & \langle t^\alpha \rangle = 19.5862 \, \quad \forall \alpha, \quad \langle {\cal V} \rangle = 15027.3, \quad \langle V \rangle = 3.74709\cdot10^{-30} \nonumber\\
& & {\rm Eigen}(V_{ij}) = \{6.81793\cdot 10^{-18},\, 4.68145\cdot 10^{-19}, \, 4.68145\cdot 10^{-19}\}. \nonumber
\eea
We note that the typical values for the uplifting parameter can be  around ${\cal C}_{up} \simeq {\cal O}(0.1)$, e.g. see \cite{Cicoli:2017shd,Cicoli:2021dhg}, and for that reason we consider $g_s = 0.3$, given that smaller values of $g_s$ would demand a smaller tuned value for ${\cal C}_{up}$.


\section{Inflating in perturbative LVS}
\label{sec_revisit}


In this section we present a concrete global embedding of the perturbative LVS models and the inflationary proposal as presented in a series of papers \cite{Antoniadis:2018hqy,Antoniadis:2018ngr,Antoniadis:2019doc,Antoniadis:2019rkh,Antoniadis:2020ryh,Antoniadis:2020stf, Antoniadis:2021lhi}. For that purpose we start by presenting the inflationary proposal of \cite{Antoniadis:2018ngr,Antoniadis:2020stf} in  light of model dependent parameters relevant for getting the various cosmological observables. Subsequently we will present an explicit CY orientifold examples with very similar properties to those of the toroidal case, used to build the inflationary proposal of \cite{Antoniadis:2020stf}, and we will derive the inflationary potential as a leading order effects of the generic scalar potential induced in this concrete model.


\subsection{Moduli stabilisation and mass hierarchy}

In the context of perturbative LVS, an inflationary model driven by the overall volume (${\cal V}$) of the compactifying sixfold has been proposed in \cite{Antoniadis:2020stf}. The underlying idea is to begin with GKP \cite{Giddings:2001yu} type of solution assuming that the complex-structure moduli as well as the axio-dilaton are supersymmetrically stabilized via the standard GVW flux superpotential \cite{Gukov:1999ya}. Subsequently, one can {\it self-consistently} stabilize the three K\"ahler moduli (corresponding to a toroidal-like model) using sub-leading scalar potential corrections sourced form the {\it log-loop} effects in the K\"ahler potential and the D-term contributions, which appear at order of ${\cal V}^{-2}$ in volume scaling. However the flux dependent coefficients can be argued to be tuned so that to remain in the GKP solutions realised earlier \cite{Giddings:2001yu}. Finally one has the following scalar potential for the three K\"ahler moduli at the leading order,
\bea
\label{eq:V-task1-0}
& & V_{0} =  {\cal C}_1 \left(\frac{\hat\xi - 4\,\hat\eta + 2\,\hat\eta \, \ln{\cal V}}{{\cal V}^3}\right) + \sum_{\alpha =1}^3 \frac{\hat{d}_\alpha}{\tau_\alpha^{3}},
\eea
where ${\cal C}_1 = \frac{3}{4}\,\kappa\,|W_0|^2,$ as seen from Eq.~(\ref{eq:pheno-potV2}). The perturbative LVS \cite{Antoniadis:2018hqy} has been proposed in a toroidal based model with the following volume-form written in terms of the two-cycle and the four-cycle volumes,
\bea
\label{eq:volume-form}
& & {\cal V} = n_0\, \, \, t^1\, t^2\, t^3 = \frac{1}{\sqrt{n_0}}\, \sqrt{\tau_1\, \tau_2\, \tau_3}, \quad {\rm where} \quad \tau_\alpha = \partial_{t^\alpha}{\cal V}.
\eea
Here we have introduced a parameter $n_0$ which is the triple intersection number of the CY threefold for which the overall volume takes the above toroidal-like form. Explicit examples of such volume forms have been presented for K3-fibred CY threefolds in \cite{Gao:2013pra,Leontaris:2022rzj}. Subsequently, considering the tree-level K\"ahler metric arising from this volume form, one obtains the following canonical normalized fields $\varphi^\alpha$ corresponding to the $4$-cycle volume moduli $\{\tau_1, \tau_2, \tau_3\}$,
\bea
\label{eq:cononical-varphi0}
& & \varphi^\alpha = \frac{1}{\sqrt{2}} \ln \tau_\alpha, \qquad \forall \, \alpha \in \{1, 2, 3\}.
\eea
Following the conventions of \cite{Antoniadis:2020stf},
in order to investigate inflationary aspects we now define another canonical basis $\phi^\alpha$, given as below
\bea
\label{eq:cononical-varphi1}
& & \phi^1 = \frac{1}{\sqrt{3}} \left(\varphi^1+ \varphi^2 + \varphi^3 \right) = \sqrt{\frac{2}{3}} \ln(\sqrt{n_0}\,{\cal V}) , \\
& & \phi^2 = \frac{1}{\sqrt{2}} \left(\varphi^1- \varphi^2 \right), \nonumber\\
& & \phi^3 = \frac{1}{\sqrt{6}} \left(\varphi^1+ \varphi^2  -2 \varphi^3 \right). \nonumber
\eea
In this basis the field $\phi^1$ is entirely aligned along the overall volume ${\cal V}$ of the CY threefold. Using these canonical fields ($\phi^i$) the scalar potential (\ref{eq:V-task1-0}) can be rewritten in the following form,
\bea
\label{eq:V-task1-1}
& & V_{0} = e^{-\sqrt{6}\phi^1} \left(d_1 e^{-3 \, \phi^2 - \sqrt{3} \, \phi^3} + d_2 \, e^{3 \, \phi^2 - \sqrt{3} \, \phi^3} + d_3 \, e^{2 \sqrt{3}\, \phi^3}\right) \\
& & \hskip1.5cm + 2\, \hat\eta\, e^{-3\sqrt{\frac{3}{2}} \phi^1}\, n_0^{3/2} \, {\cal C}_1\, \left(\frac{\hat\xi}{2\hat\eta} + \sqrt{\frac{3}{2}} \, \phi^1 - 2 - \frac{1}{2}\ln n_0 \right). \nonumber
\eea
Let us note that the extremization conditions for the potential (\ref{eq:V-task1-1}) can be reshuffled such that the VEVs of the three canonical moduli $\langle \phi^\alpha \rangle$ at the respective extrema of the uplifted potential can be generically determined by the following relations:
\bea
\label{eq:dSextrema}
& & \hskip-1.5cm a_1 = \, e^{-\sqrt{\frac{3}{2}} \langle\phi^1\rangle} \left(\sqrt{\frac{3}{2}} \, \langle \phi^1 \rangle - a_2\right), \quad \langle \phi^2 \rangle = \frac{1}{6} \ln\left(\frac{d_1}{d_2}\right), \quad \langle \phi^3 \rangle = \frac{1}{6 \sqrt{3}} \ln\left(\frac{d_1d_2}{d_3^2}\right)~,
\eea
where
\bea
\label{eq:a1-a2}
& & \hskip-1cm a_1 \equiv - \, \frac{(d_1\, d_2\, d_3)^{1/3}}{n_0^{3/2}\, \hat\eta \, {\cal C}_1} \, \geq \,0, \qquad a_2 = -\frac{\hat\xi}{2\,\hat\eta} + \frac{7}{3} + \frac{1}{2} \, \ln n_0 \, > \, 0~.
\eea
It is worth emphasizing   that the perturbative LVS minimum can be recovered from the first relation in Eq.~(\ref{eq:dSextrema}) via setting $a_1 = 0$, which corresponds to $d_\alpha = 0$, i.e. without having any D-term contributions to the scalar potential which means uplifting term is absent. This leads to the AdS solution of the perturbative LVS and the following relation determining the VEV of the volume modulus,
\bea
& & \sqrt{\frac{3}{2}} \, \langle \phi^1 \rangle = a_2 \quad \Longrightarrow \quad \langle {\cal V} \rangle \equiv \frac{1}{\sqrt{n_0}} \, e^{\sqrt{\frac{3}{2}} \langle\phi^1\rangle} = e^{-\,\frac{\hat\xi}{2\hat\eta} + \frac{7}{3}} \simeq e^{0.37/g_s^2},
\eea
which we have derived earlier. For the uplifted scenario, one has to tune the D-term coefficients $d_\alpha$ or $a_1$ such that the VEV of the AdS minimum does not significantly change while uplifting it delicately to some dS vacuum. At the same time, the first relation in Eq.~(\ref{eq:dSextrema}) also suggests that one has to tune the $d_\alpha$ parameter to the order of ${\cal V}^{-1}$ in order to compensate the exponential factor for achieving this goal.

Moreover, the first relation of (\ref{eq:dSextrema}) determines the VEV of the overall volume ${\cal V}$ in terms of a product logarithmic function $y = {\cal W}_k(z)$ for some $k \in {\mathbb Z}$, which generically appears as a solution to the equation of the form $y \, e^y = z$. In particular, for $\{y, z\} \in {\mathbb R}$, the equation $y \, e^y = z$ can be solved only for $z \geq -\frac{1}{e}$. For such cases, there exists only a single solution $y = {\cal W}_0[z]$ when $z \geq 0$, and there are two solutions $y = {\cal W}_0[z]$ and $y = {\cal W}_{-1}[z]$ when $-\frac{1}{e} \leq z < 0$. Comparing the first relation of (\ref{eq:dSextrema}) with equation $y \, e^y = z$ we find that,
\bea
\label{eq:cond0}
& & z = -\, a_1\, e^{a_2} \leq 0,
\eea
where as we argued earlier, for $a_1 = 0$, i.e., the absence of the uplifting piece corresponds to the perturbative LVS case with AdS minimum which is consistent with the argument that $z = 0$ results in a single solution. For non-zero uplifting case, $z<0$ will lead to two solutions; one corresponding to the minimum and another one to the maximum or a saddle point.
This means that, to ensure the existence of a de Sitter minimum, one will have a constraint on the model-dependent parameters given as below:
\bea
\label{eq:cond1}
& & \frac{1}{e} \geq a_1\, e^{a_2} > 0\,.
\eea
This simply dictates that  $a_1$ should be tuned  in conjunction with the uplifting coefficients $d_\alpha$'s in order to compensate the exponential increase of the form $e^{a_2} \propto e^{1/g_s^2}$.

In addition, to ensure that this is a tachyon-free minimum, one has to see the Hessian eigenvalues. Using the extremization conditions in Eq.~(\ref{eq:dSextrema}), the non-trivial components of the Hessian $V_{\alpha\beta}$ are simplified to take the following form,
\bea
\label{eq:HessdS}
& & \langle {(V_0)}_{11} \rangle =  -\,9\, \hat\eta\,n_0^{3/2}\, {\cal C}_1 \, e^{-3\,\sqrt{\frac{3}{2}} \langle\phi^1\rangle} \left(1 + a_2 - \sqrt{\frac{3}{2}}\, \langle \phi^1 \rangle \right), \\
& & \langle {(V_0)}_{22} \rangle = -\,18\, \hat\eta\, n_0^{3/2} \,{\cal C}_1 \, e^{-3\,\sqrt{\frac{3}{2}} \langle\phi^1\rangle} \left(\sqrt{\frac{3}{2}}\, \langle \phi^1 \rangle - a_2\right) = \langle {(V_0)}_{33} \rangle. \nonumber
\eea
Given that ${\cal C}_1 > 0$ and $\hat\xi/\hat\eta < 0$, this shows that the Hessian (or mass matrix) may not be generically positive semi-definite, and a minimum is ensured only when the following holds,
\bea
\label{eq:cond2}
& & a_2 \, \,  < \, \,  \sqrt{\frac{3}{2}} \, \langle \phi^1 \rangle \, \, < \, \, 1 + a_2.
\eea
In fact the mass hierarchy between the overall volume modulus $\phi^1$ and the remaining two moduli ($\phi^2, \phi^3$) which is needed for justifying the single-field inflation, can be ensured if the following holds,
\bea
\label{eq:mass-hierarchy0}
& & {\cal R}_{\rm hierarchy} \equiv \frac{m^2_{\phi^1}}{m^2_{\phi^\alpha}} = \, \frac{\left(1 + a_2 - \sqrt{\frac{3}{2}}\, \langle \phi^1 \rangle \right)}{2\, \left(\sqrt{\frac{3}{2}}\, \langle \phi^1 \rangle - a_2\right)} \ll 1\,, \qquad \alpha \in \{2, 3\};
\eea
Moreover we can impose another condition by demanding an uplifting to Minkowskian/dS vacuum by the presence of $D$-term effects. For that purpose, considering Eq.~(\ref{eq:dSextrema}), we find that the following holds at this minimum,
\bea
\label{eq:pot-with-a1-a2}
& & \langle V_{0} \rangle = - \hat\eta\, n_0^{3/2}\,{\cal C}_1 \, e^{-3\,\sqrt{\frac{3}{2}} \langle\phi^1\rangle} \left(\sqrt{\frac{3}{2}}\, \langle \phi^1 \rangle - a_2 - \frac{2}{3}\right).
\eea
Now, assuming that the desired tuned values of the uplifting coefficients $d_\alpha$ can be achieved, we can solve the condition $\langle V_{0} \rangle \geq 0$ to get the following additional constraint,
\bea
\label{eq:cond3}
& & \frac{2}{3} + a_2 \leq \sqrt{\frac{3}{2}}\, \langle \phi^1 \rangle.
\eea
Subsequently, the bounds for having a Minkowskian/dS minimum is given as,
\bea
\label{eq:cond4}
& & \frac{2}{3} + a_2 \leq \sqrt{\frac{3}{2}}\, \langle \phi^1 \rangle < 1 + a_2,
\eea
where the equal sign holds for the Minkowskian case, for which the Hessian takes the following form,
\bea
& & \langle {(V_0)}_{11} \rangle =  -3 \, \hat\eta \, n_0^{3/2}\, {\cal C}_1 \, e^{-3\,\sqrt{\frac{3}{2}} \langle\phi^1\rangle} > 0, \\
& & \langle {(V_0)}_{22} \rangle =  -\,12\, \hat\eta\,n_0^{3/2}\, {\cal C}_1 \, e^{-3\,\sqrt{\frac{3}{2}} \langle\phi^1\rangle} = \langle {(V_0)}_{33} \rangle > 0, \nonumber
\eea
ensuring  a tachyon-free minimum. This analysis also shows that the overall volume mode $\phi^1$ is the lightest modulus with masses equal to,
\bea
& & M_{1} = \frac{1}{2} M_{2} = \frac{1}{2} M_{3} = \frac{\sqrt{3} \sqrt{|\hat\eta|}\, \sqrt{\cal C}_1}{\langle{\cal V}\rangle^{3/2}}.
\eea
Although we do not see a huge mass-hierarchy among the three moduli for the Minkowskian case (or for a dS with extremely small cosmological constant), however the Hessian components (\ref{eq:HessdS}) at the minimum suggest that one can create a mass-hierarchy by considering the parameters such that the following relation holds:
\bea
\label{eq:cond5}
& & \left(\sqrt{\frac{3}{2}}\, \langle \phi^1 \rangle - a_ 2 \right) \to 1.
\eea
For such cases it should be still quite justified to consider the single field effective potential for the lightest modulus $\phi^1$ while assuming that the other two sit at their respective minimum. Eliminating the VEVs for $\phi^2$ and $\phi^3$ using $d_\alpha$'s from the first line of (\ref{eq:dSextrema}), one has the following single field effective potential for the $\phi^1$ modulus,
\bea
\label{eq:Vinf}
& & \hskip-1.5cm V_0(\phi^1) = -\, {\cal B} \, e^{-3\sqrt{\frac{3}{2}} \phi^1} \left(\sqrt{\frac{3}{2}} \phi^1 - \frac{3}{2}\, e^{\sqrt{\frac{3}{2}} \phi^1}\, a_1 \, -a_2 + \frac{1}{3} \right), \quad {\cal B} = -\, 2\, \hat\eta\, n_0^{3/2} \, {\cal C}_1 > 0.
\eea
The overall exponential pre-factor in (\ref{eq:Vinf}) ensures that $\langle V \rangle$ is of the order of ${\cal V}^{-3}$, similar to the standard LVS. We note that the above form of the inflationary potential is the same as the one analysed in \cite{Antoniadis:2020stf}, and the two effective potentials can be matched by simply using the following identifications in their respective model dependent parameters,
\bea
\label{eq:OLDmodel}
& & \hskip-1cm q + \delta = \frac{1}{3} - a_2, \qquad \sigma = -a_1 < 0.
\eea
The single-field potential (\ref{eq:Vinf}) results in the following derivatives,
\bea
\label{eq:derivatives-Vinf}
& & V^\prime_{0} = \partial_{\phi^1} V_{0} = \frac{\sqrt{3}}{2\sqrt{2}}\, {\cal B} \, e^{-3\sqrt{\frac{3}{2}} \phi^1} \left(\sqrt{\frac{3}{2}} \phi^1 - \, e^{\sqrt{\frac{3}{2}} \phi^1}\, a_1 \, -a_2 \right), \\
& & V^{\prime\prime}_{0} = \partial^2_{\phi^1} V_{0} = -\,\frac{27}{2}\, {\cal B} \, e^{-3\sqrt{\frac{3}{2}} \phi^1} \left(\sqrt{\frac{3}{2}} \phi^1 - \frac{2}{3}\, e^{\sqrt{\frac{3}{2}} \phi^1}\, a_1 \, -a_2 - \frac{1}{3} \right). \nonumber
\eea
We note that the inflationary potential (\ref{eq:Vinf}) has the following set of model dependent parameters,
\bea
& & \{W_0, \, \, g_s, \, \, d_0\}, \qquad {\rm or} \qquad \{{\cal B}, \, \, a_1, \, \, a_2 \}.
\eea
Now the main idea is to choose these parameters such that the de-Sitter minimum is realised along with the consistent experimental values for the cosmological observables such as the scalar perturbation amplitude $(P_s)$ and the spectral index $(n_s -1)$, at least 60 e-foldings, etc. Note that the parameter $a_2$ depends only on $g_s$, and determines the VEV of the overall volume modulus in perturbative LVS. Furthermore, given that one has to respect the condition (\ref{eq:cond1}) for having a dS minimum, we define a slightly refined parameter $x$ to take care of the uplifting in the following way,
\bea
\label{eq:condx}
& & a_1 \equiv e^{-a_2 - 1- x} \quad \Longrightarrow \quad 0 < e^{-x} \leq 1.
\eea
Another motivation for defining this parameter $x$ is the fact that it quantifies the separation between the two extrema such that the two extrema collapse to a single one for $x \to 0$. In other words, this parameter governs the ``length" of the inflationary plateau as we will elaborate later on. Given that, in the large volume limit the potential goes to zero, it can result in a dS solution if there are at least two extrema, and therefore there exists a critical value $x_c$ determined by (\ref{eq:cond1}) and (\ref{eq:condx}) beyond which only AdS solutions can be possible. 

Subsequently, using Eqs.~(\ref{eq:derivatives-Vinf}) and (\ref{eq:condx}), the two extrema corresponding to the local minimum/maximum, and the point of inflection are determined in the form of product-log functions as below,
\bea
\label{eq:extrema-and-inflationPoint}
& & \phi^1_{\rm min} = \sqrt{\frac{2}{3}} \left(a_2 - {\cal W}_0[-e^{-1-x}] \right), \qquad \phi^1_{\rm max} = \sqrt{\frac{2}{3}} \left(a_2 - {\cal W}_{-1}[-e^{-1-x}] \right), \\
& & \phi^1_{\rm inflec1} = \sqrt{\frac{2}{3}} \left(\frac{1}{3} + a_2 - {\cal W}_0[-e^{-1-x}] \right), \quad \phi^1_{\rm inflec2} = \sqrt{\frac{2}{3}} \left(\frac{1}{3} + a_2 - {\cal W}_{-1}[-e^{-1-x}] \right). \nonumber
\eea
However one of the inflection points $\phi^1_{\rm inflec2}$ lies on the RH side of maximum, i.e., out of the region where the inflaton rolls towards the minimum, and we are interested in exploring inflationary possibilities in which the inflaton slowly rolls down into the minimum on a nearly flat track to result in the desired cosmological observables.

From (\ref{eq:extrema-and-inflationPoint}) one also finds that the difference $\Delta \phi^1 = \phi^1_{\rm max}  -  \phi^1_{\rm min}$ does not depend on the parameter $a_2$ as
\bea
\label{eq:shift-in-phi}
& & \hskip-1.5cm \Delta\phi^1 \equiv \phi^1_{\rm max}  -  \phi^1_{\rm min} = \sqrt{\frac{2}{3}} \left({\cal W}_0[-e^{-1-x}] - \, {\cal W}_{-1}[-e^{-1-x}] \right),
\eea
and hence it does not depend on the string coupling $g_s$ as well. In fact $\Delta\phi^1$ can be solely controlled by the uplifting parameter $x$ and one finds that $\Delta\phi^1=0$ for $x = 0$ as the two extrema collapse, showing that the parameter $x$ is crucial in determining the  shape of the potential.


\subsection{Volume modulus inflation}
\begin{figure}[H]
\centering
\includegraphics[scale=0.80]{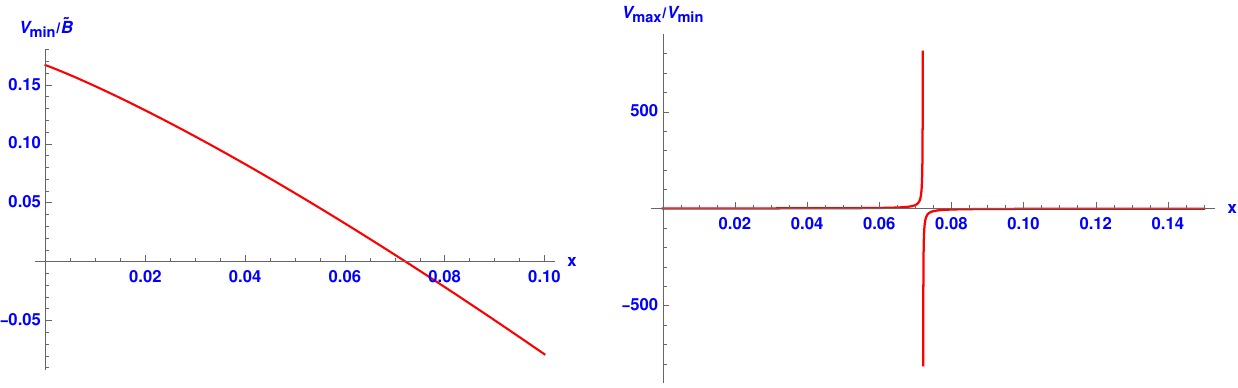}
\caption{On the left side, using (\ref{eq:extrema-and-inflationPoint1}), the factor $V_{\rm min}/\tilde{\cal B}$ evaluated at the minimum is plotted for $x$ which reflects that a Minkowskian minimum corresponds to $x \simeq 0.0721318$ as shown in (\ref{eq:xmink}). The plot in the right shows that for values of $x > x_{\rm Mink}$, the ratio of the two scales flips sign as well.}
\label{fig1}
\end{figure}

\begin{figure}[H]
\centering
\includegraphics[scale=0.90]{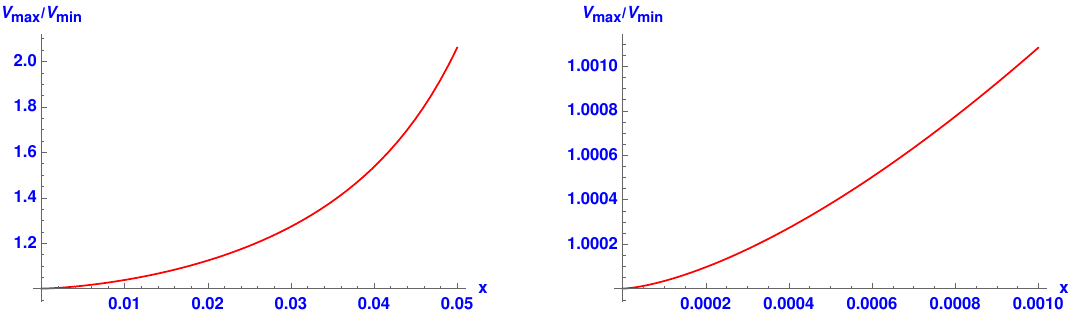}
\caption{The ratio $\left(V_{\rm max}/V_{\rm min}\right)$ is plotted for $x$ which shows that for sufficiently small value of $x$, say $x \lesssim 10^{-2}$, there is no scale separation between the respective values of the potential at the two extrema. }
\label{fig2}
\end{figure}

In this section we discuss the inflationary aspects of the single field potential. Motivated  by the Hessian analysis in (\ref{eq:HessdS}) leading to the mass-hierarchy expression (\ref{eq:mass-hierarchy0}) we define the following shifted field $\phi$,
\bea
\label{eq:shiftedphi}
& & \sqrt{\frac{3}{2}}\,  \phi^1  - a_ 2  - 1 \equiv \sqrt{\frac{3}{2}}\, \phi,
\eea
where we still keep the same canonical normalisation factor to maintain the same normalisation as in the previous mass matrix. The idea is to only keep track of the minimisation such that $\langle \phi \rangle$ is small enough for justifying the single field approximations as given in (\ref{eq:cond5}). Now the mass hierarchy for justifying the single-field approach can be ensured if
\bea
\label{eq:mass-hierarchy}
& & {\cal R}_{\rm hierarchy} \equiv \frac{m^2_{\phi}}{m^2_{\phi^\alpha}} = - \, \frac{\sqrt{\frac{3}{2}}\,\langle \phi \rangle}{2\, \left(1 + \sqrt{\frac{3}{2}}\, \langle \phi \rangle \right)} \ll 1\,, \qquad \alpha \in \{2, 3\};
\eea
which implies that one needs to consistently realise $\langle \phi \rangle \lesssim 0$ such that:
\bea
\label{eq:cond6}
& & \hskip-1.5cm \langle \phi \rangle < -\sqrt{\frac{2}{3}} \simeq -0.816497, \qquad {\rm or} \qquad -\frac{2}{3}\sqrt{\frac{2}{3}} \simeq -0.544331 < \langle \phi \rangle < 0.
\eea
In addition, respecting the constraint (\ref{eq:cond4}) along with the above implies that,
\bea
\label{eq:cond7}
& & -\frac{1}{3} \sqrt{\frac{2}{3}} \leq \langle \phi \rangle < 0,
\eea
where equality corresponds to the Minkowskian solution. This is a very crucial bound which fixes the range of $\phi$ modulus at the minimum.
Having the shift (\ref{eq:shiftedphi}) and the introduction of $x$ parameter in (\ref{eq:condx}), the scalar potential (\ref{eq:Vinf}) takes the following form,
\bea
\label{eq:Vinf-shifted}
& & \hskip-1.5cm V_{\rm inf}(\phi) = -\, \tilde{\cal B} \, e^{-3\sqrt{\frac{3}{2}} \phi} \left(\sqrt{\frac{3}{2}} \phi - \frac{3}{2}\, e^{-x\, + \sqrt{\frac{3}{2}} \phi} + \frac{4}{3} \right),
\eea
where the overall coefficient $\tilde{\cal B}$, which depends on the parameters $g_s$ and $|W_0|$- is given as below,
\bea
\label{eq:tildeB}
& & \tilde{\cal B} \equiv \tilde{\cal B}(|W_0|,g_s) = -\, \frac{\chi({\rm CY})\, \sqrt{g_s}\, \, |W_0|^2\,e^{-10-\frac{9 \zeta[3]}{g_s^2\, \pi^2}}}{64\pi} > 0,
\eea
where we assume that $\chi({\rm CY}) < 0$ which is a typical choice for the CY threefold used for model building in the type IIB superstring compactifications. Subsequently, the derivatives and the Hessian (\ref{eq:derivatives-Vinf}) take the following respective forms,
\bea
\label{eq:derivatives-Vinf-shifted}
& & V^\prime_{\rm inf} = \partial_{\phi} V_{\rm inf} = \frac{3\sqrt{3}}{\sqrt{2}}\, \tilde{\cal B} \, e^{-3\sqrt{\frac{3}{2}} \phi} \left(\sqrt{\frac{3}{2}} \phi - \, e^{-\,x\,+ \sqrt{\frac{3}{2}} \phi}\, + 1 \right), \\
& & V^{\prime\prime}_{\rm inf} = \partial^2_{\phi} V_{\rm inf} = -\,\frac{27}{2}\, \tilde{\cal B} \, e^{-3\sqrt{\frac{3}{2}} \phi} \left(\sqrt{\frac{3}{2}} \phi - \frac{2}{3}\, e^{- x\,+ \sqrt{\frac{3}{2}} \phi}\, + \frac{2}{3} \right). \nonumber
\eea
Now the four points of interest are two extrema corresponding to the minimum and maximum of the potential and two inflection points,
\bea
\label{eq:extrema-and-inflationPoint1}
& & \hskip-1.5cm \phi_{\rm min} = -\sqrt{\frac{2}{3}} \left(1+{\cal W}_0[-e^{-1-x}] \right), \qquad \phi_{\rm max} = - \sqrt{\frac{2}{3}} \left(1 + {\cal W}_{-1}[-e^{-1-x}] \right), \\
& & \hskip-1.5cm \phi_{\rm inflec1} = -\sqrt{\frac{2}{3}} \left(\frac{2}{3} +  {\cal W}_0\left[-\frac{2}{3}\,e^{-\frac{2}{3}-x}\right] \right), \quad \phi_{\rm inflec2} = -\sqrt{\frac{2}{3}} \left(\frac{2}{3} +  {\cal W}_{-1}\left[-\frac{2}{3}\,e^{-\frac{2}{3}-x}\right] \right). \nonumber
\eea
As observed earlier in Eq.~(\ref{eq:shift-in-phi}) we find that $\Delta \phi \equiv \phi_{\rm max} - \phi_{\rm min}$ which corresponds to the length between the maximum and the minimum solely depends on a single parameter $x$, and the two extrema collapse for $x = 0$. Moreover, using the expression for $\phi_{\rm min}$ from Eq.~(\ref{eq:extrema-and-inflationPoint1}), the VEV of the potential (\ref{eq:Vinf-shifted}) at the minimum is given as,
\bea
& & \langle V_{\rm inf} \rangle = -\frac{1}{6}\, \tilde{\cal B} \, e^{3 + 3\, {\cal W}_0[-e^{-1-x}]}\, \left(2 + 3\, {\cal W}_0[-e^{-1-x}] \right).
\eea
Thus we can see that working with the shifted modulus $\phi$  as introduced in (\ref{eq:shiftedphi}) results in a potential such that one can determine the VEV of $\phi$ as well as the uplifting by a single parameter $x$. Further, demanding a Minkowskian minimum results in a unique value of $x$ which does not depend on any other parameters as we see below,
\bea
\label{eq:xmink}
& & \hskip-1.5cm {\cal W}_0[-e^{-1-x}] = -\frac{2}{3} \qquad \Longrightarrow \qquad x_{\rm Mink} = -\frac{1}{3} + \ln3 - \ln 2 \simeq 0.0721318,
\eea
which has been also mentioned in \cite{Antoniadis:2020stf}. It turns out that the behaviour of the potential drastically changes for displacements around this critical value of $x$ as can be seen  in Fig.~\ref{fig1}. Further, in Fig.~\ref{fig2} we have plotted the ratio of the values the scalar potential takes at its maximum and  minimum with respect to $x$, and we find that for sufficiently small value of $x$, the scale separation between the two extrema of the potential tends to attain negligibly small values.

The slow-roll parameters are generically defined through the derivatives of the Hubble parameter as below,
\bea
\label{eq:slow-roll-H}
& & \hskip-1.5cm \epsilon_H = -\frac{\dot{H}}{H^2} = \frac{1}{H} \frac{d H}{dN}, \qquad \qquad \eta_H = \frac{\dot\epsilon_H}{\epsilon_H \, H} = \frac{1}{\epsilon_H} \frac{d\epsilon_H}{dN},
\eea
where dot denotes the time derivative while $N$ denotes the number of e-foldings determined by,
\bea
\label{eq:e-fold-def}
& & N(\phi) = \int H \, dt = \int_{\phi_{\rm end}}^{\phi_\ast} \, \frac{1}{\sqrt{2 \epsilon_H}}\, d\phi \, \simeq \, \int_{\phi_{\rm end}}^{\phi_\ast} \, \frac{V_{\rm inf}}{V^\prime_{\rm inf}}\, d\phi\,\,,
\eea
where $\phi_\ast$ is  the point of horizon exit at which the cosmological observables are to be matched with the experimentally observed values. However, the slow-roll inflationary parameters can  also be defined through the derivatives of the potential
\bea
\label{eq:slow-roll-V-def}
& & \epsilon_V \equiv \frac{1}{2} \left(\frac{V^\prime_{\rm inf}}{V_{\rm inf}}\right)^2 , \qquad \eta_V \equiv \frac{V^{\prime\prime}_{\rm inf}}{V_{\rm inf}}. \nonumber
\eea
In fact, for single field inflation, the two sets of slow-roll parameters, namely $(\epsilon_H, \eta_H)$ and $(\epsilon_V, \eta_V)$ can be correlated as \cite{Achucarro:2018vey},
\bea
\label{eq:slow-roll-relation}
& & \epsilon_V = \epsilon_H\left(1 + \frac{\eta_H}{2\left(3 - \epsilon_H\right)}\right)^2 \simeq \epsilon_H, \quad \eta_H \simeq -2\, \eta_V + 4\, \epsilon_V
\eea
and subsequently the cosmological observables such as the scalar perturbation amplitude, the spectral index, its running are also controlled by the parameter $x$ given that these observables are correlated with the slow-roll parameters $\epsilon_V$ and $\eta_V$ as below \cite{Planck:2018jri},
\bea
\label{eq:cosmo-observables}
& & P_s \equiv \frac{V_{\rm inf}^\ast}{24 \pi^2 \, \epsilon_H^\ast} \simeq 2.2 \times 10^{-9}, \qquad {\rm or} \qquad \frac{V_{\rm inf}^{\ast3}}{V_{\rm inf}^{\prime\ast^2}} \simeq 2.6\times 10^{-7},\\
& & n_s - 1 = -2 \epsilon_H^\ast - \eta_H^\ast \simeq 2 \, \eta_V^\ast - 6\, \epsilon_V^\ast \simeq -0.04,\nonumber\\
& & r = 16 \epsilon_H^\ast \simeq 16 \epsilon_V^\ast,\nonumber
\eea
where all the cosmological observables are evaluated at the horizon exit $\phi = \phi^\ast$ and one also has sufficient e-foldings: $N(\phi^\ast) \gtrsim 60$. For an inflationary model in which one gets more  e-foldings towards the minimum, it is better to use $\epsilon_H$ as compared to $\epsilon_V$ for which the integrand has a singularity located at $\phi = \phi_{\rm min}$.

Now, using the inflationary potential (\ref{eq:Vinf-shifted}), the slow-roll parameters (\ref{eq:slow-roll-V-def}) are given as,
\bea
\label{eq:slow-roll-V}
& & \hskip-2.5cm \epsilon_V = \frac{27\left(\sqrt{\frac{3}{2}} \phi - \, e^{-\,x\,+ \sqrt{\frac{3}{2}} \phi}\, + 1 \right)^2}{4\left(\sqrt{\frac{3}{2}} \phi - \frac{3}{2}\, e^{-x\, + \sqrt{\frac{3}{2}} \phi} + \frac{4}{3} \right)^2}, \qquad \eta_V = \frac{27\left(\sqrt{\frac{3}{2}} \phi - \frac{2}{3}\, e^{- x\,+ \sqrt{\frac{3}{2}} \phi}\, + \frac{2}{3}\right)}{2\left(\sqrt{\frac{3}{2}} \phi - \frac{3}{2}\, e^{-x\, + \sqrt{\frac{3}{2}} \phi} + \frac{4}{3}\right)}.
\eea
This shows that the slow-roll parameters are also solely controlled by the parameter $x$. However as one can easily anticipate, these expressions for $(\epsilon_V, \eta_V)$ have various zeros and singularities, which may influence the cosmological observables.

For the inflationary model building one usually uses uplifting effects in order to delicately uplift an AdS minimum to a dS minimum with tiny cosmological constant. For that purpose one would need to set $x$ close to its Minkowskian value $x \simeq 0.0721318$ as can be seen from the Fig.~\ref{fig1}. However, upon close examination we find that the spectral index values generally fall outside the experimental limits for the possible range of the typical inflationary plateau. The inflaton could roll down towards the minimum, from a point near the maximum or the inflection point, 
and therefore one is left with the possibility of constructing models in which $\langle V_{\rm inf} \rangle$ is relatively much larger than the cosmological constant.


Given that the dynamics of the shifted modulus $\phi$ depends only on a single parameter $x$, it would be useful  to present some numerics to have an estimate about the ranges of the various possible ingredients involved in the inflationary dynamics. These are presented in Table \ref{tab1} which shows that the the first inflection point indeed lies between the maxima and the minima of the respective models while the second one is outside on the right of the potential. Moreover Table \ref{tab1} also shows that for having a single field model the smaller values of $x$ could be preferred as the same leads to more hierarchy among the $\phi$ (or the $\phi^1$) modulus as compared to the other two moduli, namely $\phi^2$ and $\phi^3$, which are stabilised by a leading order D-term effect.

\begin{table}[h]
\centering
\begin{tabular}{|c|c|cccc|c|}
\hline
\cellcolor[gray]{0.9} &\cellcolor[gray]{0.9} &\cellcolor[gray]{0.9} &\cellcolor[gray]{0.9} &\cellcolor[gray]{0.9} &\cellcolor[gray]{0.9} &\cellcolor[gray]{0.9} \\
\cellcolor[gray]{0.9}Model & \cellcolor[gray]{0.9}$x$ &\cellcolor[gray]{0.9} $\cellcolor[gray]{0.9}\phi_{\rm min}$  &\cellcolor[gray]{0.9} $\phi_{\rm max}$  & \cellcolor[gray]{0.9}$\phi_{\rm inflect1}$  &\cellcolor[gray]{0.9} $\phi_{\rm inflect2}$  &\cellcolor[gray]{0.9} ${\cal R}_{\rm hierarchy}$ \\
\cellcolor[gray]{0.9} &\cellcolor[gray]{0.9} &\cellcolor[gray]{0.9} &\cellcolor[gray]{0.9} &\cellcolor[gray]{0.9} &\cellcolor[gray]{0.9} &\cellcolor[gray]{0.9} \\
\hline
 \cellcolor[gray]{0.9}& & & & & & \\
\cellcolor[gray]{0.9}S1 & 0.0721318 & -0.272166 & 0.350567 & -0.09166 & 0.792544 & 0.25 \\
\cellcolor[gray]{0.9}S2 & 0.07 & -0.26865 & 0.344739 & -0.0894826 & 0.788061 & 0.245188 \\
\cellcolor[gray]{0.9}S3 & 0.06 & -0.25117 & 0.316403 & -0.0789419 & 0.766699 & 0.222146 \\
\cellcolor[gray]{0.9}S4 & 0.05 & -0.23173 & 0.286103 & -0.0678149 & 0.744747 & 0.198139 \\
\cellcolor[gray]{0.9}S5 & 0.04 & -0.2097 & 0.253208 & -0.0560263 & 0.722128 & 0.172792 \\
\cellcolor[gray]{0.9}S6 & 0.03 & -0.184014 & 0.216653 & -0.043484 & 0.698750 & 0.145470 \\
\cellcolor[gray]{0.9}S7 & 0.02 & -0.152599 & 0.174363 & -0.0300721 & 0.674499 & 0.114927 \\
\cellcolor[gray]{0.9}S8 & 0.01 & -0.110092 & 0.120976 & -0.0156425 & 0.649224 & 0.077924 \\
\cellcolor[gray]{0.9}S9 & 0.001 & -0.0359725 & 0.0370612 & -0.0016257 & 0.625443 & 0.023044 \\
\cellcolor[gray]{0.9}{\bf S10} &\cellcolor[gray]{0.93} 0.00033 &\cellcolor[gray]{0.93} -0.0207969 &\cellcolor[gray]{0.93} 0.0211562 & \cellcolor[gray]{0.93}-0.000538089 &\cellcolor[gray]{0.93} 0.623629 &\cellcolor[gray]{0.93} 0.013068 \\
\cellcolor[gray]{0.9}{\bf S11} &\cellcolor[gray]{0.93} 0.0001 &\cellcolor[gray]{0.93} -0.0114926 &\cellcolor[gray]{0.93} 0.0116015 &\cellcolor[gray]{0.93} -0.000163226 &\cellcolor[gray]{0.93} 0.623004 &\cellcolor[gray]{0.93} 0.007138 \\
\cellcolor[gray]{0.9}S12 & 0.00001 & -0.00364604 & 0.00365693 & -0.0000163292 & 0.62276 & 0.002243 \\
\cellcolor[gray]{0.9}S13 & $10^{-6}$ & -0.00115416 & 0.00115524 & -1.633$\times 10^{-6}$ & 0.622735 & 0.000708 \\
\cellcolor[gray]{0.9}S14 & $10^{-7}$ & -0.000365094 & 0.000365203 & -1.633$\times 10^{-7}$ & 0.622733 & 0.000224 \\
\hline
\end{tabular}
\caption{Fourteen possible candidate models are presented corresponding to their respective $x$ values. We observe that decreasing the $x$ increases the mass hierarchy as defined in Eq.~(\ref{eq:mass-hierarchy}). However, these candidate models need to be tested to see if they can consistently reproduce the cosmological observables.}
\label{tab1}
\end{table}

\begin{table}[h]
\centering
\begin{tabular}{|c|c c|cccccc|}
\hline
\cellcolor[gray]{0.9}&\cellcolor[gray]{0.9} &\cellcolor[gray]{0.9} &\cellcolor[gray]{0.9} &\cellcolor[gray]{0.9} &\cellcolor[gray]{0.9} &\cellcolor[gray]{0.9} &\cellcolor[gray]{0.9} &\cellcolor[gray]{0.9}  \\
\cellcolor[gray]{0.9}Model &\cellcolor[gray]{0.9} $x$ &\cellcolor[gray]{0.9} $\phi^\ast$  & \cellcolor[gray]{0.9}  &\cellcolor[gray]{0.9}   &\cellcolor[gray]{0.9} $\sqrt{n_0}\, \langle {\cal V} \rangle$  &\cellcolor[gray]{0.9}  &\cellcolor[gray]{0.9}  &\cellcolor[gray]{0.9}  \\
\cellcolor[gray]{0.9}& \cellcolor[gray]{0.9} &\cellcolor[gray]{0.9}  &\cellcolor[gray]{0.97}  $a_2 = 4$  &\cellcolor[gray]{0.97}  $a_2 = \frac{13}{3}$  & \cellcolor[gray]{0.97}$a_2 = 5$  &\cellcolor[gray]{0.97} $a_2 = 6$ &\cellcolor[gray]{0.97} $a_2 = 7$ &\cellcolor[gray]{0.97} $a_2 = 8$  \\
\hline
& & & & & & & & \\
S1 & 0.0721318 & - & 106.343 & 148.413 & 289.069 & 785.772 & 2135.95 & 5806.11 \\
S2 & 0.07 & -0.232103 & 106.801 & 149.053 & 290.317 & 789.162 & 2145.16 & 5831.16 \\
S3 & 0.06 & -0.168771 & 109.113 & 152.279 & 296.599 & 806.239 & 2191.59 & 5957.35 \\
S4 & 0.05 & -0.126763 & 111.742 & 155.948 & 303.745 & 825.665 & 2244.39 & 6100.89 \\
S5 & 0.04 & -0.092541 & 114.798 & 160.213 & 312.052 & 848.247 & 2305.77 & 6267.74 \\
S6 & 0.03 & -0.063491 & 118.466 & 165.333 & 322.025 & 875.355 & 2379.46 & 6468.05 \\
S7 & 0.02 & -0.0386125 & 123.113 & 171.818 & 334.657 & 909.691 & 2472.80 & 6721.76 \\
S8 & 0.01 & -0.0173722 & 129.692 & 181.0 & 352.54 & 958.304 & 2604.94 & 7080.96 \\
S9 & 0.001 & -0.00105389 & 142.016 & 198.2 & 386.041 & 1049.37 & 2852.48 & 7753.83 \\
\cellcolor[gray]{0.97}{\bf S10} &\cellcolor[gray]{0.97} 0.00033 & \cellcolor[gray]{0.97}0.000061912 &\cellcolor[gray]{0.97} 144.681 &\cellcolor[gray]{0.97} {\bf 201.918} &\cellcolor[gray]{0.97} 393.283 &\cellcolor[gray]{0.97} 1069.05 & \cellcolor[gray]{0.97}2905.99 & \cellcolor[gray]{0.97}7899.30 \\
\cellcolor[gray]{0.97}{\bf S11} &\cellcolor[gray]{0.97} 0.0001 &\cellcolor[gray]{0.97} 0.000441890 &\cellcolor[gray]{0.97} 146.339 & \cellcolor[gray]{0.97}204.232 &\cellcolor[gray]{0.97} 397.79 &\cellcolor[gray]{0.97} {\bf 1081.31} &\cellcolor[gray]{0.97} 2939.29 & \cellcolor[gray]{0.97}7989.83 \\
S12 & 0.00001 & 0.000590154 & 147.752 & 206.204 & 401.631 & 1091.75 & 2967.68 & 8066.98 \\
S13 & 1$\times10^{-6}$ & 0.000604967 & 148.204 & 206.835 & 402.859 & 1095.08 & 2976.75 & 8091.64 \\
S14 & 1$\times10^{-7}$ & 0.000606448 & 148.347 & 207.035 & 403.248 & 1096.14 & 2979.63 & 8099.46 \\
\hline
\end{tabular}
\caption{For the possible candidate models corresponding to the respective $x$ values, the  values of Horizon exit $\phi^\ast$ is calculated for $n_s - 1 = -0.04$. In addition, the VEV of the overall volume modulus ${\cal V}$ corresponding to a given $x$ is presented for a range of $a_2$ values. }
\label{tab2}
\end{table}

However, let us mention that all the possible candidate models presented in Table \ref{tab1} need to be tested further  to see if they can consistently reproduce the cosmological observables. In this regard, the spectral index $(n_s)$ is a crucial observable which has to satisfy $n_s - 1 \simeq -0.04$, and for this purpose we present the corresponding values for horizon exit $\phi^\ast$ by solving the relation $n_s - 1 = -0.04$ as presented in Table \ref{tab2}. Moreover, Table \ref{tab2} presents a range of values for the $a_2$ parameter, which for a given concrete model, can generically determine the corresponding value of the string coupling $g_s$ for a given triple intersection number $n_0$ of the CY threefold. It turns out that one has the following relation:
\bea
\label{eq:gs-a2-n0}
& & g_s = \frac{\sqrt{3}\, \zeta[3]}{\pi\, \sqrt{a_2 -\frac{7}{3} - \frac{1}{2} \, \ln n_0}},
\eea
which results in typical values of $g_s$ as mentioned in Table \ref{tab3}.

\begin{table}[h]
\centering
\begin{tabular}{|c|cccccc|}
\hline
\cellcolor[gray]{0.9}$g_s$ &\cellcolor[gray]{0.9} $a_2$ & \cellcolor[gray]{0.9}$a_2$ &\cellcolor[gray]{0.9} $a_2$ &\cellcolor[gray]{0.9} $a_2$ & \cellcolor[gray]{0.9}$a_2$ & \cellcolor[gray]{0.9}$a_2$  \\
\cellcolor[gray]{0.9} &\cellcolor[gray]{0.97} $4$ &\cellcolor[gray]{0.97} $13/3$ &\cellcolor[gray]{0.97} $5$ &\cellcolor[gray]{0.97} $6$ & \cellcolor[gray]{0.97}$7$ &\cellcolor[gray]{0.97} $8$  \\
 \hline
\cellcolor[gray]{0.97}$n_0 = 1$ & 0.468219 & 0.427423 & 0.370160 & 0.315673 & 0.279814 & 0.253927  \\
\cellcolor[gray]{0.97}$n_0 = 2$ & 0.526103 & 0.470090 & 0.396845 & 0.331740 & 0.290822 & 0.262068 \\
\hline
\end{tabular}
\caption{A set of values for the string-coupling $g_s$ corresponding to $n_0$ and $a_2$ parameters listed in Table \ref{tab2}.}
\label{tab3}
\end{table}

In addition, Table \ref{tab2} shows that the VEV of the overall volume modulus $\langle {\cal V} \rangle$ can be read-off once the $n_0$ parameter associated with the CY threefold is known in a given concrete model. Also we observe that for a fixed value of the $a_2$ parameter, the overall volume $\langle {\cal V} \rangle$ does not change significantly throughout the entire allowed range of the $x$ values. In fact we note that for $x \leq 10^{-4}$, VEV of the overall volume $\langle {\cal V} \rangle $ typically remains the same for the fixed $a_2$ values. For $x \simeq 3.3 \times 10^{-4}$ and $a_2 = \frac{13}{3}$ we recover the model presented in \cite{Antoniadis:2020stf} which we denote as the candidate model S10 in our collection. As seen from Table \ref{tab2} this model results in $\langle {\cal V} \rangle \simeq 200$ which may not be considered to be large enough to ensure viability of the model against various possible corrections. Moreover, the mass-hierarchy being $m_{\phi^1}/m_{\phi^{2}} = m_{\phi^1}/m_{\phi^{3}} = 0.114$.

Finally, let us mention that apart from the spectral index, the other requirement for inflationary dynamics is to have the sufficient number of e-foldings. It turns out that for $x > 0.05$ one does not have more than a few e-foldings. Even for $x \sim 0.001$ one get around 20 e-foldings. However, for $x < 0.001$ the number of e-folds increase significantly and one gets $N_e(\phi^\ast) \sim 75$ for $x = 10^{-4}$ and if one decreases the value of $x$ further one gets too many e-foldings. For example, $x = \{10^{-5}, 10^{-6}, \10^{-7}\}$ results in around $\{406, 1456, 7278\}$ e-folds respectively. This also suggests that having too low $x$ values may dilute things too much in the post inflationary epoch and therefore one should preferably use $x \sim 10^{-4}$ for any typical model building purpose. We find that for $x = 3.28 \times 10^{-4}$ one gets $N_e(\phi^\ast) \sim 70$.

\subsubsection*{A benchmark model}
\begin{figure}
\centering
\includegraphics[scale=0.91]{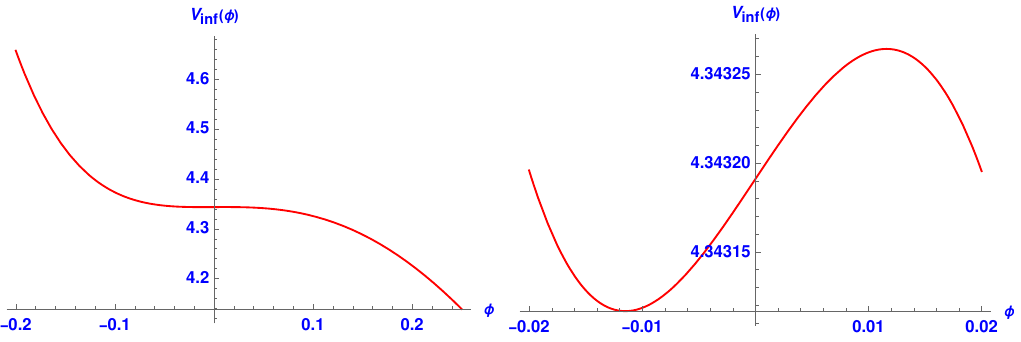}
\caption{Inflationary potential (\ref{eq:Vinf-shifted}) ($10^{13}\times V_{\rm inf}(\phi)$) plotted for the benchmark model $x = 10^{-4}$ and $a_2 = 6$.}
\label{fig3}
\end{figure}

\begin{figure}
\centering
\includegraphics[scale=0.91]{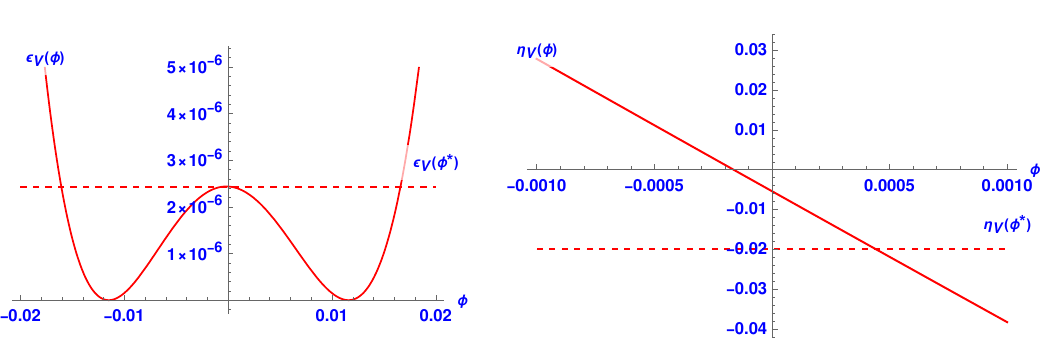}
\caption{The slow-roll parameters $\epsilon_V$ and $\eta_V$ plotted for the benchmark model $x = 10^{-4}$ and $a_2 = 6$. The dotted horizontal lines in the two plots correspond to $\epsilon_V^\ast = 2.42 \times 10^{-6}$ and $\eta_V^\ast = -0.02$.}
\label{fig4}
\end{figure}

Based on the discussion so far, we consider the following benchmark model of inflation driven by the overall volume modulus in perturbative LVS,
\bea
\label{eq:model-before-global}
& & \hskip-1.5cm x = 0.0001, \qquad a_2 = 6, \qquad \tilde{\cal B} = 7.56\times 10^{-12},
\eea
which can appropriately produce the cosmological observables within the experimental bounds as mentioned in Eq.~(\ref{eq:cosmo-observables}). Also, the above parameters $a_2$ and $x$ lead to $a_1 \simeq 0.00128947/\sqrt{n_0}$. In addition, the requirements (\ref{eq:a1-a2}) correspond to the following model dependent stringy parameters,
\bea
\label{eq:stringy-parameters}
& & \hskip-1cm -\chi({\rm CY})|W_0|^2 \simeq 1.23, \qquad d = (d_1 d_2 d_3)^{1/3}= 2.2735 \times 10^{-6}\, n_0,
\eea
where for the typical CY threefolds with $\chi(\rm CY) = -{\cal O}(100)$, one would have $|W_0| \simeq {\cal O}(0.1)$. Subsequently the moduli VEVs in this scheme of moduli stabilization are,
\bea
\label{eq:VEVs1}
& & \hskip-1cm \langle \phi^1 \rangle = 5.70398, \qquad \langle \phi^2 \rangle = \frac{1}{6} \ln\left(\frac{d_1}{d_2}\right), \quad \langle \phi^3 \rangle = \frac{1}{6 \sqrt{3}} \ln\left(\frac{d_1d_2}{d_3^2}\right),
\eea
and if we take $d_1 = d_2 = d_3 = d$, and for $n_0 = 1$ in (\ref{eq:volume-form}) we have an isotropic moduli stabilization with the following,
\bea
\label{eq:VEVs2}
& & \hskip-1.5cm g_s = 0.316, \qquad \langle \tau_\alpha \rangle = 105.349, \qquad \langle {\cal V} \rangle \simeq 1081.31, \qquad \frac{m_{\phi^1}}{m_{\phi^{2}}} = 0.0844882 =\frac{m_{\phi^1}}{m_{\phi^{3}}}, \nonumber
\eea
which corresponds to $\langle t^\alpha \rangle = 10.264$ for the two-cycle moduli in this isotropic moduli stabilisation. The inflationary potential is shown in Fig.~\ref{fig3} while the corresponding slow-roll parameters are plotted in Fig.~\ref{fig4}. Further, the inflaton shift during inflation $\Delta\phi = \phi^\ast - \phi_{min} = 0.0119345$ suggesting that it is a small-field inflation. In this regard it may be worth mentioning that several concerns appearing for the large field models are naturally avoided for the current model, e.g. see \cite{Cicoli:2017axo,Cicoli:2018tcq} along the lines of swampland implications \cite{Ooguri:2006in,Grimm:2018ohb,Blumenhagen:2018nts,Corvilain:2018lgw}. Moreover, it turns out that $\epsilon_V^\ast \simeq 2.42 \times 10^{-6}$ and therefore leads to the tensor-to-scalar ratio being $r = 16 \epsilon_V^\ast \simeq 3.88 \times 10^{-5}$ which is a typical outcome of the small-field inflation.

To conclude, the model realised in perturbative LVS is indeed a large volume and weak coupling model, which usually guarantees the robustness of the single-field inflationary potential against the various possible (un-)known corrections, which we will discuss in the concrete global setting in the next subsection.


\subsection{Global embedding}
\label{sec_global}
In this subsection we aim to provide a global embedding of the inflationary model discussed in the previous section. The main idea is to use a concrete CY orientifold with toroidal-like volume form \cite{Leontaris:2022rzj}, which could result in an inflationary potential at the leading order and  we explore all the possible sub-leading corrections.  Subsequently we investigate their impact on the robustness of the inflationary dynamics in the next section.

\subsubsection*{Finding the CY threefold}
Let us start by presenting an explicit CY threefold example which possesses a toroidal-like volume as given in Eq.~(\ref{eq:volume-form}). The main motivation for the same follows from the proposal of \cite{Antoniadis:2018hqy,Antoniadis:2018ngr,Antoniadis:2019doc,Antoniadis:2019rkh,Antoniadis:2020ryh,Antoniadis:2020stf, Antoniadis:2021lhi} where some symmetries between the various volume moduli were needed for the setting of the overall mechanism. For this purpose, we have scanned the CY dataset of Kreuzer-Skarke \cite{Kreuzer:2000xy} with $h^{1,1} = 3$ and have found that there are at least two geometries which could suitably give this volume form \cite{Gao:2013pra,Leontaris:2022rzj}.  One such CY threefold corresponding to the polytope Id: 249 in the CY database of \cite{Altman:2014bfa} can be defined by the following toric data:
\begin{center}
\begin{tabular}{|c|ccccccc|}
\hline
\cellcolor[gray]{0.9}Hyp &\cellcolor[gray]{0.9} $x_1$  &\cellcolor[gray]{0.9} $x_2$  &\cellcolor[gray]{0.9} $x_3$  &\cellcolor[gray]{0.9} $x_4$  &\cellcolor[gray]{0.9} $x_5$ & \cellcolor[gray]{0.9}$x_6$  &\cellcolor[gray]{0.9} $x_7$       \\
\hline
\cellcolor[gray]{0.9}4 & 0  & 0 & 1 & 1 & 0 & 0  & 2   \\
\cellcolor[gray]{0.9}4 & 0  & 1 & 0 & 0 & 1 & 0  & 2   \\
\cellcolor[gray]{0.9}4 & 1  & 0 & 0 & 0 & 0 & 1  & 2   \\
\hline
& $K3$  & $K3$ & $K3$ &  $K3$ & $K3$ & $K3$  &  SD  \\
\hline
\end{tabular}
\end{center}
The Hodge numbers are $(h^{2,1}, h^{1,1}) = (115, 3)$, the Euler number is $\chi=-224$ and the Stanley-Reisner ideal is:
\be
{\rm SR} =  \{x_1 x_6, \, x_2 x_5, \, x_3 x_4 x_7 \} \,. \nn
\ee
This CY threefold was also considered for exploring odd-moduli and exchange of non-trivially identical divisors in \cite{Gao:2013pra}. Moreover, a del-Pezzo upgraded version of this example which corresponds to a CY threefold with $h^{1,1}=4$ has been considered in for chiral global embedding of Fibre inflation model in \cite{Cicoli:2017axo}.

The analysis of the divisor topologies using {\it cohomCalg} \cite{Blumenhagen:2010pv, Blumenhagen:2011xn} shows that they can be represented by the following Hodge diamonds:
\bea
K3 &\equiv& \begin{tabular}{ccccc}
    & & 1 & & \\
   & 0 & & 0 & \\
  1 & & 20 & & 1 \\
   & 0 & & 0 & \\
    & & 1 & & \\
  \end{tabular}, \qquad \quad {\rm SD} \equiv \begin{tabular}{ccccc}
    & & 1 & & \\
   & 0 & & 0 & \\
  27 & & 184 & & 27 \\
   & 0 & & 0 & \\
    & & 1 & & \\
  \end{tabular}.
\eea
Considering the basis of smooth divisors $\{D_1, D_2, D_3\}$ we get the following intersection polynomial which has just one non-zero classical triple intersection number \footnote{There is another CY threefold in the database of \cite{Altman:2014bfa} which has the intersection polynomial of the form $I_3 = D_1 D_2 D_3$, however that CY threefold (corresponding to the polytope Id: 52) has non-trivial fundamental group.}:
\bea
& & I_3 = 2\, D_1\, D_2\, D_3,
\eea
while the second Chern-class of the CY is given by,
\bea
c_2({\rm CY}) = 5 D_3^2+12 D_1 D_2 + 12 D_2 D_3+12 D_1 D_3.
\eea
Subsequently, considering the K\"ahler form $J = \sum_{\alpha =1}^3 t^\alpha D_\alpha$, the overall volume and the 4-cycle volume moduli can be given as follows:
\bea
& & \hskip-1cm {\cal V} = 2\, t^1\, t^2\, t^3, \qquad \qquad \tau_1 = 2\, t^2 t^3,  \quad  \tau_2 = 2\, t^1 t^3, \quad  \tau_3 = 2 \,t^1 t^2 \,.
\label{Taus}
\eea
This volume form can also be expressed in the following form:
\bea
& & \hskip-1cm {\cal V} = 2 \, t^1\, t^2\, t^3 = t^1 \tau_1 = t^2 \tau_2 = t^3 \tau_3 = \frac{1}{\sqrt{2}}\,\sqrt{\tau_1 \, \tau_2\, \tau_3}~.
\eea
This confirms that the volume form ${\cal V}$ is indeed like a toroidal case with an exchange symmetry $1 \leftrightarrow 2 \leftrightarrow 3$ under which all the three $K3$ divisors which are part of the basis are exchanged. Further, the K\"ahler cone for this setup is described by the conditions below,
\bea
\label{KahCone}
\text{K\"ahler cone:} &&  t^1 > 0\,, \quad t^2 > 0\,, \quad t^3 > 0\,.
\eea
We note that the volume form can also be given expressed as,
\bea
\label{eq:t-tau-vol}
{\cal V} = t^1 \, \tau_1 = t^2 \, \tau_2  = t^3 \, \tau_3,
\eea
which means that the transverse distance for the stacks of $D7$-branes wrapping the divisor $D_1$ is given by $t^1$ and similarly $t^2$ is the transverse distance for $D7$-branes wrapping the divisor $D_2$ and so on. These properties about the transverse distances and all the K3 divisors interesting with one another on A ${\mathbb T}^2$ is perfectly like what one has for the toroidal case. These symmetries are consistent with the basis requirement for generating logarithmic string-loop effects as elaborated in \cite{Antoniadis:2018hqy,Antoniadis:2018ngr,Antoniadis:2019doc,Antoniadis:2019rkh,Antoniadis:2020ryh,Antoniadis:2020stf,Antoniadis:2021lhi}.

Moreover, using the general K\"ahler metric expression in Eq. (\ref{eq:simpinvKij-1}) and the classical triple intersection numbers, the tree-level metric takes the following form,
\bea
\label{eq:Kij-tree}
& & K_{\alpha\beta}^0 = \frac{1}{4\, {\cal V}^2} \left(
\begin{array}{ccc}
 (t^1)^2 & 0 & 0 \\
 0 & (t^2)^2 & 0 \\
 0 & 0 & (t^3)^2 \\
\end{array}
\right) = \frac{1}{4} \left(
\begin{array}{ccc}
 (\tau_1)^{-2} & 0 & 0 \\
 0 & (\tau_2)^{-2} & 0 \\
 0 & 0 & (\tau_3)^{-2} \\
\end{array}
\right),
\eea
where we have used (\ref{eq:t-tau-vol}) in the second step.


\subsubsection*{Orientifold involution, fluxes and brane setting}
For a given holomorphic involution, one needs to introduce D3/D7-branes and fluxes in order to cancel all the charges. For example, one can nullify the D7-tadpoles via introducing stacks of $N_a$ D7-branes wrapped around suitable divisors (say $D_a$) and their orientifold images ($D_a^\prime$) such that the following relation holds \cite{Blumenhagen:2008zz}:
\bea
\label{eq:D7tadpole}
& & \sum_a\, N_a \left([D_a] + [D_a^\prime] \right) = 8\, [{\rm O7}]\,.
\eea
Moreover, the presence of D7-branes and O7-planes also contributes to the D3-tadpoles, which, in addition, receive contributions from  $H_3$ and $F_3$ fluxes, D3-branes and O3-planes. The D3-tadpole cancellation condition is given as \cite{Blumenhagen:2008zz}:
\be
N_{\rm D3} + \frac{N_{\rm flux}}{2} + N_{\rm gauge} = \frac{N_{\rm O3}}{4} + \frac{\chi({\rm O7})}{12} + \sum_a\, \frac{N_a \left(\chi(D_a) + \chi(D_a^\prime) \right) }{48}\,,
\label{eq:D3tadpole}
\ee
where $N_{\rm flux} = (2\pi)^{-4} \, (\alpha^\prime)^{-2}\int_X H_3 \wedge F_3$ is the contribution from background fluxes and $N_{\rm gauge} = -\sum_a (8 \pi)^{-2} \int_{D_a}\, {\rm tr}\, {\cal F}_a^2$ is due to D7 worldvolume fluxes. However, for the simple case where D7-tadpoles are cancelled by placing 4 D7-branes (plus their images) on top of an O7-plane, (\ref{eq:D3tadpole}) reduces to the following form:
\be
N_{\rm D3} + \frac{N_{\rm flux}}{2} + N_{\rm gauge} =\frac{N_{\rm O3}}{4} + \frac{\chi({\rm O7})}{4}\,.
\label{eq:D3tadpole1}
\ee
For our CY threefold, we note that there are six equivalent reflection involutions corresponding to flipping first six coordinates, i.e. $x_i \to - x_i$ for each $i \in \{1, 2, .., 6\}$. Each of these involutions result in two O7-planes however there is not much scope of considering D7-stacks wrapping all the divisors of the basis. In addition, the D3 tadpole conditions are quite strict in the sense that RHS of (\ref{eq:D3tadpole1}) results in 12, leaving very little scope for choosing the $F_3/H_3$ fluxes. However considering the involution $x_7 \to - x_7$ leads to the better possibilities for brane setting. This results in only one fixed point set with $\{O7 = D_7\}$ along with no $O3$-planes, and subsequently one can consider the following brane setting having 3 stacks of $D7$-branes wrapping each of the three divisors $\{D_1, D_2, D_3\}$ in the basis,
\bea
& & 8\, [O_7] = 8 \left([D_1] + [D_1^\prime] \right) + 8 \left([D_2] + [D_2^\prime] \right) + 8 \left([D_3] + [D_3^\prime] \right)\,,
\eea
along with the $D3$ tadpole cancellation condition being given as
\be
N_{\rm D3} + \frac{N_{\rm flux}}{2} + N_{\rm gauge} = 0 + \frac{240}{12} + 8 + 8 + 8 = 44\,,
\ee
which unlike the other six reflection involutions results in some flexibility with turning on background flux as well as the gauge flux. In fact, if the D7-tadpole cancellation condition is satisfied by placing four D7-branes on top of the O7-plane, the string loop corrections to the scalar potential may turn out to be very simple. We shall therefore focus on a slightly more complicate D7-brane setup which gives rise also to winding loop effects. This can be achieved by placing D7-branes not entirely on top of the O7-plane

In order to obtain a chiral visible sector on the D7-brane stacks wrapping $D_1$, $D_2$ and $D_3$ we need to turn on worldvolume gauge fluxes of the form:
\be
{\cal F}_i = \sum_{j=1}^{h^{1,1}} f_{ij}\hat{D}_j + \frac12 \hat{D}_i - \iota_{D_i}^*B \quad\text{with}\quad f_{ij}\in \mathbb{Z} \quad\text{and}\quad i=1,2,3\,,
\ee
where the half-integer contribution is due to Freed-Witten anomaly cancellation \cite{Minasian:1997mm,Freed:1999vc}
. However, given that the three stacks of D7-branes are wrapping an spin divisor K3 with $c_1({\rm K3}) = 0$, and given that all the intersection numbers for this CY threefolds are even, the pullback of the $B$-field on any divisor $D_\alpha$ does not generate an half-integer flux, and therefore one can appropriately adjust fluxes such that ${\cal F}_\alpha \in {\mathbb Z}$ for all $\alpha = 1, 2, 3$. We shall therefore consider a non-vanishing gauge flux on the worldvolume of $D_\alpha$ of the form:
\be
{\cal F}_\alpha = \sum_{j=1}^{3} f_{\alpha j}\hat{D}_j \quad\text{with}\quad f_{\alpha j}\in \mathbb{Z}\,.
\ee
Let us note that in our present concrete CY construction the choice of orientifold involution which leads to having three stacks of $D7$-branes intersecting at three ${\mathbb T}^2$'s is such that there are no $O3$-planes present, and therefore anti-$D3$ uplifting proposal of \cite{Cicoli:2015ylx,Crino:2020qwk, AbdusSalam:2022krp} is not directly applicable   to our case. However, $D$-term uplifting and the $T$-brane uplifting processes are applicable to this model. in fact, turning on non-trivial gauge flux appropriately on each of the three stacks of D7-branes, one generates the following FI-parameter,
\be
\xi_\alpha = \frac{1}{4\pi {\cal V}}\, \int_{D_\alpha} {\cal F} \wedge J = - \frac{i}{2\pi} \sum_\beta q_{\alpha\beta} \,\partial_{T_\beta}K, \quad {\rm where} \quad q_{\alpha\beta} = \int_{\rm CY} D_\alpha \wedge D_\beta \wedge {\cal F}.
\ee
Subsequently, for our CY threefold which has $\kappa_{123} = 2$ as the only non-trivial triple intersection number, the D-term induced contributions to the scalar potential can be given as,
\bea
\label{eq:VDup1}
& & \hskip-2cm V_{\rm D} \propto \sum_{\alpha =1}^3 \left[\frac{1}{\tau_\alpha} \left(\sum_{\beta \neq \alpha}\, q_{\alpha\beta}\, \frac{\partial K}{\partial \tau_\beta}\right)^2 \right] \simeq \sum_{\alpha =1}^3 \frac{\hat{d}_\alpha}{f_\alpha^{(3)}},
\eea
where $f_\alpha^{(3)}$ denotes some homogeneous cubic polynomial in generic four-cycle volume $\tau_\beta$. This can be further simplified to take the following explicit form,
\bea
\label{eq:VDup2}
& & V_{\rm D} = \frac{d_1}{\tau_1} \left(\frac{q_{12}}{\tau_2} + \frac{q_{13}}{\tau_3}\right)^2 + \frac{d_2}{\tau_2} \left(\frac{q_{21}}{\tau_1} + \frac{q_{23}}{\tau_3}\right)^2 + \frac{d_3}{\tau_3} \left(\frac{q_{31}}{\tau_1} + \frac{q_{32}}{\tau_2}\right)^2.
\eea
Notice that this form of D-term potential is quite complicated as compared to the simple form in Eq.~(\ref{eq:VDup0}), and therefore it would be interesting to see if one can still manage to extract an effective single-field potential by minimising two of the three-moduli using these D-term effects.


\subsubsection*{Scalar potential with sub-leading corrections}
Given that there are no rigid divisors present, a priori this setup will not receive non-perturbative superpotential contributions from instanton or gaugino condensation. In fact because of the very same reason this CY could be naively considered to be not well suited for doing phenomenology in the conventional sense, given that both the popular moduli stabilisation schemes (namely KKLT and LVS) which are available, make use of non-perturbative correction in the superpotential for stabilising the K\"ahler moduli.

Moreover the divisor intersection curves are given in table \ref{Tab1} which shows that all the three $D7$-brane stacks intersect at ${\mathbb T}^2$ while each of those intersect the $O7$-plane on a curve ${\cal H}_9$ defined by $h^{0,0} = 1$ and $h^{1,0} = 9$. These properties about the transverse distances and the divisor interesting on ${\mathbb T}^2$ is perfectly like what one has for the toroidal case, though the divisors are $K3$ for the current situation unlike ${\mathbb T}^4$ divisors of the six-torus. These symmetries are consistent with the basic requirement for generating logarithmic string-loop effects as elaborated in \cite{Antoniadis:2018hqy,Antoniadis:2018ngr,Antoniadis:2019doc,Antoniadis:2019rkh,Antoniadis:2020ryh,Antoniadis:2020stf,Antoniadis:2021lhi}.

Further we note that there are no non-intersection $D7$-brane stacks and the $O7$-planes along with no $O3$-planes present as well, and therefore this model does not induce the KK-type string-loop corrections to the K\"ahler potential. However, given the fact that each of the three $D7$-brane stacks as well as $O7$-plane intersect one another on non-contractible curves (e.g. see Table \ref{Tab1}), one will have string-loop effects of the winding-type to be given as below,
\bea
\label{eq:Vgs-Winding-globalmodel}
& & V_{g_s}^{\rm W} = - \frac{\kappa\,|W_0|^2}{{\cal V}^3} \left(\frac{C_1^W}{t^1} + \frac{C_2^W}{t^2} +\frac{C_3^W}{t^3} +\frac{C_4^W}{2(t^1+t^2)} +\frac{C_5^W}{2(t^2+t^3)} +\frac{C_6^W}{2(t^3+t^1)} \right)\,,
\eea
where $C_i^W$'s are complex-structure moduli dependent quantities and can be taken as parameter for the moduli dynamics of the sub-leading effects. Note that we have used the volume of a given two-torus $t^\alpha_\cap$ at the intersection locus of any two $D7$-brane stacks as given below,
\bea
& & \hskip-1.5cm \int_{CY} J \wedge D_1 \wedge D_2 = 2t^3, \quad \int_{CY} J \wedge D_2 \wedge D_3 = 2t^1, \quad \int_{CY} J \wedge D_3 \wedge D_1 = 2t^2,
\eea
where the K\"ahler form is taken as $J = t^1 D_1 + t^2 D_2 + t^3 D_3$. Further, the size of curves at the intersection of $O7$-plane with the 3 $D7$-brane stacks are given as,
\bea
& & \hskip-1cm \int_{CY} J \wedge [O7] \wedge D_1= 4 (t^2 + t^3),\int_{CY} J \wedge [O7] \wedge D_2 = 4 (t^1 + t^3), \int_{CY} J \wedge [O7] \wedge D_3= 4 (t^1 + t^2). \nonumber
\eea

\begin{table}[h]
  \centering
 \begin{tabular}{|c|c|c|c|c|c|c|c|}
\hline
\cellcolor[gray]{0.9}  &\cellcolor[gray]{0.9} $D_1$  &\cellcolor[gray]{0.9} $D_2$  &\cellcolor[gray]{0.9} $D_3$  & \cellcolor[gray]{0.9}$D_4$  & \cellcolor[gray]{0.9}$D_5$ &\cellcolor[gray]{0.9} $D_6$  & \cellcolor[gray]{0.9}$D_7$  \\
    \hline
		\hline
\cellcolor[gray]{0.9}$D_1$ & $\emptyset$  &  ${\mathbb T}^2$      &  ${\mathbb T}^2$        &  ${\mathbb T}^2$   &  ${\mathbb T}^2$  &  $\emptyset$   &  ${\cal H}_9$  \\
\cellcolor[gray]{0.9}$D_2$ &  ${\mathbb T}^2$ & $\emptyset$        &  ${\mathbb T}^2$        &  ${\mathbb T}^2$   &    $\emptyset$   & ${\mathbb T}^2$  & ${\cal H}_9$
\\
\cellcolor[gray]{0.9}$D_3$  &  ${\mathbb T}^2$      &  ${\mathbb T}^2$        & $\emptyset$ & $\emptyset$ &  ${\mathbb T}^2$   &  ${\mathbb T}^2$   &  ${\cal H}_9$
\\
\cellcolor[gray]{0.9}$D_4$  &  ${\mathbb T}^2$      &  ${\mathbb T}^2$        & $\emptyset$ & $\emptyset$ &  ${\mathbb T}^2$   &  ${\mathbb T}^2$   &  ${\cal H}_9$
\\
\cellcolor[gray]{0.9}$D_5$ &  ${\mathbb T}^2$ & $\emptyset$        &  ${\mathbb T}^2$        &  ${\mathbb T}^2$   &    $\emptyset$   & ${\mathbb T}^2$  & ${\cal H}_9$
\\
\cellcolor[gray]{0.9}$D_6$ & $\emptyset$  &  ${\mathbb T}^2$      &  ${\mathbb T}^2$        &  ${\mathbb T}^2$   &  ${\mathbb T}^2$  &  $\emptyset$   &  ${\cal H}_9$  \\
\cellcolor[gray]{0.9}$D_7$ & ${\cal H}_9$  &  ${\cal H}_9$      &  ${\cal H}_9$        &  ${\cal H}_9$   &  ${\cal H}_9$  &  ${\cal H}_9$   &  ${\cal H}_{97}$
\\
    \hline
  \end{tabular}
  \caption{Intersection curves of the two coordinate divisors. Here ${\cal H}_g$ denotes a curve with Hodge numbers $h^{0,0} = 1$ and $h^{1,0} = g$, and hence ${\cal H}_1 \equiv {\mathbb T}^2$, while ${\cal H}_0 \equiv {\mathbb P}^1$.}
\label{Tab1}
\end{table}

\begin{table}[h]
  \centering
 \begin{tabular}{|c|c|c|c|c|c|c|c|}
\hline
 \cellcolor[gray]{0.9} &\cellcolor[gray]{0.9} $D_1$  &\cellcolor[gray]{0.9} $D_2$  & \cellcolor[gray]{0.9}$D_3$  &\cellcolor[gray]{0.9} $D_4$  &\cellcolor[gray]{0.9} $D_5$ & \cellcolor[gray]{0.9}$D_6$  &\cellcolor[gray]{0.9} $D_7$  \\
    \hline
		\hline
\cellcolor[gray]{0.9}$D_1$  & 0 & 2 ${t^3}$ & 2 ${t^2}$ & 2 ${t^2}$ & 2 ${t^3}$ & 0 & 4 ${t^2}$+4 ${t^3}$ \\
\cellcolor[gray]{0.9}$D_2$  &  & 0 & 2 ${t^1}$ & 2 ${t^1}$ & 0 & 2 ${t^3}$ & 4 ${t^1}$+4 ${t^3}$ \\
\cellcolor[gray]{0.9}$D_3$  & & & 0 & 0 & 2 ${t^1}$ & 2 ${t^2}$ & 4 ${t^1}$+4 ${t^2}$ \\
\cellcolor[gray]{0.9}$D_4$  & & &  & 0 & 2 ${t^1}$ & 2 ${t^2}$ & 4 ${t^1}$+4 ${t^2}$ \\
\cellcolor[gray]{0.9}$D_5$  & && & & 0 & 2 ${t^3}$ & 4 ${t^1}$+4 ${t^3}$ \\
\cellcolor[gray]{0.9}$D_6$  & &&& & & 0 & 4 ${t^2}$+4 ${t^3}$ \\
\cellcolor[gray]{0.9}$D_7$  & &&&&& & 16 $(t^1 + t^2 + t^3)$ \\
    \hline
  \end{tabular}
  \caption{Volume of the two-cycles at the intersection local of the two coordinate divisors $D_i$ presented in Table \ref{Tab1}. This shows, for example, that the curve intersecting at divisors $D_1$ and $D_2$ has a volume along $t^3$, like in the usual toroidal scenarios. Also, this table is symmetrical and lower left entries can be read-off from the right upper sector.}
\label{Tab2}
\end{table}

\noindent
Moreover let us note that the topological quantities $\Pi_\alpha$'s appearing in the higher derivative $F^4$ corrections are given as,
\bea
& & \Pi_\alpha = 24 \quad \forall \, \alpha \in \{1, 2,..,6\}; \quad \Pi_7 = 124.
\eea
Thus,  we observe that although this CY have several properties like a toroidal case, the divisor being $K3$ implies their corresponding $\Pi = 24$ unlike the ${\mathbb T}^4$ case which has a vanishing $\Pi$, and hence no such higher derivative effects. Subsequently we find the following corrections to the scalar potential ,
\bea
\label{eq:F^4-term-globalmodel}
& & V_{{\rm F}^4} = - \frac{\lambda\,\kappa^2\,|W_0|^4}{g_s^{3/2} {\cal V}^4}\, 24 \, \left(t^1 + t^2 + t^3\right).
\eea

\subsubsection*{Summarising the scalar potential pieces}

Finally, combining all the perturbative effects collected so far, namely the BBHL's $(\alpha^\prime)^3$ corrections \cite{Becker:2002nn}, the perturbative string-loop effects of \cite{Antoniadis:2018hqy} as well as the higher derivative $F^4$ corrections of \cite{Ciupke:2015msa}, a master formula for perturbative scalar potential using Gukov-Vafa-Witten's flux superpotential $W_0$ can be generically obtained as the following,
\bea
\label{eq:masterVgen}
& & \hskip-0.75cm V_{\rm tot} = V_{\rm D} + V_{\alpha^\prime + {\rm log} \, g_s} + V_{g_s}^{\rm KK} + V_{g_s}^{\rm W} +V_{{\rm F}^4} + \dots \\
& & \simeq V_{\rm D} + \frac{\kappa}{{\cal Y}^2} \biggl[\frac{3\, {\cal V} }{2\, {\cal Y}^2} \left(1 + \frac{\partial {\cal Y}_1}{\partial{\cal V}}\right)^2 \left(6 {\cal V} \tilde{\cal P}_3 - \tilde{\cal P}_4 \right) - 3 \biggr]\, |W_0|^2 \nonumber\\
& & +\kappa \, g_s^2 \, \frac{|W_0|^2}{16\,{\cal V}^4} \sum_{\alpha,\beta} C_\alpha^{\rm KK} C_\beta^{\rm KK} \left(2\,t^\alpha t^\beta - 4\, {\cal V} \,k^{\alpha\beta}\right) -2 \kappa \frac{|W_0|^2}{{\cal V}^3} \, \sum_\alpha \frac{C_\alpha^W}{t^\alpha_\cap} \nonumber\\
& &  -  \frac{\lambda\,\kappa^2\,|W_0|^4}{g_s^{3/2} {\cal V}^4} \Pi_\alpha \, t^\alpha + \dots \,, \nonumber
\eea
Now for the purpose of moduli stabilisation and subsequently exploring the possibility of inflation, we use the master formula (\ref{eq:masterVgen}) to get a simplified version of the scalar potential given as below,
\bea
\label{eq:Vfinal-simp}
& & \hskip-1cm V_{\rm tot} = \biggl[\frac{d_1}{\tau_1} \left(\frac{q_{12}}{\tau_2} + \frac{q_{13}}{\tau_3}\right)^2 + \frac{d_2}{\tau_2} \left(\frac{q_{21}}{\tau_1} + \frac{q_{23}}{\tau_3}\right)^2 + \frac{d_3}{\tau_3} \left(\frac{q_{31}}{\tau_1} + \frac{q_{32}}{\tau_2}\right)^2\biggr] + \, {\cal C}_1 \, \left(\frac{\hat\xi - 4\,\hat\eta + 2\,\hat\eta \, \ln{\cal V}}{{\cal V}^3}\right) \nonumber\\
& &  \hskip-0.5cm + \frac{{\cal C}_2}{{\cal V}^4} \left(\tau_1 + \tau_2 + \tau_3 + \frac{\tau_1 \tau_2}{2(\tau_1 + \tau_2)} + \frac{\tau_2 \tau_3}{2(\tau_2 + \tau_3)} + \frac{\tau_3 \tau_1}{2(\tau_3 + \tau_1)}\right) \, +  \frac{{\cal C}_3}{{\cal V}^3}\,\left(\frac{1}{\tau_1} + \frac{1}{\tau_2}+\frac{1}{\tau_3} \right) \\
& & \hskip-0.5cm + \, 6 \, {\cal C}_1 \left(\frac{3 \hat{\eta } \hat{\xi }+4 \hat{\eta }^2+\hat{\xi }^2-2 \hat{\eta } \hat{\xi} \ln{\cal V} -2 \hat{\eta }^2 \, \ln{\cal V}}{{\cal V}^4}\right)   + {\cal O}({\cal V}^{-n}) + \dots \,, \qquad n > 4; \nonumber
\eea
where we used the relation (\ref{eq:t-tau-vol}) in the second line, and the various coefficients ${\cal C}_i$'s are given as,
\bea
\label{eq:calCis}
& & \hskip-1cm {\cal C}_1 = \frac{3\,\kappa\, |W_0|^2}{4}, \quad {\cal C}_2 = \frac{4\, {\cal C}_1\,  {\cal C}_w}{3}, \quad {\cal C}_3 = - \frac{24\, \lambda\,\kappa^2\, |W_0|^4}{g_s^{3/2}}, \quad |\lambda| =  \, {\cal O}(10^{-4}), \quad \kappa = \frac{g_s}{8\pi}\cdot
\eea
Note that we have set $e^{K_{cs}} =1$, and for simplicity we set the $C_i^W$ parameters as $C_1^W = C_2^W = C_3^W = C_4^W = C_5^W = C_6^W = -\, {\cal C}_w$ which is compatible with the  our current interest of isotropic moduli stabilisation. For the current global model candidate, the Euler characteristic is: $\chi({\rm CY}) = -224$, and $\Pi_\alpha = 24\, \, \forall \alpha \in \{1,2,3\}$ corresponding to the three K3 divisors of the underlying CY threefold. Further, using Eq.~(\ref{eq:def-xi-eta}) we have,
\bea
& & \hskip-1cm \hat\xi 
= \frac{14\, \zeta[3]}{\pi^3\, g_s^{3/2}}, \qquad \hat\eta 
= - \frac{14 \sqrt{g_s} \zeta[2]}{\pi^3}, \qquad \frac{\hat\xi}{\hat\eta} = -\frac{\zeta[3]}{\zeta[2]\, g_s^2}\,.
\eea
With simplified version of the scalar potential (\ref{eq:Vfinal-simp}) we are now in a position to perform the study of moduli stabilisation. However, before doing that let us note the following points about the collection of terms presented in (\ref{eq:Vfinal-simp}):

\begin{itemize}

\item
The first block in the first line captures the leading most contributions from the D-term which appear at ${\cal O}({\cal V}^{-2})$ in the terms of the volume scaling.

\item
The second block of the first line captures the leading order contributions from the piece $V_{\alpha^\prime + {\rm log} \, g_s}$ which appears at ${\cal O}({\cal V}^{-3})$ in the large volume expansion.

\item
The first piece of the second line presents the typical winding type string-loop effects which appears at ${\cal O}({\cal V}^{-10/3})$ in the large volume expansion. In fact there can be additional loop corrections motivated by the field theoretic computations \cite{vonGersdorff:2005bf,Gao:2022uop}, however we do not include those corrections in the current analysis.

\item
The second piece of the second line presents the higher derivative F$^4$ corrections which appears at ${\cal O}({\cal V}^{-11/3})$ in the large volume expansion.

\item
The third line represents corrections of order ${\cal O}({\cal V}^{-4})$ and smaller. If the dominant sub-leading corrects do not spoil the inflation, one may expect that the corrections of ${\cal O}({\cal V}^{-4})$ and lower in volume scalings should not affect the inflationary dynamics.

\end{itemize}


\section{Robustness against sub-leading corrections}
\label{sec_robustness}


The single-field inflationary dynamics we have discussed so far corresponds to the overall volume modulus, and is driven by a combination of {\it log-loop} effects \cite{Antoniadis:2018hqy} and the BBHL corrections \cite{Becker:2002nn}. Given that additional contributions to the scalar potential exist, e.g. those arising from the  higher derivative F$^4$-corrections \cite{Ciupke:2015msa} as well as the other types of string-loop effects \cite{Berg:2005ja,Berg:2007wt,vonGersdorff:2005bf,Cicoli:2007xp}, after our global embedding proposal it would be interesting to investigate the robustness of the inflationary dynamics, at least against the known sub-leading corrections. For this purpose we will consider the following formulation of the scalar potential (\ref{eq:Vfinal-simp}) expressed in terms of the canonical normalised fields $\phi^\alpha$ defined through the Eqs.~(\ref{eq:cononical-varphi0})-(\ref{eq:cononical-varphi1}),
\bea
\label{eq:Vgen1}
& & \hskip-1cm V = V_{1} + V_{2} + V_{3} + V_4 + \dots,
\eea
where $\dots$ denotes additional sub-leading corrections appearing at ${\cal O}({\cal V}^{-4})$ or more suppressed, while the four pieces are explicitly given as below,
\bea
\label{eq:Vgen2}
& & \hskip-1cmV_{1} = \frac{d_1}{\tau_1} \left(\frac{q_{12}}{\tau_2} + \frac{q_{13}}{\tau_3}\right)^2 + \frac{d_2}{\tau_2} \left(\frac{q_{21}}{\tau_1} + \frac{q_{23}}{\tau_3}\right)^2 + \frac{d_3}{\tau_3} \left(\frac{q_{31}}{\tau_1} + \frac{q_{32}}{\tau_2}\right)^2 \\
& & \hskip-0.5cm = e^{-\sqrt{6}\phi^1}\, \biggl[d_1 e^{-\, \phi^2 - \sqrt{3} \, \phi^3} \left(q_{12} \,e^{\phi^2} + q_{13} \, e^{\sqrt{3} \phi^3} \right)^2 \nonumber\\
& & + d_2 e^{-\, \phi^2 - \sqrt{3} \, \phi^3} \left(q_{21} + q_{23} \, e^{\phi^2 + \sqrt{3} \phi^3} \right)^2 + d_3 e^{-2\, \phi^2} \left(q_{31} \,+ q_{32} \, e^{2\phi^2} \right)^2 \biggr], \nonumber\\
& & \hskip-1cm V_2 = {\cal C}_1 \left(\frac{\hat\xi - 4\,\hat\eta + 2\,\hat\eta \, \ln{\cal V}}{{\cal V}^3}\right) = 2\, \hat\eta\, e^{-3\sqrt{\frac{3}{2}} \phi^1}\, n_0^{3/2} \, {\cal C}_1\, \left(\sqrt{\frac{3}{2}} \, \phi^1 -a_2 + \frac{1}{3} \right), \nonumber\\
& & \hskip-1cm V_3 = \frac{{\cal C}_2}{{\cal V}^4} \left(\tau_1 + \tau_2 + \tau_3 + \frac{\tau_1 \tau_2}{2(\tau_1 + \tau_2)} + \frac{\tau_2 \tau_3}{2(\tau_2 + \tau_3)} + \frac{\tau_3 \tau_1}{2(\tau_3 + \tau_1)}\right) \nonumber\\
& & \hskip-0.5cm = n_0^2 \, {\cal C}_2 \,e^{-5\sqrt{\frac{2}{3}} \phi^1} \biggl(e^{-\frac{2}{\sqrt{3}} \phi^3}+ e^{-\phi^2 + \frac{1}{\sqrt{3}}\phi^3} + e^{\phi^2 + \frac{1}{\sqrt{3}}\phi^3} + \frac{e^{\phi^2 + \frac{1}{\sqrt{3}}\phi^3}}{2(1+\,e^{2\phi^2})} \nonumber\\
& & \hskip-0.0cm + \frac{e^{\frac{1}{\sqrt{3}}\phi^3}}{2(e^{\phi^2}+e^{\sqrt{3} \phi^3})} + \frac{e^{\phi^2 + \frac{1}{\sqrt{3}}\phi^3}}{2(1+\,e^{\phi^2+\sqrt3 \phi^3})}  \biggr), \nonumber\\
& & \hskip-1cm V_4 =  \frac{{\cal C}_3}{{\cal V}^3} \, \left(\frac{1}{\tau_1} + \frac{1}{\tau_2} + \frac{1}{\tau_3}\right) = n_0^{3/2} \, {\cal C}_3 \,e^{-\frac{11}{\sqrt{6}} \phi^1}  \left(e^{-\phi^2 - \frac{1}{\sqrt{3}}\phi^3} + e^{\phi^2 - \frac{1}{\sqrt{3}}\phi^3} + e^{\frac{2}{\sqrt{3}} \phi^3}\right). \nonumber
\eea
Here we have used the definition of $a_2$ as defined earlier in Eq.~(\ref{eq:a1-a2}). Using the generic scalar potential in Eqs.~(\ref{eq:Vgen1})-(\ref{eq:Vgen2}), the three extremisation conditions arising from $\partial_{\phi^\alpha} V = 0$ can be consistently solved to create a desired mass-hierarchy leading to a single-field effective potential.


\subsection{Revisiting moduli stabilisation and mass hierarchy}


As seen from Eqs.~(\ref{eq:Vgen1})-(\ref{eq:Vgen2}), unlike the previous case where the D-term contributions were given by a very simple expression as shown in the Eq.~(\ref{eq:VDup0}), the same turn out to be quite complicated in our explicit global model, and therefore it is not obvious whether a single-field potential with the overall volume-modulus unfixed can be consistently extracted out of a collection of pieces in $V_1$ of (\ref{eq:Vgen2}) which heavily depends on all the three-fields to begin with. However it has been shown in \cite{Leontaris:2022rzj} that utilising the symmetries of the underlying CY threefold, it is possible to self-consistently perform an {\it isotropic moduli stabilisation} by considering appropriate flux dependent parameters $d_\alpha$ and  $q_{\alpha\beta}$. The form of the potential $V_1$ in Eq.~(\ref{eq:Vgen2}) is suggestive that the volume modulus $\phi^1$ which appears just as an overall factor and remains unfixed, can be used as an inflaton candidate. Moreover, the functional dependence of $V_1$ on the $\phi^1$ modulus remains the same as what we had earlier for the D-terms in Eq.~(\ref{eq:VDup0}). So the only task to reproduce the previous scenario is to fix the $\phi^2$ and $\phi^3$ moduli at their minimum such that the qualitative form of the single field inflationary potential is recovered. It turns out that the following set of constraints on the model dependent parameters
\bea
\label{eq:d-qs1}
& & d_2 = d_1 \frac{q_{12}^2-q_{13}^2}{q_{23}^2-q_{21}^2} > 0, \qquad d_3 = d_1 \frac{q_{12}^2-q_{13}^2}{q_{31}^2-q_{32}^2} > 0
\eea
consistently solve the three extremisation conditions such that moduli VEVs are determined from the following relations,
\bea
\label{eq:V-task2-3}
& & \hskip-1cm d_1 =  - {\cal Q}\,\biggl[\hat\eta \, {\cal C}_1\, n_0^{3/2}\, e^{-\sqrt{\frac{3}{2}} \langle\phi^1\rangle} \left(\sqrt{\frac{3}{2}} \, \langle \phi^1 \rangle -a_2\right) +\frac{25}{12} \, n_0^2 \, {\cal C}_2 \, e^{-2\sqrt{\frac{2}{3}} \langle\phi^1\rangle} \\
& & \hskip-0.5cm  + \frac{11}{6} \, n_0^{3/2} \, {\cal C}_3 \, e^{-\frac{5}{\sqrt6} \langle\phi^1\rangle}\biggr], \qquad \langle \phi^2 \rangle = 0 = \langle \phi^3 \rangle, \nonumber
\eea
where ${\cal Q} \neq 0$ is a ratio depending on the flux parameters $q_{\alpha\beta}$'s which can be given as,
\bea
\label{eq:Rq}
& & {\cal Q}^{-1} = \frac{(q_{12}+q_{13})}{3(q_{21}-q_{23})(q_{31}-q_{32})}\biggl(q_{13}q_{21}(q_{31}-3q_{32})+q_{12}q_{23}(q_{32}-3q_{31})\\
& & \hskip2cm + (q_{12}q_{21} + q_{13}q_{23}) (q_{31}+q_{32}) \biggr). \nonumber
\eea
Notice that the constraints in (\ref{eq:V-task2-3}) qualitatively reduce to the previous case as in (\ref{eq:dSextrema}), i.e. without the sub-leading corrections ${\cal C}_2 = 0 = {\cal C}_3$.

However, let us also note that imposing the constraint (\ref{eq:d-qs1})  results in the off-diagonal terms in the Hessian, and hence generically leads to complicated eigenvalue expressions. For simplicity arguments and  in connection with the previous analysis we impose another set of constraint on the $q_{\alpha\beta}$ parameters :
\bea
\label{eq:q-cond}
& & q_{12} = q_{23} = q_{31}, \qquad q_{21} = q_{13} = q_{32}, \qquad q_{12} \neq \pm q_{21},
\eea
which results in following isotropic-like conditions,
\bea
\label{eq:Rq-d-qs2}
& & {\cal Q} = \frac{1}{(q_{12} + q_{21})^2} \neq 0, \qquad d_2 = d_1 > 0, \qquad d_3 = d_1 > 0.
\eea
Subsequently, the Hessian is diagonal and the non-trivial components evaluated at the minimum are given as,
\bea
\label{eq:HessdS-gen}
& & \hskip-2cm \langle {V}_{11} \rangle =  -\,9\, \hat\eta\,n_0^{3/2}\, {\cal C}_1 \, e^{-3\,\sqrt{\frac{3}{2}} \langle\phi^1\rangle} \left(1 + a_2 - \sqrt{\frac{3}{2}}\, \langle \phi^1 \rangle \right) \\
& & + 25\, n_0^2\, {\cal C}_2 \, e^{-5\,\sqrt{\frac{2}{3}} \langle\phi^1\rangle} + \frac{55}{2}\, n_0^{3/2}\, {\cal C}_3 \, e^{-\frac{11}{\sqrt6} \langle\phi^1\rangle}, \nonumber\\
& & \hskip-2cm \langle {V}_{22} \rangle = -\,6\, \hat\eta\, {\cal Q}\, n_0^{3/2} \,{\cal C}_1 \, e^{-3\,\sqrt{\frac{3}{2}} \langle\phi^1\rangle} \left(\sqrt{\frac{3}{2}}\, \langle \phi^1 \rangle - a_2\right) (q_{12}^2+q_{21}^2) \nonumber\\
& & - \frac{1}{4}\, n_0^2 \, {\cal Q} \,{\cal C}_2 \, e^{-5\,\sqrt{\frac{2}{3}} \langle\phi^1\rangle} (43q_{12}^2 - 14q_{12}q_{21} + 43 q_{21}^2) \nonumber\\
& & - \, n_0^{3/2} \, {\cal Q} \,{\cal C}_3 \, e^{-\frac{11}{\sqrt6} \langle\phi^1\rangle} (9q_{12}^2 - 4 q_{12}q_{21} + 9 q_{21}^2) = \langle {V}_{33} \rangle, \nonumber
\eea
where the first piece corresponds to the leading order inflationary potential while the pieces with parameters ${\cal C}_2$ and ${\cal C}_3$ are due to the inclusion of sub-leading corrections. Now, the mass ratio between the overall volume modulus $\phi^1$ and the remaining two moduli ($\phi^2, \phi^3$) turns out to be rather a complicated expression, however the leading order piece looks like
\bea
\label{eq:mass-hierarchy2}
& & {\cal R}_{\rm hierarchy} \equiv \frac{m^2_{\phi^1}}{m^2_{\phi^\alpha}} = \, \frac{3\, (q_{12}+q_{21})^2\left(1 + a_2 - \sqrt{\frac{3}{2}}\, \langle \phi^1 \rangle \right)}{2\, (q_{12}^2 + q_{21}^2)\left(\sqrt{\frac{3}{2}}\, \langle \phi^1 \rangle - a_2\right)} + \dots \,, \qquad \alpha \in \{2, 3\};
\eea
which is slightly different from the previous condition (\ref{eq:mass-hierarchy0}) due to the richer structure in the D-term potential. The VEV of the potential is given as,
\bea
\label{eq:Vmin-global}
& & \hskip-1.5cm \langle V \rangle = - \hat\eta\, n_0^{3/2}\,{\cal C}_1 \, e^{-3\,\sqrt{\frac{3}{2}} \langle\phi^1\rangle} \left(\sqrt{\frac{3}{2}}\, \langle \phi^1 \rangle - a_2 - \frac{2}{3}\right) - \frac{5}{2}\, n_0^2\,{\cal C}_2 \, e^{-5\sqrt{\frac{2}{3}} \langle \phi^1 \rangle} -\frac{5}{2}\, n_0^{3/2}\,{\cal C}_3 \, e^{-\frac{11}{\sqrt6} \langle \phi^1 \rangle},
\eea
which generalises the earlier expression (\ref{eq:pot-with-a1-a2}). Nevertheless after setting the two moduli at their minimum $\langle \phi^2 \rangle = 0 = \langle \phi^3 \rangle$, the single field effective inflationary potential takes the following form,
\bea
\label{eq:V-phi1-global}
& & \hskip-1.5cm V(\phi^1) = -\, {\cal B} \, e^{-3\sqrt{\frac{3}{2}} \phi^1} \left(\sqrt{\frac{3}{2}} \phi^1 - \frac{3}{2}\, e^{\sqrt{\frac{3}{2}} \phi^1}\, a_1\, (q_{12}+q_{21})^2 \, -a_2 + \frac{1}{3} \right) \\
& & + \frac{15}{4}\, n_0^2\,{\cal C}_2 \, e^{-5\sqrt{\frac{2}{3}} \phi^1} + 3\, n_0^{3/2}\,{\cal C}_3 \, e^{-\frac{11}{\sqrt6} \phi^1}, \quad {\cal B} = -\, 2\, \hat\eta\, n_0^{3/2} \, {\cal C}_1 > 0. \nonumber
\eea
which at leading order is indeed similar to the single-field inflationary potential as (\ref{eq:Vinf}) with the same definitions of $a_1$ and $a_2$ as introduced in (\ref{eq:a1-a2}). In order to fully match the leading order potential with our previous case in (\ref{eq:Vinf}) we set $q_{12} = 1$ and $q_{21} = 0$ which give ${\cal Q} = 1$. Subsequently, using (\ref{eq:V-task2-3}) all the moduli VEVs are given as,
\bea
\label{eq:d-vs-phi1}
& & \hskip-1cm d_1 =  - \hat\eta \, {\cal C}_1\, n_0^{3/2}\, e^{-\sqrt{\frac{3}{2}} \langle\phi^1\rangle} \left(\sqrt{\frac{3}{2}} \, \langle \phi^1 \rangle -a_2\right) +\frac{25}{12} \, n_0^2 \, {\cal C}_2 \, e^{-2\sqrt{\frac{2}{3}} \langle\phi^1\rangle} \\
& & \hskip-0.5cm  + \frac{11}{6} \, n_0^{3/2} \, {\cal C}_3 \, e^{-\frac{5}{\sqrt6} \langle\phi^1\rangle}, \qquad \langle \phi^2 \rangle = 0 = \langle \phi^3 \rangle, \nonumber
\eea


\subsection{Revisiting the inflationary dynamics}


Considering the shifted field $\phi$ as defined in Eq.~(\ref{eq:shiftedphi}), we can rewrite the single-field potential in Eq.~(\ref{eq:V-phi1-global}) in the following form,
\bea
\label{eq:Vinf-global}
& & \hskip-1.5cm V_{\rm inf}(\phi) = -\, \tilde{\cal B} \, e^{-3\sqrt{\frac{3}{2}} \phi} \left(\sqrt{\frac{3}{2}} \phi - \frac{3}{2}\, e^{-x\, + \sqrt{\frac{3}{2}} \phi} + \frac{4}{3} \right) + \tilde{\cal C}_2 \, e^{-5\sqrt{\frac{2}{3}} \phi} + \,\tilde{\cal C}_3 \, e^{-\frac{11}{\sqrt6} \phi},
\eea
where the various coefficients depending on the model dependent parameters $g_s$ and $|W_0|$ and $\lambda$ are given as below,
\bea
\label{eq:tildeBC2C3}
& & \tilde{\cal B} \equiv \tilde{\cal B}(|W_0|,g_s) = -\,\kappa \frac{\chi({\rm CY})\, \sqrt{g_s}\, \, |W_0|^2\,e^{-10-\frac{9 \zeta[3]}{g_s^2\, \pi^2}}}{64\pi} > 0,\\
& & \tilde{\cal C}_2 = \frac{15}{4}\, \kappa \, {\cal C}_w\,|W_0|^2 n_0^{1/3}\, e^{-\frac{100}{9}-\frac{10 \zeta[3]}{g_s^2\, \pi^2}}, \qquad \tilde{\cal C}_3 = -\frac{72\,\kappa^2\, \lambda\, |W_0|^4\, }{g_s^{3/2}\,n_0^{1/3}}\, e^{-\frac{110}{9}-\frac{11 \zeta[3]}{g_s^2\, \pi^2}},\nonumber
\eea
where we set $\kappa \equiv e^{K_{cs}} g_s/(8 \pi) = 1$ and $\lambda$ is typically given as $|\lambda| \simeq {\cal O)}(10^{-4}-10^{-3})$ \cite{Grimm:2017pid,Cicoli:2023njy}. Further, the derivatives and the Hessian (\ref{eq:derivatives-Vinf}) take the following respective forms,
\bea
\label{eq:derivatives-Vinf-global}
& & \hskip-1.5cm \partial_{\phi} V_{\rm inf} = \frac{3\sqrt{3}}{\sqrt{2}}\, \tilde{\cal B} \, e^{-3\sqrt{\frac{3}{2}} \phi} \left(\sqrt{\frac{3}{2}} \phi - \, e^{-\,x\,+ \sqrt{\frac{3}{2}} \phi}\, + 1 \right) -\, 5\sqrt{\frac{2}{3}} \tilde{\cal C}_2 \, e^{-5\sqrt{\frac{2}{3}} \phi} -\frac{11}{\sqrt6} \,\tilde{\cal C}_3 \, e^{-\frac{11}{\sqrt6} \phi}, \\
& & \hskip-1.5cm \partial^2_{\phi} V_{\rm inf} = -\,\frac{27}{2}\, \tilde{\cal B} \, e^{-3\sqrt{\frac{3}{2}} \phi} \left(\sqrt{\frac{3}{2}} \phi - \frac{2}{3}\, e^{- x\,+ \sqrt{\frac{3}{2}} \phi}\, + \frac{2}{3} \right) + \, \frac{50}{3} \tilde{\cal C}_2 \, e^{-5\sqrt{\frac{2}{3}} \phi} +\frac{121}{6} \,\tilde{\cal C}_3 \, e^{-\frac{11}{\sqrt6} \phi}, \nonumber
\eea
where the first terms in both of the pieces of (\ref{eq:derivatives-Vinf-global}) are the same as previously found in (\ref{eq:derivatives-Vinf-shifted}) whereas  the additional terms are due to the sub-leading corrections. Now the four points of interest, namely the two extrema corresponding to the minimum and maximum of the potential and the two inflection points, are to be determined numerically. Moreover the modified expressions of the slow-roll parameters $(\epsilon_V, \eta_V)$ can be easily read-off from the expression of the derivatives (\ref{eq:derivatives-Vinf-global}) and the potential (\ref{eq:Vinf-global}), and we do not aim to write it as the same can be useful only for numerical solutions.

Before coming to the numerical analysis, let us mention that the inflationary potential (\ref{eq:Vinf-global}) basically involves a total of four parameters which control the dynamics of the inflaton modulus $\phi$ and can be relevant for realising cosmological observables, with/without the sub-leading corrections. These parameters are:
\bea
\label{eq:4parameters}
& & x, \qquad \tilde{\cal B}, \qquad \tilde{\cal C}_2, \qquad \tilde{\cal C}_3,
\eea
where we recall that $\tilde{B}$ controls the leading order BBHL and log-loop effects while $\tilde{\cal C}_2$ parameter controls the winding-loop effects, and the $\tilde{\cal C}_3$ parameter determines the higher derivative F$^4$-corrections. In addition, the parameter $x$ controls the uplifting and solely determines the VEV of the $\phi$ modulus in the absence of sub-leading effects, and therefore it also controls the inflaton shift during inflation. These four parameters (\ref{eq:4parameters}) generically depend on the various model dependent `stringy ingredients' such as $\left\{g_s,\, W_0,\, \chi({\rm CY}), \, n_0, \,{\cal C}_w, \, \lambda\right\}$ as seen from Eq.~(\ref{eq:tildeBC2C3}). Moreover, recall that the origin of the $x$ parameter lies in the parameter $a_1$ correlated through $a_1 = e^{-a_2 - x - 1}$. Note that $a_1$ and $a_2$ parameters are defined through (\ref{eq:a1-a2}) which involves the uplifting parameter $d$ that for isotropic moduli stabilisation has been considered to be $d = d_1 = d_2 = d_3$. Given that the parameter $x$ turns out to be very sensitive, we will take a fixed value $x = 10^{-4}$ as we mentioned for the benchmark model in the absence of sub-leading corrections, and subsequently we will  readjust the $W_0$ parameter in order to compensate the minor effects induced from considering the non-trivial values of the $\tilde{\cal C}_2$ and $\tilde{\cal C}_3$ parameters. So instead of starting with fixed values of stringy ingredients $\{W_0,\, C_w, \, \lambda\}$ as independent parameters and adjusting the uplifting through $x$, we will begin with $\{x,\, C_w, \, \lambda\}$ as independent parameters and absorb/readjust the uplifting demand via tuning the $W_0$ parameter.

Having said the above, for demonstrating  the various insights of the inflationary dynamics we take the following model dependent parameters,
\bea
\label{eq:global-model1}
& & \hskip-0.3cm \chi({\rm CY}) = -224, \quad n_0 =2, \quad g_s = \frac{1}{3}, \quad x = 10^{-4},
\eea
which using (\ref{eq:def-xi-eta}) and (\ref{eq:a1-a2}) results in the following,
\bea
\label{eq:global-model2}
& & \hskip-2cm \hat\xi = 2.82024, \quad \hat\eta = -0.428811, \quad a_2 = 5.96834, \quad a_1 \equiv e^{-a_2 -x -1} = 0.00094112. \nonumber 
\eea
Building on this, subsequently we will construct benchmark models via finding suitable values for three parameters $W_0, \, {\cal C}_w$ and $\lambda$ such that the cosmological observables are appropriately produced. Note that the choice of parameters in (\ref{eq:global-model1}) have been made in line with the previous  benchmark model presented in Eq.~(\ref{eq:model-before-global}) corresponding to S11 in Table \ref{tab1}-Table \ref{tab2}.

\begin{figure}
\centering
\includegraphics[scale=0.86]{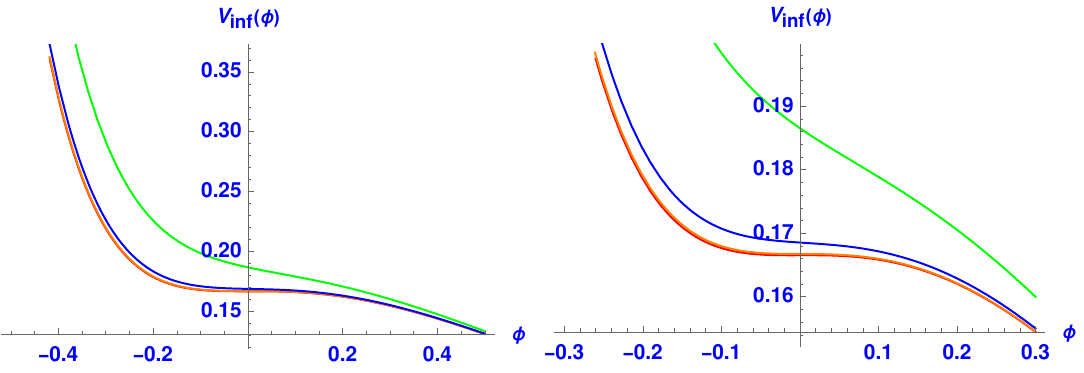}
\caption{Inflationary potential $V_{\rm inf}(\phi)$ given in Eq. (\ref{eq:Vinf-global}) is plotted for $x = 10^{-4}$, $\tilde{\cal B} = 1$ and four sets of $\{{\cal R}_1, {\cal R}_2 \}$ values controlling the sub-leading corrections: $\{0, 0\}, \{0.0001, 0.0001\}, \{0.001, 0.001\}$ and $\{0.01, 0.01\}$. In fact, the plots corresponding to the first two sets almost overlap on each other while the fourth curve on the right side corresponds to the fourth set which looses its minimum around $\phi \simeq 0$, something which is needed for single field approximation estimated in (\ref{eq:mass-hierarchy2}).}
\label{fig5}
\end{figure}

With the strategy as discussed above, we will explore numerical models by setting the string parameters as in (\ref{eq:global-model1}) which further results in,
\bea
\label{eq:global-model3}
& & \tilde{\cal B} = 1.51694 \times 10^{-9}\, |W_0|^2, \qquad \tilde{\cal C}_2 = 1.22570 \times 10^{-9}\, {\cal C}_w\, |W_0|^2, \\
& & \tilde{\cal C}_3 = - 8.47389 \times 10^{-9}\, \lambda \, |W_0|^4. \nonumber
\eea
Thus we need to choose just three parameters, namely $W_0$, ${\cal C}_w$ and $\lambda$ for our model building. Also, for the choice $g_s = 1/3$ which we have set, the ratio of the two coefficients corresponding to the sub-leading corrections included through the coefficients ${\cal C}_2$ and ${\cal C}_3$ are estimated as follows,
\bea
\label{eq:ratio-R1-R2}
& & {\cal R}_1 = \frac{\tilde{\cal C}_2}{\tilde{\cal B}} = 0.80801\, {\cal C}_w, \qquad {\cal R}_2 = \frac{\tilde{\cal C}_3}{\tilde{\cal B}} = -5.58619\,|W_0|^2 \lambda.
\eea
For having some estimates on the range of parameters ${\cal R}_1$ and ${\cal R}_2$ which keeps the minimum of the potential near $\phi\simeq 0$ as needed for the single-field approximation estimated in (\ref{eq:mass-hierarchy2}), we present the behaviour of the inflationary potential $V_{\rm inf}(\phi)$ of Eq.~ (\ref{eq:Vinf-global}) in Figure \ref{fig5}.

Therefore we will need ${\cal C}_w \ll 1$ for ensuring control over the second correction arising from the winding-type string loop effects, and given that they generically depend on the complex structure moduli, one may expect that it can practically be possible via flux tuning. However for the higher derivative F$^4$ corrections we need smaller values for ($W_0^2|\lambda|$) and given that we expect to have $\lambda \simeq -10^{-4}$ in typical models \cite{Grimm:2017pid,Cicoli:2023njy}, such corrections should also be naturally under controlled. We will investigate and demonstrate the relevance of these arguments in explicit numerical models.


\subsection{Numerical analysis}


First let us present a benchmark model without including the corrections, subsequently we will add corrections to revisit the inflationary dynamics. We consider the following benchmark model
\bea
\label{eq:global-M1}
& \hskip-2.5cm {\bf M1:} &  \quad W_0 = 0.07, \qquad {\cal C}_w = 0, \qquad \lambda = 0,
\eea
for which the various VEVs, the cosmological observables and other details are given as below,
\bea
\label{eq:global-M1cosmo}
& \hskip-1.5cm {\bf M1:} & \tilde{\cal B} = 7.43299 \times 10^{-12}, \qquad \tilde{\cal C}_2 = 0 , \qquad \tilde{\cal C}_3 = 0, \\
& & \langle \phi \rangle = -0.0114926, \quad \langle \tau_\alpha \rangle = 103.149, \quad \langle {\cal V} \rangle = 740.771, \quad \langle V \rangle = 1.2377 \times 10^{-12},\nonumber\\
& & m_\phi^2 = 0.0214147 \, m^2_{\phi^\alpha}, \qquad m^2_{\phi^\alpha} = 2.29333 \times 10^{-11}; \quad {\rm for} \quad \alpha \in\{ 2, 3\},\nonumber\\
& & \phi^\ast = 0.00044189, \quad \epsilon_V^\ast = 2.42176\times 10^{-6}, \qquad \eta_V^\ast = -0.0199927, \qquad N_e \simeq 72,\nonumber\\
& & P_s = 2.56 \times 10^{-7},\qquad n_s = 0.96, \qquad r = 3.88 \times 10^{-5}. \nonumber
\eea
The various parameters, cosmological observables and the VEVs of moduli in (\ref{eq:global-M1cosmo}) show that this model is similar to the previous benchmark model presented in Eq.~(\ref{eq:model-before-global}). Now we analyze  the effects of correction and see under what circumstances this models remains robust.

\subsubsection*{Inclusion of winding-loop corrections}
For investigating the cases when the higher derivative F$^4$ corrections are absent, the ratio (\ref{eq:ratio-R1-R2}) shows that ${\cal R}_1 \simeq 0.81 \, {\cal C}_w$, and hence one would need quite small values of ${\cal C}_w$ to keep inflationary dynamics still working. However, we also note that the coefficient ${\cal C}_w$ in the winding-loop corrections is generically a complex structure moduli dependent quantity, and hence can be argued to be possibly tuned to small values, e.g. see cases in \cite{Cicoli:2016xae,Cicoli:2017axo}.
\bea
\label{eq:global-M2}
& \hskip-1.5cm {\bf M2:} & W_0 = 0.039, \qquad {\cal C}_w = 5\cdot10^{-5}, \qquad \lambda = 0,\nonumber\\
& & \\
& & \tilde{\cal B} = 2.30726 \times 10^{-12}, \qquad \tilde{\cal C}_2 = 9.32148 \times 10^{-17}, \qquad \tilde{\cal C}_3 = 0,\nonumber\\
& & \langle \phi \rangle = -0.00849328, \quad \langle \tau_\alpha \rangle = 103.402, \quad \langle {\cal V} \rangle = 743.497, \quad \langle V \rangle = 3.84288 \times 10^{-13},\nonumber\\
& & m_\phi^2 = 0.0158405 \, m^2_{\phi^\alpha}, \qquad m^2_{\phi^\alpha} = 7.0667 \times 10^{-12} \quad {\rm for} \quad \alpha \in\{ 2, 3\},\nonumber\\
& & \phi^\ast = 0.000564736, \quad \epsilon_V^\ast = 7.31547\times 10^{-7}, \qquad \eta_V^\ast = -0.0199978, \qquad N_e \simeq 105,\nonumber\\
& & P_s = 2.63 \times 10^{-7},\qquad n_s = 0.96, \qquad r = 1.17 \times 10^{-5}. \nonumber
\eea

\subsubsection*{Inclusion of higher derivative corrections}
In the absence of winding-type string-loop effects, the parameter ${\cal R}_2$ which controls the higher derivative F$^4$ corrections turns out to be ${\cal R}_2 = -5.586 |W_0|^2 \lambda$ as seen from (\ref{eq:ratio-R1-R2}), which unlike the case of winding-loop effects depends on the $W_0$ parameter and for fractional values of $W_0$ it can help ensuring the robustness, even for larger values of $|\lambda|$. For the typical value of the $\lambda$ parameter as $|\lambda| ={\cal O}\left(10^{-4} -10^{-2}\right)$, where we note that even $\lambda \sim10^{-4}$ could be treated as natural values as argued in \cite{Grimm:2017pid,Cicoli:2023njy}, we present the following two benchmark models with various VEVs, the cosmological observables and other details given as below:
\bea
\label{eq:global-M3}
& \hskip-1.5cm {\bf M3:} & W_0 = 0.068, \qquad {\cal C}_w = 0, \qquad \lambda = -10^{-4},\nonumber\\
& & \\
& & \tilde{\cal B} = 7.01431 \times 10^{-12}, \qquad \tilde{\cal C}_2 = 0, \qquad \tilde{\cal C}_3 = 1.81184 \times 10^{-17},\nonumber\\
& & \langle \phi \rangle = -0.0113063, \quad  \langle \tau_\alpha \rangle = 103.165, \quad \langle {\cal V} \rangle = 740.94, \quad \langle V \rangle = 1.168 \times 10^{-12},\nonumber\\
& & m_\phi^2 = 0.0210708 \, m^2_{\phi^\alpha}, \qquad m^2_{\phi^\alpha} = 2.16317 \times 10^{-11} \quad {\rm for} \quad \alpha \in\{ 2, 3\},\nonumber\\
& & \phi^\ast = 0.000451391, \quad \epsilon_V^\ast = 2.2707\times 10^{-6}, \qquad \eta_V^\ast = -0.0199932, \qquad N_e \simeq 83,\nonumber\\
& & P_s = 2.57 \times 10^{-7},\qquad n_s = 0.96, \qquad r = 3.63 \times 10^{-5}. \nonumber
\eea
\bea
\label{eq:global-M4}
& \hskip-1.2cm {\bf M4:} & W_0 = 0.056, \qquad {\cal C}_w = 0, \qquad \lambda = -10^{-3},\nonumber\\
& & \\
& & \tilde{\cal B} = 4.75711 \times 10^{-12}, \qquad \tilde{\cal C}_2 = 0, \qquad \tilde{\cal C}_3 = 8.3337 \times 10^{-17} ,\nonumber\\
& & \langle \phi \rangle = -0.0101646, \quad \langle \tau_\alpha \rangle = 103.261, \quad \langle {\cal V} \rangle = 741.977, \quad  \langle V \rangle = 7.92212 \times 10^{-13},\nonumber\\
& & m_\phi^2 = 0.0189639 \, m^2_{\phi^\alpha}, \qquad m^2_{\phi^\alpha} = 1.46297 \times 10^{-11} \quad {\rm for} \quad \alpha \in\{ 2, 3\},\nonumber\\
& & \phi^\ast = 0.00050631, \quad \epsilon_V^\ast = 1.4928\times 10^{-6}, \qquad \eta_V^\ast = -0.0199955, \qquad N_e \simeq 87,\nonumber\\
& & P_s = 2.65 \times 10^{-7},\qquad n_s = 0.96, \qquad r = 2.389 \times 10^{-5}. \nonumber
\eea
Let us also recall that the F$^4$ corrections to the scalar potential depends on the second Chern number $c_2(D_i)$ of a divisor, and for some particularly specific cases, some of those can be identically zero depending on the divisor topologies \cite{Cicoli:2023njy,Shukla:2022dhz}.

For the sake of simplicity and clarity, we have presented the impact of the two corrections by adding one at a time. However the argument goes through when  both corrections are included as seen from the model {\bf M5}.
\bea
\label{eq:global-M5}
& \hskip-1.5cm {\bf M5:} & W_0 = 0.038, \qquad {\cal C}_w = 5\cdot10^{-5}, \qquad \lambda = -10^{-4},\nonumber\\
& & \\
& & \tilde{\cal B} = 2.19046 \times 10^{-12}, \qquad \tilde{\cal C}_2 = 8.84958 \times 10^{-17}, \qquad \tilde{\cal C}_3 = 1.76692 \times 10^{-18},\nonumber\\
& & \langle \phi \rangle = -0.00841545, \quad  \langle \tau_\alpha \rangle = 103.409, \quad \langle {\cal V} \rangle = 743.568, \quad \langle V \rangle = 3.64835 \times 10^{-13},\nonumber\\
& & m_\phi^2 = 0.015697 \, m^2_{\phi^\alpha}, \qquad m^2_{\phi^\alpha} = 6.70767 \times 10^{-12} \quad {\rm for} \quad \alpha \in\{ 2, 3\},\nonumber\\
& & \phi^\ast = 0.000567702, \quad \epsilon_V^\ast = 7.05464 \times 10^{-7}, \qquad \eta_V^\ast = -0.0199979, \qquad N_e \simeq 97,\nonumber\\
& & P_s = 2.59 \times 10^{-7},\qquad n_s = 0.96, \qquad r = 1.13 \times 10^{-5}. \nonumber
\eea
In fact, the model {\bf M5} remains similar even for $\lambda = -10^{-3}$ resulting in $W_0 = 0.0334$ giving similar cosmological observables. This shows that the model is robust against higher derivative F$^4$ corrections with the ratio ${\cal R}_2 \sim (10^{-6}-10^{-5})$ which can be thought to be natural values due to a $(2\pi)^4$ factor appearing in the definition of $\lambda$ \cite{Grimm:2017pid,Cicoli:2023njy}. However we find that considering $\tilde{\cal C}_3 \gtrsim 10^{-4}$ may significantly shift the minimum away from being $\langle \phi \rangle \sim 0$, and hence single-field approximation may not be remain valid.

Finally, let us conclude by mentioning that we have used a set of sub-leading corrections in our analysis of realising the inflationary potential, and subsequently for checking its robustness. These corrections are basically the perturbative string-loop effects, and the higher derivative $\alpha^\prime$ corrections induced at the F$^2$-order and the F$^4$-order in the F-term series. Further, in our explicit CY orientifold construction we have also ensured that the typical non-perturbative corrections, for example those arising from the ED3 instantons or gaugino condensation effects on D7-brane via wrapping appropriate rigid divisors, can be naturally forbidden as there are no rigid divisors in the global construction. Here it may be worth mentioning that the prescription of rigidifying the non-rigid divisors by using magnetic fluxes is a delicately designed mechanism; e.g. see \cite{Bianchi:2011qh,Bianchi:2012pn,Louis:2012nb}, and therefore generically one does not expect or find it likely to make the non-rigid divisors automatically contribute to the holomorphic superpotential via ED3 instantons. In addition, generically there can be other sub-leading corrections, e.g. non-perturbative worldsheet instanton corrections to the K\"ahler potential \cite{RoblesLlana:2006is} which can be also motivated by the modular completion arguments \cite{Grimm:2007xm}. However these corrections are expected to appear with an exponential suppression factor in terms of the two-cycle volume moduli $t^\alpha$, and for the current case all the three moduli are `large' size moduli unlike the blow-up moduli in the standard LVS \cite{Balasubramanian:2005zx}, and therefore we expect these corrections to be sub-leading and not affecting the robustness of the inflationary dynamics. Also, these worldsheet instanton corrections are mostly relevant for the models with the so-called odd-moduli arising from the odd (1,1)-cohomology \cite{Grimm:2007xm} which is trivial in our current CY orientifold. Nevertheless there may arise other non-perturbative string-loop corrections to the K\"ahler potential, e.g. having an exponential dependence ($e^{-S}$) on axio-dilaton modulus \cite{Barreiro:1997rp}, however for our approach we still assume that the complex structure moduli and the axio-dilaton are stabilized supersymmetrically via the tree-level flux superpotential, and such corrections, if allowed, should create only a small shift in the VEV of the axio-dilaton. Moreover, the K\"ahler moduli dependence of such non-perturbative corrections to the K\"ahler potential is not clearly understood so far. In fact once the explicit forms of such corrections are known it will be interesting to exploit them for stabilizing the axionic moduli which in the current model remain massless due to the K\"ahler potential respecting the axionic shift symmetry. Subsequently these axionic moduli may have interesting implications for the post-inflationary aspects such as reheating or addressing the issues like dark radiation and dark matter, e.g. on the lines of \cite{Cicoli:2012aq, Cicoli:2023opf}. However addressing these issues in a concrete global model is beyond the scope of the current work.


\section{Summary and conclusions}
\label{sec_conclusions}

In this article, we have presented a global embedding of the inflationary model \cite{Antoniadis:2020stf} in the context of perturbative LVS. In addition we have investigated its robustness against the possible sub-leading corrections to the scalar potential in a given concrete global model. This inflationary model has been originally proposed in the framework of the toroidal orientifold setup \cite{Antoniadis:2018ngr,Antoniadis:2019rkh,Antoniadis:2020stf} in which inflaton field corresponds to the overall volume of the six-torus ${\mathbb T}^6$, and the inflationary potential is induced via a combination of BBHL's perturbative ${\alpha^\prime}^3$ correction \cite{Becker:2002nn} and the so-called {\it log-loop} corrections appearing at the string 1-loop level \cite{Antoniadis:2018hqy,Antoniadis:2019rkh,Antoniadis:2020stf}. These corrections lead to the so-called perturbative LVS in which the volume of the compactifying sixfold is stabilized to exponentially large values in terms of weak string coupling. A global embedding of the perturbative LVS by using a K3-fibred CY threefold has been proposed in \cite{Leontaris:2022rzj}

Continuing with our global embedding proposal of the perturbative LVS in \cite{Leontaris:2022rzj}, first we have revisited the inflationary model of \cite{Antoniadis:2018ngr,Antoniadis:2019rkh} in some good detail, and have presented more insights of the original proposal. In this regard, it is worth noting that while seeking the global embedding we find that the inflationary dynamics is controlled by a single parameter $x$ defined though $a_1 = e^{-a_2-x-1}$ where the two parameters $a_1$ and $a_2$ depend on the stringy parameters, as given in Eq.~(\ref{eq:a1-a2}), which include string-coupling $g_s$, the uplifting parameter $d_\alpha$, magnitude of the flux superpotential $W_0$, Euler character $\chi(\rm CY)$ and the triple intersection number $n_0$ of the compactifying CY threefold. With a constant shift in the canonical field $\phi^1$ corresponding to the overall volume modulus ${\cal V}$, we have observed that the parameter $x$ determines not only the minimum of the potential but also crucially controls the various cosmological observables through the so-called slow-roll parameters $\epsilon_V$ and $\eta_V$. In order to illustrate these main interesting features we have presented a class of candidate models considering a range of $x$ parameter as given in Table \ref{tab1} - Table \ref{tab3}. This collection shows how changing the values of $x$ along with $a_2$ can result in a wide range of $(g_s, \langle {\cal V} \rangle)$ values as well as the hierarchy parameter ${\cal R}_{\rm hierarchy}$ which controls the mass-hierarchy between the inflaton modulus and the remaining two moduli. Moreover, in our collection of candidate models presented in Table \ref{tab1} - Table \ref{tab3}, the model {\bf S10} with $x \simeq 3.3 \times 10^{-4}$ represents the one proposed in \cite{Antoniadis:2020stf} which, in terms of stringy realisation, corresponds to $\langle {\cal V} \rangle \simeq 200$ and $g_s \simeq 0.43$ assuming $n_2 = 1$, i.e. ${\cal V} = t^1 t^2 t^3 = \sqrt{\tau_1\tau_2\tau_3}$. This observation has also suggested that apart from seeking the global embedding, another crucial  aspect for this model could be to investigate its stability against the apparent sub-leading corrections. In this regard, one of the many appealing things about having a concrete global model at hand is the fact that it facilitates the computation of explicit expressions of such sub-leading corrections which can be subsequently used for investigating their impact on the inflationary dynamics.

Having the above strategy in mind, we have initiated to look for the global embedding of the volume modulus inflation in a concrete global construction for perturbative LVS as proposed in \cite{Leontaris:2022rzj}. For that we reviewed the relevant pieces of information corresponding to a K3-fibred CY threefold \cite{Gao:2013pra,Leontaris:2022rzj} which has similar volume-form (\ref{eq:volume-form}) as the one which appears in the conventional toroidal orientifold setup proposed in \cite{Antoniadis:2018ngr,Antoniadis:2019rkh,Antoniadis:2020stf}.  Subsequently we discussed the orientifold involution and the possible brane setting by considering three stacks of D7-brane wrapping the three K3 divisors in the basis of two-forms. In this concrete setting, the D-term contributions to the scalar potential have a quite rich structure depending on all the three moduli in a complicated manner, and it is not a priory clear if the overall volume remains unstabilized by such leading order D-term effects. However we have found that the D-term effects still have just an overall scaling ${\cal V}^2$ in terms of the volume scaling, and subsequently the overall volume modulus can be stabilised after including the BBHL correction along with the log-loop string-loop effects, resulting in perturbative LVS with a tachyon-free dS minimum.

Once the global model is constructed with all the necessary ingredients, we find that there is a series of explicit corrections to the scalar potential arising from various sources. For example, one can ensure the presence of generic winding-type string loop corrections even though there are no KK-type string-loop corrections due to the criteria presented in \cite{Berg:2004ek,Berg:2007wt}. Moreover, the higher derivative F$^4$-corrections \cite{Ciupke:2015msa} are inevitable as unlike the ${\mathbb T}^4$ divisors of the toroidal construction, the three K3 divisors have non-zero second Chern numbers, namely $\Pi_\alpha = 24$ for each of the three K3 divisors wrapping the three stacks of D7-branes. To test the robustness of the inflationary model we conducted a thorough re-examination of the details
regarding the analytics of the moduli stabilization and the single-field approach.  Subsequently we performed a meticulous numerical analysis and  found that the simple inflationary proposal of \cite{Antoniadis:2020stf} can be quite robust against these two types of corrections, modulo certain conditions are fulfilled.

For the winding-loop corrections one needs some tuned values of the coefficients ${\cal C}_w \lesssim 10^{-4}$ which might be achieved given that ${\cal C}_w$ can generically depend on the complex structure moduli. Further, given the fact that for our isotropic moduli stabilisation, we have $\langle {\cal V} \rangle \simeq 740$ and $g_s = 1/3$, which means to have, not too large volume and not too weak string coupling, and therefore it is not surprising that winding-type string loop corrections may get important for larger values of ${\cal C}_w$ parameter. This can also be understood from the fact that they appear at 1-loop level, similar to the {\it log-loop} correction which is among the leading order pieces and is used for realising perturbative LVS. However, unlike the winding-type loop corrections, one could argue that F$^4$-corrections are part of a different series expanded in terms of F-terms and hence are likely to be sub-leading as compared to the leading order F$^2$-contributions, and the inflationary model remain robust against such F$^4$-corrections because the coefficient $\lambda$ controlling such corrections can be argued to be suppressed by a factor $(2\pi)^4$ \cite{Grimm:2017pid,Cicoli:2023njy}, similar to the BBHL's ${\alpha^\prime}^3$ corrections being suppressed by a factor $(2 \pi)^3$ appearing in the explicit expression as $\hat\xi = - \frac{\chi({\rm CY})\, \zeta[3]}{2 (2 \pi)^3\, g_s^3}$.

Although one can consistently realise the necessary inflationary observables in this simple construction, the cosmological constant is relatively much higher than what is the current observational value, and for that purpose an additional open-string modulus has been used as waterfall direction \cite{Antoniadis:2021lhi}. It would be interesting to investigate the possibility of embedding this idea in our current global model by considering open-string moduli. In addition, let us mention that there are additional string-loop effects, including another one of {\it log-loop} type, which has been recently computed through a field theoretic approach in \cite{Gao:2022uop} and it would be interesting to test the robustness of this inflationary proposal against such corrections as well. Given the nice features of this global toroidal-like construction, it would be interesting to realise Fibre inflation by supporting a chiral visible sector, and we plan to present this analysis in a companion work \cite{Bera:2024sbx}.


\section*{Acknowledgments}
We would like to thank Xin Gao and Arthur Hebecker for useful discussions. PS would like to thank the {\it Department of Science and Technology (DST), India} for the kind support.


\appendix




\bibliographystyle{JHEP}
\bibliography{reference}

\providecommand{\href}[2]{#2}\begingroup\raggedright\begin{thebibliography}{10}

\bibitem{Cicoli:2023opf}
M.~Cicoli, J.~P. Conlon, A.~Maharana, S.~Parameswaran, F.~Quevedo and
  I.~Zavala, \emph{{String cosmology: From the early universe to today}},
  \href{https://doi.org/10.1016/j.physrep.2024.01.002}{\emph{Phys. Rept.}
  {\bfseries 1059} (2024) 1--155},
  [\href{https://arxiv.org/abs/2303.04819}{{\ttfamily 2303.04819}}].

\bibitem{McAllister:2023vgy}
L.~McAllister and F.~Quevedo, \emph{{Moduli Stabilization in String Theory}},
  \href{https://arxiv.org/abs/2310.20559}{{\ttfamily 2310.20559}}.

\bibitem{Witten:1996bn}
E.~Witten, \emph{{Nonperturbative superpotentials in string theory}},
  \href{https://doi.org/10.1016/0550-3213(96)00283-0}{\emph{Nucl. Phys. B}
  {\bfseries 474} (1996) 343--360},
  [\href{https://arxiv.org/abs/hep-th/9604030}{{\ttfamily hep-th/9604030}}].

\bibitem{Green:1997di}
M.~B. Green and P.~Vanhove, \emph{{D instantons, strings and M theory}},
  \href{https://doi.org/10.1016/S0370-2693(97)00785-5}{\emph{Phys. Lett. B}
  {\bfseries 408} (1997) 122--134},
  [\href{https://arxiv.org/abs/hep-th/9704145}{{\ttfamily hep-th/9704145}}].

\bibitem{Blumenhagen:2009qh}
R.~Blumenhagen, M.~Cvetic, S.~Kachru and T.~Weigand, \emph{{D-Brane Instantons
  in Type II Orientifolds}},
  \href{https://doi.org/10.1146/annurev.nucl.010909.083113}{\emph{Ann. Rev.
  Nucl. Part. Sci.} {\bfseries 59} (2009) 269--296},
  [\href{https://arxiv.org/abs/0902.3251}{{\ttfamily 0902.3251}}].

\bibitem{Becker:2002nn}
K.~Becker, M.~Becker, M.~Haack and J.~Louis, \emph{{Supersymmetry breaking and
  alpha-prime corrections to flux induced potentials}},
  \href{https://doi.org/10.1088/1126-6708/2002/06/060}{\emph{JHEP} {\bfseries
  06} (2002) 060}, [\href{https://arxiv.org/abs/hep-th/0204254}{{\ttfamily
  hep-th/0204254}}].

\bibitem{Kachru:2002sk}
S.~Kachru, M.~B. Schulz, P.~K. Tripathy and S.~P. Trivedi, \emph{{New
  supersymmetric string compactifications}}, {\emph{JHEP} {\bfseries 0303}
  (2003) 061}, [\href{https://arxiv.org/abs/hep-th/0211182}{{\ttfamily
  hep-th/0211182}}].

\bibitem{Balasubramanian:2005zx}
V.~Balasubramanian, P.~Berglund, J.~P. Conlon and F.~Quevedo,
  \emph{{Systematics of moduli stabilisation in Calabi-Yau flux
  compactifications}},
  \href{https://doi.org/10.1088/1126-6708/2005/03/007}{\emph{JHEP} {\bfseries
  03} (2005) 007}, [\href{https://arxiv.org/abs/hep-th/0502058}{{\ttfamily
  hep-th/0502058}}].

\bibitem{Conlon:2005ki}
J.~P. Conlon, F.~Quevedo and K.~Suruliz, \emph{{Large-volume flux
  compactifications: Moduli spectrum and D3/D7 soft supersymmetry breaking}},
  \href{https://doi.org/10.1088/1126-6708/2005/08/007}{\emph{JHEP} {\bfseries
  08} (2005) 007}, [\href{https://arxiv.org/abs/hep-th/0505076}{{\ttfamily
  hep-th/0505076}}].

\bibitem{Blanco-Pillado:2009dmu}
J.~J. Blanco-Pillado, D.~Buck, E.~J. Copeland, M.~Gomez-Reino and N.~J. Nunes,
  \emph{{Kahler Moduli Inflation Revisited}},
  \href{https://doi.org/10.1007/JHEP01(2010)081}{\emph{JHEP} {\bfseries 01}
  (2010) 081}, [\href{https://arxiv.org/abs/0906.3711}{{\ttfamily 0906.3711}}].

\bibitem{Cicoli:2017shd}
M.~Cicoli, I.~Garcìa-Etxebarria, C.~Mayrhofer, F.~Quevedo, P.~Shukla and
  R.~Valandro, \emph{{Global Orientifolded Quivers with Inflation}},
  \href{https://doi.org/10.1007/JHEP11(2017)134}{\emph{JHEP} {\bfseries 11}
  (2017) 134}, [\href{https://arxiv.org/abs/1706.06128}{{\ttfamily
  1706.06128}}].

\bibitem{Cicoli:2008gp}
M.~Cicoli, C.~P. Burgess and F.~Quevedo, \emph{{Fibre Inflation: Observable
  Gravity Waves from IIB String Compactifications}},
  \href{https://doi.org/10.1088/1475-7516/2009/03/013}{\emph{JCAP} {\bfseries
  0903} (2009) 013}, [\href{https://arxiv.org/abs/0808.0691}{{\ttfamily
  0808.0691}}].

\bibitem{Cicoli:2016chb}
M.~Cicoli, D.~Ciupke, S.~de~Alwis and F.~Muia, \emph{{$\alpha'$ Inflation:
  moduli stabilisation and observable tensors from higher derivatives}},
  \href{https://doi.org/10.1007/JHEP09(2016)026}{\emph{JHEP} {\bfseries 09}
  (2016) 026}, [\href{https://arxiv.org/abs/1607.01395}{{\ttfamily
  1607.01395}}].

\bibitem{Cicoli:2016xae}
M.~Cicoli, F.~Muia and P.~Shukla, \emph{{Global Embedding of Fibre Inflation
  Models}}, \href{https://doi.org/10.1007/JHEP11(2016)182}{\emph{JHEP}
  {\bfseries 11} (2016) 182},
  [\href{https://arxiv.org/abs/1611.04612}{{\ttfamily 1611.04612}}].

\bibitem{Cicoli:2017axo}
M.~Cicoli, D.~Ciupke, V.~A. Diaz, V.~Guidetti, F.~Muia and P.~Shukla,
  \emph{{Chiral Global Embedding of Fibre Inflation Models}},
  \href{https://doi.org/10.1007/JHEP11(2017)207}{\emph{JHEP} {\bfseries 11}
  (2017) 207}, [\href{https://arxiv.org/abs/1709.01518}{{\ttfamily
  1709.01518}}].

\bibitem{Cicoli:2011ct}
M.~Cicoli, F.~G. Pedro and G.~Tasinato, \emph{{Poly-instanton Inflation}},
  \href{https://doi.org/10.1088/1475-7516/2011/12/022}{\emph{JCAP} {\bfseries
  12} (2011) 022}, [\href{https://arxiv.org/abs/1110.6182}{{\ttfamily
  1110.6182}}].

\bibitem{Blumenhagen:2012ue}
R.~Blumenhagen, X.~Gao, T.~Rahn and P.~Shukla, \emph{{Moduli Stabilization and
  Inflationary Cosmology with Poly-Instantons in Type IIB Orientifolds}},
  \href{https://doi.org/10.1007/JHEP11(2012)101}{\emph{JHEP} {\bfseries 11}
  (2012) 101}, [\href{https://arxiv.org/abs/1208.1160}{{\ttfamily 1208.1160}}].

\bibitem{Gao:2013hn}
X.~Gao and P.~Shukla, \emph{{On Non-Gaussianities in Two-Field Poly-Instanton
  Inflation}}, \href{https://doi.org/10.1007/JHEP03(2013)061}{\emph{JHEP}
  {\bfseries 03} (2013) 061},
  [\href{https://arxiv.org/abs/1301.6076}{{\ttfamily 1301.6076}}].

\bibitem{Gao:2014fva}
X.~Gao, T.~Li and P.~Shukla, \emph{{Cosmological observables in multi-field
  inflation with a non-flat field space}},
  \href{https://doi.org/10.1088/1475-7516/2014/10/008}{\emph{JCAP} {\bfseries
  10} (2014) 008}, [\href{https://arxiv.org/abs/1403.0654}{{\ttfamily
  1403.0654}}].

\bibitem{Bansal:2024uzr}
S.~Bansal, L.~Brunelli, M.~Cicoli, A.~Hebecker and R.~Kuespert, \emph{{Loop
  Blow-up Inflation}},  \href{https://arxiv.org/abs/2403.04831}{{\ttfamily
  2403.04831}}.

\bibitem{Bianchi:2011qh}
M.~Bianchi, A.~Collinucci and L.~Martucci, \emph{{Magnetized E3-brane
  instantons in F-theory}},
  \href{https://doi.org/10.1007/JHEP12(2011)045}{\emph{JHEP} {\bfseries 12}
  (2011) 045}, [\href{https://arxiv.org/abs/1107.3732}{{\ttfamily 1107.3732}}].

\bibitem{Bianchi:2012pn}
M.~Bianchi, A.~Collinucci and L.~Martucci, \emph{{Freezing E3-brane instantons
  with fluxes}}, \href{https://doi.org/10.1002/prop.201200030}{\emph{Fortsch.
  Phys.} {\bfseries 60} (2012) 914--920},
  [\href{https://arxiv.org/abs/1202.5045}{{\ttfamily 1202.5045}}].

\bibitem{Louis:2012nb}
J.~Louis, M.~Rummel, R.~Valandro and A.~Westphal, \emph{{Building an explicit
  de Sitter}}, \href{https://doi.org/10.1007/JHEP10(2012)163}{\emph{JHEP}
  {\bfseries 10} (2012) 163},
  [\href{https://arxiv.org/abs/1208.3208}{{\ttfamily 1208.3208}}].

\bibitem{Blumenhagen:2007sm}
R.~Blumenhagen, S.~Moster and E.~Plauschinn, \emph{{Moduli Stabilisation versus
  Chirality for MSSM like Type IIB Orientifolds}},
  \href{https://doi.org/10.1088/1126-6708/2008/01/058}{\emph{JHEP} {\bfseries
  01} (2008) 058}, [\href{https://arxiv.org/abs/0711.3389}{{\ttfamily
  0711.3389}}].

\bibitem{Blumenhagen:2008zz}
R.~Blumenhagen, V.~Braun, T.~W. Grimm and T.~Weigand, \emph{{GUTs in Type IIB
  Orientifold Compactifications}},
  \href{https://doi.org/10.1016/j.nuclphysb.2009.02.011}{\emph{Nucl.Phys.}
  {\bfseries B815} (2009) 1--94},
  [\href{https://arxiv.org/abs/0811.2936}{{\ttfamily 0811.2936}}].

\bibitem{Cvetic:2012ts}
M.~Cvetic, R.~Donagi, J.~Halverson and J.~Marsano, \emph{{On Seven-Brane
  Dependent Instanton Prefactors in F-theory}},
  \href{https://doi.org/10.1007/JHEP11(2012)004}{\emph{JHEP} {\bfseries 11}
  (2012) 004}, [\href{https://arxiv.org/abs/1209.4906}{{\ttfamily 1209.4906}}].

\bibitem{Blumenhagen:2012kz}
R.~Blumenhagen, X.~Gao, T.~Rahn and P.~Shukla, \emph{{A Note on Poly-Instanton
  Effects in Type IIB Orientifolds on Calabi-Yau Threefolds}},
  \href{https://doi.org/10.1007/JHEP06(2012)162}{\emph{JHEP} {\bfseries 06}
  (2012) 162}, [\href{https://arxiv.org/abs/1205.2485}{{\ttfamily 1205.2485}}].

\bibitem{Berg:2004ek}
M.~Berg, M.~Haack and B.~Kors, \emph{{Loop corrections to volume moduli and
  inflation in string theory}},
  \href{https://doi.org/10.1103/PhysRevD.71.026005}{\emph{Phys. Rev.}
  {\bfseries D71} (2005) 026005},
  [\href{https://arxiv.org/abs/hep-th/0404087}{{\ttfamily hep-th/0404087}}].

\bibitem{vonGersdorff:2005bf}
G.~von Gersdorff and A.~Hebecker, \emph{{Kahler corrections for the volume
  modulus of flux compactifications}},
  \href{https://doi.org/10.1016/j.physletb.2005.08.024}{\emph{Phys. Lett. B}
  {\bfseries 624} (2005) 270--274},
  [\href{https://arxiv.org/abs/hep-th/0507131}{{\ttfamily hep-th/0507131}}].

\bibitem{Berg:2005ja}
M.~Berg, M.~Haack and B.~Kors, \emph{{String loop corrections to Kahler
  potentials in orientifolds}},
  \href{https://doi.org/10.1088/1126-6708/2005/11/030}{\emph{JHEP} {\bfseries
  11} (2005) 030}, [\href{https://arxiv.org/abs/hep-th/0508043}{{\ttfamily
  hep-th/0508043}}].

\bibitem{Berg:2005yu}
M.~Berg, M.~Haack and B.~Kors, \emph{{On volume stabilization by quantum
  corrections}},
  \href{https://doi.org/10.1103/PhysRevLett.96.021601}{\emph{Phys. Rev. Lett.}
  {\bfseries 96} (2006) 021601},
  [\href{https://arxiv.org/abs/hep-th/0508171}{{\ttfamily hep-th/0508171}}].

\bibitem{Cicoli:2007xp}
M.~Cicoli, J.~P. Conlon and F.~Quevedo, \emph{{Systematics of String Loop
  Corrections in Type IIB Calabi-Yau Flux Compactifications}},
  \href{https://doi.org/10.1088/1126-6708/2008/01/052}{\emph{JHEP} {\bfseries
  01} (2008) 052}, [\href{https://arxiv.org/abs/0708.1873}{{\ttfamily
  0708.1873}}].

\bibitem{Gao:2022uop}
X.~Gao, A.~Hebecker, S.~Schreyer and G.~Venken, \emph{{Loops, local corrections
  and warping in the LVS and other type IIB models}},
  \href{https://doi.org/10.1007/JHEP09(2022)091}{\emph{JHEP} {\bfseries 09}
  (2022) 091}, [\href{https://arxiv.org/abs/2204.06009}{{\ttfamily
  2204.06009}}].

\bibitem{Ciupke:2015msa}
D.~Ciupke, J.~Louis and A.~Westphal, \emph{{Higher-Derivative Supergravity and
  Moduli Stabilization}},
  \href{https://doi.org/10.1007/JHEP10(2015)094}{\emph{JHEP} {\bfseries 10}
  (2015) 094}, [\href{https://arxiv.org/abs/1505.03092}{{\ttfamily
  1505.03092}}].

\bibitem{Aldazabal:2006up}
G.~Aldazabal, P.~G. Camara, A.~Font and L.~Ibanez, \emph{{More dual fluxes and
  moduli fixing}},
  \href{https://doi.org/10.1088/1126-6708/2006/05/070}{\emph{JHEP} {\bfseries
  0605} (2006) 070}, [\href{https://arxiv.org/abs/hep-th/0602089}{{\ttfamily
  hep-th/0602089}}].

\bibitem{deCarlos:2009fq}
B.~de~Carlos, A.~Guarino and J.~M. Moreno, \emph{{Flux moduli stabilisation,
  Supergravity algebras and no-go theorems}},
  \href{https://doi.org/10.1007/JHEP01(2010)012}{\emph{JHEP} {\bfseries 01}
  (2010) 012}, [\href{https://arxiv.org/abs/0907.5580}{{\ttfamily 0907.5580}}].

\bibitem{deCarlos:2009qm}
B.~de~Carlos, A.~Guarino and J.~M. Moreno, \emph{{Complete classification of
  Minkowski vacua in generalised flux models}},
  \href{https://doi.org/10.1007/JHEP02(2010)076}{\emph{JHEP} {\bfseries 1002}
  (2010) 076}, [\href{https://arxiv.org/abs/0911.2876}{{\ttfamily 0911.2876}}].

\bibitem{Blumenhagen:2015kja}
R.~Blumenhagen, A.~Font, M.~Fuchs, D.~Herschmann, E.~Plauschinn, Y.~Sekiguchi
  et~al., \emph{{A Flux-Scaling Scenario for High-Scale Moduli Stabilization in
  String Theory}},
  \href{https://doi.org/10.1016/j.nuclphysb.2015.06.003}{\emph{Nucl. Phys. B}
  {\bfseries 897} (2015) 500--554},
  [\href{https://arxiv.org/abs/1503.07634}{{\ttfamily 1503.07634}}].

\bibitem{Shukla:2016xdy}
P.~Shukla, \emph{{Revisiting the two formulations of Bianchi identities and
  their implications on moduli stabilization}},
  \href{https://doi.org/10.1007/JHEP08(2016)146}{\emph{JHEP} {\bfseries 08}
  (2016) 146}, [\href{https://arxiv.org/abs/1603.08545}{{\ttfamily
  1603.08545}}].

\bibitem{Plauschinn:2020ram}
E.~Plauschinn, \emph{{Moduli Stabilization with Non-Geometric Fluxes
  \textemdash{} Comments on Tadpole Contributions and de-Sitter Vacua}},
  \href{https://doi.org/10.1002/prop.202100003}{\emph{Fortsch. Phys.}
  {\bfseries 69} (2021) 2100003},
  [\href{https://arxiv.org/abs/2011.08227}{{\ttfamily 2011.08227}}].

\bibitem{Damian:2023ote}
C.~Damian and O.~Loaiza-Brito, \emph{{Galois groups of uplifted de Sitter
  vacua}},  \href{https://arxiv.org/abs/2307.08468}{{\ttfamily 2307.08468}}.

\bibitem{Antoniadis:2018hqy}
I.~Antoniadis, Y.~Chen and G.~K. Leontaris, \emph{{Perturbative moduli
  stabilisation in type IIB/F-theory framework}},
  \href{https://doi.org/10.1140/epjc/s10052-018-6248-4}{\emph{Eur. Phys. J.}
  {\bfseries C78} (2018) 766},
  [\href{https://arxiv.org/abs/1803.08941}{{\ttfamily 1803.08941}}].

\bibitem{Antoniadis:2019rkh}
I.~Antoniadis, Y.~Chen and G.~K. Leontaris, \emph{{Logarithmic loop
  corrections, moduli stabilisation and de Sitter vacua in string theory}},
  \href{https://arxiv.org/abs/1909.10525}{{\ttfamily 1909.10525}}.

\bibitem{Antoniadis:2020ryh}
I.~Antoniadis, Y.~Chen and G.~K. Leontaris, \emph{{String loop corrections and
  de Sitter vacua}}, \href{https://doi.org/10.22323/1.376.0099}{\emph{PoS}
  {\bfseries CORFU2019} (2020) 099}.

\bibitem{Antoniadis:2018ngr}
I.~Antoniadis, Y.~Chen and G.~K. Leontaris, \emph{{Inflation from the internal
  volume in type IIB/F-theory compactification}},
  \href{https://doi.org/10.1142/S0217751X19500428}{\emph{Int. J. Mod. Phys.}
  {\bfseries A34} (2019) 1950042},
  [\href{https://arxiv.org/abs/1810.05060}{{\ttfamily 1810.05060}}].

\bibitem{Antoniadis:2019doc}
I.~Antoniadis, Y.~Chen and G.~K. Leontaris, \emph{{Moduli stabilisation and
  inflation in type IIB/F-theory}},
  \href{https://doi.org/10.22323/1.347.0068}{\emph{PoS} {\bfseries CORFU2018}
  (2019) 068}, [\href{https://arxiv.org/abs/1901.05075}{{\ttfamily
  1901.05075}}].

\bibitem{Antoniadis:2020stf}
I.~Antoniadis, O.~Lacombe and G.~K. Leontaris, \emph{{Inflation near a
  metastable de Sitter vacuum from moduli stabilisation}},
  \href{https://doi.org/10.1140/epjc/s10052-020-08581-9}{\emph{Eur. Phys. J. C}
  {\bfseries 80} (2020) 1014},
  [\href{https://arxiv.org/abs/2007.10362}{{\ttfamily 2007.10362}}].

\bibitem{Antoniadis:2021lhi}
I.~Antoniadis, O.~Lacombe and G.~K. Leontaris, \emph{{Hybrid inflation and
  waterfall field in string theory from D7-branes}},
  \href{https://doi.org/10.1007/JHEP01(2022)011}{\emph{JHEP} {\bfseries 01}
  (2022) 011}, [\href{https://arxiv.org/abs/2109.03243}{{\ttfamily
  2109.03243}}].

\bibitem{Leontaris:2022rzj}
G.~K. Leontaris and P.~Shukla, \emph{{Stabilising all K\"ahler moduli in
  perturbative LVS}},
  \href{https://doi.org/10.1007/JHEP07(2022)047}{\emph{JHEP} {\bfseries 07}
  (2022) 047}, [\href{https://arxiv.org/abs/2203.03362}{{\ttfamily
  2203.03362}}].

\bibitem{Cicoli:2021tzt}
M.~Cicoli, A.~Schachner and P.~Shukla, \emph{{Systematics of type IIB moduli
  stabilisation with odd axions}},
  \href{https://doi.org/10.1007/JHEP04(2022)003}{\emph{JHEP} {\bfseries 04}
  (2022) 003}, [\href{https://arxiv.org/abs/2109.14624}{{\ttfamily
  2109.14624}}].

\bibitem{Gukov:1999ya}
S.~Gukov, C.~Vafa and E.~Witten, \emph{{CFT's from Calabi-Yau four folds}},
  \href{https://doi.org/10.1016/S0550-3213(01)00289-9,
  10.1016/S0550-3213(00)00373-4}{\emph{Nucl. Phys.} {\bfseries B584} (2000)
  69--108}, [\href{https://arxiv.org/abs/hep-th/9906070}{{\ttfamily
  hep-th/9906070}}].

\bibitem{AbdusSalam:2020ywo}
S.~AbdusSalam, S.~Abel, M.~Cicoli, F.~Quevedo and P.~Shukla, \emph{{A
  systematic approach to K\"ahler moduli stabilisation}},
  \href{https://doi.org/10.1007/JHEP08(2020)047}{\emph{JHEP} {\bfseries 08}
  (2020) 047}, [\href{https://arxiv.org/abs/2005.11329}{{\ttfamily
  2005.11329}}].

\bibitem{Crino:2020qwk}
C.~Crin\`o, F.~Quevedo and R.~Valandro, \emph{{On de Sitter String Vacua from
  Anti-D3-Branes in the Large Volume Scenario}},
  \href{https://doi.org/10.1007/JHEP03(2021)258}{\emph{JHEP} {\bfseries 03}
  (2021) 258}, [\href{https://arxiv.org/abs/2010.15903}{{\ttfamily
  2010.15903}}].

\bibitem{Bento:2021nbb}
B.~V. Bento, D.~Chakraborty, S.~L. Parameswaran and I.~Zavala, \emph{{A new de
  Sitter solution with a weakly warped deformed conifold}},
  \href{https://doi.org/10.1007/JHEP12(2021)124}{\emph{JHEP} {\bfseries 12}
  (2021) 124}, [\href{https://arxiv.org/abs/2105.03370}{{\ttfamily
  2105.03370}}].

\bibitem{Burgess:2003ic}
C.~P. Burgess, R.~Kallosh and F.~Quevedo, \emph{{De Sitter string vacua from
  supersymmetric D terms}},
  \href{https://doi.org/10.1088/1126-6708/2003/10/056}{\emph{JHEP} {\bfseries
  10} (2003) 056}, [\href{https://arxiv.org/abs/hep-th/0309187}{{\ttfamily
  hep-th/0309187}}].

\bibitem{Cicoli:2015ylx}
M.~Cicoli, F.~Quevedo and R.~Valandro, \emph{{De Sitter from T-branes}},
  \href{https://doi.org/10.1007/JHEP03(2016)141}{\emph{JHEP} {\bfseries 03}
  (2016) 141}, [\href{https://arxiv.org/abs/1512.04558}{{\ttfamily
  1512.04558}}].

\bibitem{Cicoli:2021dhg}
M.~Cicoli, I.~n.~G. Etxebarria, F.~Quevedo, A.~Schachner, P.~Shukla and
  R.~Valandro, \emph{{The Standard Model quiver in de Sitter string
  compactifications}},
  \href{https://doi.org/10.1007/JHEP08(2021)109}{\emph{JHEP} {\bfseries 08}
  (2021) 109}, [\href{https://arxiv.org/abs/2106.11964}{{\ttfamily
  2106.11964}}].

\bibitem{Giddings:2001yu}
S.~B. Giddings, S.~Kachru and J.~Polchinski, \emph{{Hierarchies from fluxes in
  string compactifications}},
  \href{https://doi.org/10.1103/PhysRevD.66.106006}{\emph{Phys. Rev. D}
  {\bfseries 66} (2002) 106006},
  [\href{https://arxiv.org/abs/hep-th/0105097}{{\ttfamily hep-th/0105097}}].

\bibitem{Gao:2013pra}
X.~Gao and P.~Shukla, \emph{{On Classifying the Divisor Involutions in
  Calabi-Yau Threefolds}},
  \href{https://doi.org/10.1007/JHEP11(2013)170}{\emph{JHEP} {\bfseries 1311}
  (2013) 170}, [\href{https://arxiv.org/abs/1307.1139}{{\ttfamily 1307.1139}}].

\bibitem{Achucarro:2018vey}
A.~Achúcarro and G.~A. Palma, \emph{{The string swampland constraints require
  multi-field inflation}},
  \href{https://doi.org/10.1088/1475-7516/2019/02/041}{\emph{JCAP} {\bfseries
  1902} (2019) 041}, [\href{https://arxiv.org/abs/1807.04390}{{\ttfamily
  1807.04390}}].

\bibitem{Planck:2018jri}
{\scshape Planck} collaboration, Y.~Akrami et~al., \emph{{Planck 2018 results.
  X. Constraints on inflation}},
  \href{https://doi.org/10.1051/0004-6361/201833887}{\emph{Astron. Astrophys.}
  {\bfseries 641} (2020) A10},
  [\href{https://arxiv.org/abs/1807.06211}{{\ttfamily 1807.06211}}].

\bibitem{Cicoli:2018tcq}
M.~Cicoli, D.~Ciupke, C.~Mayrhofer and P.~Shukla, \emph{{A Geometrical Upper
  Bound on the Inflaton Range}},
  \href{https://doi.org/10.1007/JHEP05(2018)001}{\emph{JHEP} {\bfseries 05}
  (2018) 001}, [\href{https://arxiv.org/abs/1801.05434}{{\ttfamily
  1801.05434}}].

\bibitem{Ooguri:2006in}
H.~Ooguri and C.~Vafa, \emph{{On the Geometry of the String Landscape and the
  Swampland}},
  \href{https://doi.org/10.1016/j.nuclphysb.2006.10.033}{\emph{Nucl. Phys. B}
  {\bfseries 766} (2007) 21--33},
  [\href{https://arxiv.org/abs/hep-th/0605264}{{\ttfamily hep-th/0605264}}].

\bibitem{Grimm:2018ohb}
T.~W. Grimm, E.~Palti and I.~Valenzuela, \emph{{Infinite Distances in Field
  Space and Massless Towers of States}},
  \href{https://doi.org/10.1007/JHEP08(2018)143}{\emph{JHEP} {\bfseries 08}
  (2018) 143}, [\href{https://arxiv.org/abs/1802.08264}{{\ttfamily
  1802.08264}}].

\bibitem{Blumenhagen:2018nts}
R.~Blumenhagen, D.~Kl\"awer, L.~Schlechter and F.~Wolf, \emph{{The Refined
  Swampland Distance Conjecture in Calabi-Yau Moduli Spaces}},
  \href{https://doi.org/10.1007/JHEP06(2018)052}{\emph{JHEP} {\bfseries 06}
  (2018) 052}, [\href{https://arxiv.org/abs/1803.04989}{{\ttfamily
  1803.04989}}].

\bibitem{Corvilain:2018lgw}
P.~Corvilain, T.~W. Grimm and I.~Valenzuela, \emph{{The Swampland Distance
  Conjecture for K\"ahler moduli}},
  \href{https://doi.org/10.1007/JHEP08(2019)075}{\emph{JHEP} {\bfseries 08}
  (2019) 075}, [\href{https://arxiv.org/abs/1812.07548}{{\ttfamily
  1812.07548}}].

\bibitem{Kreuzer:2000xy}
M.~Kreuzer and H.~Skarke, \emph{{Complete classification of reflexive polyhedra
  in four-dimensions}},
  \href{https://doi.org/10.4310/ATMP.2000.v4.n6.a2}{\emph{Adv. Theor. Math.
  Phys.} {\bfseries 4} (2000) 1209--1230},
  [\href{https://arxiv.org/abs/hep-th/0002240}{{\ttfamily hep-th/0002240}}].

\bibitem{Altman:2014bfa}
R.~Altman, J.~Gray, Y.-H. He, V.~Jejjala and B.~D. Nelson, \emph{{A Calabi-Yau
  Database: Threefolds Constructed from the Kreuzer-Skarke List}},
  \href{https://doi.org/10.1007/JHEP02(2015)158}{\emph{JHEP} {\bfseries 02}
  (2015) 158}, [\href{https://arxiv.org/abs/1411.1418}{{\ttfamily 1411.1418}}].

\bibitem{Blumenhagen:2010pv}
R.~Blumenhagen, B.~Jurke, T.~Rahn and H.~Roschy, \emph{{Cohomology of Line
  Bundles: A Computational Algorithm}},
  \href{https://doi.org/10.1063/1.3501132, 10.1063/1.3523343}{\emph{J. Math.
  Phys.} {\bfseries 51} (2010) 103525},
  [\href{https://arxiv.org/abs/1003.5217}{{\ttfamily 1003.5217}}].

\bibitem{Blumenhagen:2011xn}
R.~Blumenhagen, B.~Jurke and T.~Rahn, \emph{{Computational Tools for Cohomology
  of Toric Varieties}}, \href{https://doi.org/10.1155/2011/152749}{\emph{Adv.
  High Energy Phys.} {\bfseries 2011} (2011) 152749},
  [\href{https://arxiv.org/abs/1104.1187}{{\ttfamily 1104.1187}}].

\bibitem{Minasian:1997mm}
R.~Minasian and G.~W. Moore, \emph{{K theory and Ramond-Ramond charge}},
  \href{https://doi.org/10.1088/1126-6708/1997/11/002}{\emph{JHEP} {\bfseries
  11} (1997) 002}, [\href{https://arxiv.org/abs/hep-th/9710230}{{\ttfamily
  hep-th/9710230}}].

\bibitem{Freed:1999vc}
D.~S. Freed and E.~Witten, \emph{{Anomalies in string theory with D-branes}},
  {\emph{Asian J. Math.} {\bfseries 3} (1999) 819},
  [\href{https://arxiv.org/abs/hep-th/9907189}{{\ttfamily hep-th/9907189}}].

\bibitem{AbdusSalam:2022krp}
S.~AbdusSalam, C.~Crin\`o and P.~Shukla, \emph{{On K3-fibred LARGE Volume
  Scenario with de Sitter vacua from anti-D3-branes}},
  \href{https://doi.org/10.1007/JHEP03(2023)132}{\emph{JHEP} {\bfseries 03}
  (2023) 132}, [\href{https://arxiv.org/abs/2206.12889}{{\ttfamily
  2206.12889}}].

\bibitem{Berg:2007wt}
M.~Berg, M.~Haack and E.~Pajer, \emph{{Jumping Through Loops: On Soft Terms
  from Large Volume Compactifications}},
  \href{https://doi.org/10.1088/1126-6708/2007/09/031}{\emph{JHEP} {\bfseries
  09} (2007) 031}, [\href{https://arxiv.org/abs/0704.0737}{{\ttfamily
  0704.0737}}].

\bibitem{Grimm:2017pid}
T.~W. Grimm, K.~Mayer and M.~Weissenbacher, \emph{{One-modulus Calabi-Yau
  fourfold reductions with higher-derivative terms}},
  \href{https://doi.org/10.1007/JHEP04(2018)021}{\emph{JHEP} {\bfseries 04}
  (2018) 021}, [\href{https://arxiv.org/abs/1712.07074}{{\ttfamily
  1712.07074}}].

\bibitem{Cicoli:2023njy}
M.~Cicoli, M.~Licheri, P.~Piantadosi, F.~Quevedo and P.~Shukla, \emph{{Higher
  derivative corrections to string inflation}},
  \href{https://doi.org/10.1007/JHEP02(2024)115}{\emph{JHEP} {\bfseries 02}
  (2024) 115}, [\href{https://arxiv.org/abs/2309.11697}{{\ttfamily
  2309.11697}}].

\bibitem{Shukla:2022dhz}
P.~Shukla, \emph{{Classifying divisor topologies for string phenomenology}},
  \href{https://doi.org/10.1007/JHEP12(2022)055}{\emph{JHEP} {\bfseries 12}
  (2022) 055}, [\href{https://arxiv.org/abs/2205.05215}{{\ttfamily
  2205.05215}}].

\bibitem{RoblesLlana:2006is}
D.~Robles-Llana, M.~Rocek, F.~Saueressig, U.~Theis and S.~Vandoren,
  \emph{{Nonperturbative corrections to 4D string theory effective actions from
  SL(2,Z) duality and supersymmetry}},
  \href{https://doi.org/10.1103/PhysRevLett.98.211602}{\emph{Phys. Rev. Lett.}
  {\bfseries 98} (2007) 211602},
  [\href{https://arxiv.org/abs/hep-th/0612027}{{\ttfamily hep-th/0612027}}].

\bibitem{Grimm:2007xm}
T.~W. Grimm, \emph{{Non-Perturbative Corrections and Modularity in N=1 Type IIB
  Compactifications}},
  \href{https://doi.org/10.1088/1126-6708/2007/10/004}{\emph{JHEP} {\bfseries
  0710} (2007) 004}, [\href{https://arxiv.org/abs/0705.3253}{{\ttfamily
  0705.3253}}].

\bibitem{Barreiro:1997rp}
T.~Barreiro, B.~de~Carlos and E.~J. Copeland, \emph{{On nonperturbative
  corrections to the Kahler potential}},
  \href{https://doi.org/10.1103/PhysRevD.57.7354}{\emph{Phys. Rev. D}
  {\bfseries 57} (1998) 7354--7360},
  [\href{https://arxiv.org/abs/hep-ph/9712443}{{\ttfamily hep-ph/9712443}}].

\bibitem{Cicoli:2012aq}
M.~Cicoli, J.~P. Conlon and F.~Quevedo, \emph{{Dark radiation in LARGE volume
  models}}, \href{https://doi.org/10.1103/PhysRevD.87.043520}{\emph{Phys. Rev.
  D} {\bfseries 87} (2013) 043520},
  [\href{https://arxiv.org/abs/1208.3562}{{\ttfamily 1208.3562}}].

\bibitem{Bera:2024sbx}
S.~Bera, D.~Chakraborty, G.~K. Leontaris and P.~Shukla, \emph{{Global Embedding
  of Fibre Inflation in Perturbative LVS}},
  \href{https://arxiv.org/abs/2406.01694}{{\ttfamily 2406.01694}}.

\end{thebibliography}\endgroup


\end{document}